\documentclass[longauth]{aa}
\usepackage{csquotes}
\usepackage{graphicx}
\usepackage{natbib}
\usepackage{scalerel}

\usepackage[table]{xcolor}

\usepackage{txfonts}
\usepackage[pdfencoding=auto,psdextra]{hyperref}
\hypersetup{
    colorlinks=true,
    linkcolor=blue,
    filecolor=magenta,      
    urlcolor=blue,
    citecolor=blue
}
\makeatletter
\renewcommand*\aa@pageof{, page \thepage{} of \pageref*{LastPage}}
\makeatother

\usepackage[utf8]{inputenc}

\usepackage[switch, modulo]{lineno}

\usepackage{euclid}

\def\photoz{photo-\textit{z}}
\def\specz{spec-\textit{z}}

\def\lephare{\texttt{LePhare}}
\def\eazy{\texttt{EAZY}}
\def\tractor{\texttt{The Tractor}}
\def\farmer{\texttt{The Farmer}}

\def\SEP{\texttt{SEP}}
\def\chOne{[3.6$\mu$m]}
\def\chTwo{[4.5$\mu$m]}

\begin{document}
%
%
   \title{\Euclid Preparation.}
    \subtitle{Cosmic Dawn Survey: Data release 1 multiwavelength catalogues for Euclid Deep Field North and Euclid Deep Field Fornax}

\newcommand{\orcid}[1]{} 
\author{Euclid Collaboration: L.~Zalesky\orcid{0000-0001-5680-2326}\thanks{\email{zalesky@hawaii.edu}}\inst{\ref{aff1}}
\and C.~J.~R.~McPartland\orcid{0000-0003-0639-025X}\inst{\ref{aff2},\ref{aff3},\ref{aff1},\ref{aff4},\ref{aff5}}
\and J.~R.~Weaver\orcid{0000-0003-1614-196X}\inst{\ref{aff6}}
\and S.~Toft\orcid{0000-0003-3631-7176}\inst{\ref{aff2},\ref{aff3}}
\and D.~B.~Sanders\orcid{0000-0002-1233-9998}\inst{\ref{aff1}}
\and B.~Mobasher\orcid{0000-0001-5846-4404}\inst{\ref{aff4}}
\and N.~Suzuki\orcid{0000-0001-7266-930X}\inst{\ref{aff7}}
\and I.~Szapudi\orcid{0000-0003-2274-0301}\inst{\ref{aff1}}
\and I.~Valdes\inst{\ref{aff1}}
\and G.~Murphree\orcid{0009-0007-7266-8914}\inst{\ref{aff1}}
\and N.~Chartab\orcid{0000-0003-3691-937X}\inst{\ref{aff8}}
\and N.~Allen\orcid{0000-0001-9610-7950}\inst{\ref{aff2}}
\and S.~Taamoli\orcid{0000-0003-0749-4667}\inst{\ref{aff4}}
\and S.~W.~J.~Barrow\orcid{0009-0001-9835-3650}\inst{\ref{aff6},\ref{aff1}}
\and O.~Ch\'{a}vez~Ortiz\orcid{0000-0003-2332-5505}\inst{\ref{aff9}}
\and S.~L.~Finkelstein\orcid{0000-0001-8519-1130}\inst{\ref{aff9}}
\and S.~Gwyn\orcid{0000-0001-8221-8406}\inst{\ref{aff10}}
\and M.~Sawicki\orcid{0000-0002-7712-7857}\inst{\ref{aff11}}
\and H.~J.~McCracken\orcid{0000-0002-9489-7765}\inst{\ref{aff12}}
\and D.~Stern\orcid{0000-0003-2686-9241}\inst{\ref{aff13}}
\and H.~Dannerbauer\orcid{0000-0001-7147-3575}\inst{\ref{aff14}}
\and B.~Altieri\orcid{0000-0003-3936-0284}\inst{\ref{aff15}}
\and S.~Andreon\orcid{0000-0002-2041-8784}\inst{\ref{aff16}}
\and N.~Auricchio\orcid{0000-0003-4444-8651}\inst{\ref{aff17}}
\and C.~Baccigalupi\orcid{0000-0002-8211-1630}\inst{\ref{aff18},\ref{aff19},\ref{aff20},\ref{aff21}}
\and M.~Baldi\orcid{0000-0003-4145-1943}\inst{\ref{aff22},\ref{aff17},\ref{aff23}}
\and S.~Bardelli\orcid{0000-0002-8900-0298}\inst{\ref{aff17}}
\and R.~Bender\orcid{0000-0001-7179-0626}\inst{\ref{aff24},\ref{aff25}}
\and C.~Bodendorf\inst{\ref{aff24}}
\and D.~Bonino\orcid{0000-0002-3336-9977}\inst{\ref{aff26}}
\and E.~Branchini\orcid{0000-0002-0808-6908}\inst{\ref{aff27},\ref{aff28},\ref{aff16}}
\and M.~Brescia\orcid{0000-0001-9506-5680}\inst{\ref{aff29},\ref{aff30},\ref{aff31}}
\and J.~Brinchmann\orcid{0000-0003-4359-8797}\inst{\ref{aff32}}
\and S.~Camera\orcid{0000-0003-3399-3574}\inst{\ref{aff33},\ref{aff34},\ref{aff26}}
\and V.~Capobianco\orcid{0000-0002-3309-7692}\inst{\ref{aff26}}
\and C.~Carbone\orcid{0000-0003-0125-3563}\inst{\ref{aff35}}
\and J.~Carretero\orcid{0000-0002-3130-0204}\inst{\ref{aff36},\ref{aff37}}
\and S.~Casas\orcid{0000-0002-4751-5138}\inst{\ref{aff38}}
\and F.~J.~Castander\orcid{0000-0001-7316-4573}\inst{\ref{aff39},\ref{aff40}}
\and M.~Castellano\orcid{0000-0001-9875-8263}\inst{\ref{aff41}}
\and G.~Castignani\orcid{0000-0001-6831-0687}\inst{\ref{aff17}}
\and S.~Cavuoti\orcid{0000-0002-3787-4196}\inst{\ref{aff30},\ref{aff31}}
\and A.~Cimatti\inst{\ref{aff42}}
\and C.~Colodro-Conde\inst{\ref{aff43}}
\and G.~Congedo\orcid{0000-0003-2508-0046}\inst{\ref{aff44}}
\and C.~J.~Conselice\orcid{0000-0003-1949-7638}\inst{\ref{aff45}}
\and L.~Conversi\orcid{0000-0002-6710-8476}\inst{\ref{aff46},\ref{aff15}}
\and Y.~Copin\orcid{0000-0002-5317-7518}\inst{\ref{aff47}}
\and L.~Corcione\orcid{0000-0002-6497-5881}\inst{\ref{aff26}}
\and F.~Courbin\orcid{0000-0003-0758-6510}\inst{\ref{aff48}}
\and H.~M.~Courtois\orcid{0000-0003-0509-1776}\inst{\ref{aff49}}
\and A.~Da~Silva\orcid{0000-0002-6385-1609}\inst{\ref{aff50},\ref{aff51}}
\and H.~Degaudenzi\orcid{0000-0002-5887-6799}\inst{\ref{aff52}}
\and G.~De~Lucia\orcid{0000-0002-6220-9104}\inst{\ref{aff19}}
\and A.~M.~Di~Giorgio\orcid{0000-0002-4767-2360}\inst{\ref{aff53}}
\and J.~Dinis\inst{\ref{aff50},\ref{aff51}}
\and F.~Dubath\orcid{0000-0002-6533-2810}\inst{\ref{aff52}}
\and C.~A.~J.~Duncan\inst{\ref{aff45},\ref{aff54}}
\and X.~Dupac\inst{\ref{aff15}}
\and S.~Dusini\orcid{0000-0002-1128-0664}\inst{\ref{aff55}}
\and M.~Farina\orcid{0000-0002-3089-7846}\inst{\ref{aff53}}
\and S.~Farrens\orcid{0000-0002-9594-9387}\inst{\ref{aff56}}
\and S.~Ferriol\inst{\ref{aff47}}
\and S.~Fotopoulou\orcid{0000-0002-9686-254X}\inst{\ref{aff57}}
\and M.~Frailis\orcid{0000-0002-7400-2135}\inst{\ref{aff19}}
\and E.~Franceschi\orcid{0000-0002-0585-6591}\inst{\ref{aff17}}
\and S.~Galeotta\orcid{0000-0002-3748-5115}\inst{\ref{aff19}}
\and B.~Garilli\orcid{0000-0001-7455-8750}\inst{\ref{aff35}}
\and W.~Gillard\orcid{0000-0003-4744-9748}\inst{\ref{aff58}}
\and B.~Gillis\orcid{0000-0002-4478-1270}\inst{\ref{aff44}}
\and C.~Giocoli\orcid{0000-0002-9590-7961}\inst{\ref{aff17},\ref{aff59}}
\and P.~G\'omez-Alvarez\orcid{0000-0002-8594-5358}\inst{\ref{aff60},\ref{aff15}}
\and A.~Grazian\orcid{0000-0002-5688-0663}\inst{\ref{aff61}}
\and F.~Grupp\inst{\ref{aff24},\ref{aff25}}
\and S.~V.~H.~Haugan\orcid{0000-0001-9648-7260}\inst{\ref{aff62}}
\and H.~Hoekstra\orcid{0000-0002-0641-3231}\inst{\ref{aff63}}
\and W.~Holmes\inst{\ref{aff13}}
\and I.~Hook\orcid{0000-0002-2960-978X}\inst{\ref{aff64}}
\and F.~Hormuth\inst{\ref{aff65}}
\and A.~Hornstrup\orcid{0000-0002-3363-0936}\inst{\ref{aff66},\ref{aff5}}
\and P.~Hudelot\inst{\ref{aff12}}
\and K.~Jahnke\orcid{0000-0003-3804-2137}\inst{\ref{aff67}}
\and B.~Joachimi\orcid{0000-0001-7494-1303}\inst{\ref{aff68}}
\and E.~Keih\"anen\orcid{0000-0003-1804-7715}\inst{\ref{aff69}}
\and S.~Kermiche\orcid{0000-0002-0302-5735}\inst{\ref{aff58}}
\and A.~Kiessling\orcid{0000-0002-2590-1273}\inst{\ref{aff13}}
\and M.~Kilbinger\orcid{0000-0001-9513-7138}\inst{\ref{aff56}}
\and B.~Kubik\orcid{0009-0006-5823-4880}\inst{\ref{aff47}}
\and K.~Kuijken\orcid{0000-0002-3827-0175}\inst{\ref{aff63}}
\and M.~K\"ummel\orcid{0000-0003-2791-2117}\inst{\ref{aff25}}
\and M.~Kunz\orcid{0000-0002-3052-7394}\inst{\ref{aff70}}
\and H.~Kurki-Suonio\orcid{0000-0002-4618-3063}\inst{\ref{aff71},\ref{aff72}}
\and R.~Laureijs\inst{\ref{aff73}}
\and S.~Ligori\orcid{0000-0003-4172-4606}\inst{\ref{aff26}}
\and P.~B.~Lilje\orcid{0000-0003-4324-7794}\inst{\ref{aff62}}
\and V.~Lindholm\orcid{0000-0003-2317-5471}\inst{\ref{aff71},\ref{aff72}}
\and I.~Lloro\inst{\ref{aff74}}
\and G.~Mainetti\inst{\ref{aff75}}
\and D.~Maino\inst{\ref{aff76},\ref{aff35},\ref{aff77}}
\and E.~Maiorano\orcid{0000-0003-2593-4355}\inst{\ref{aff17}}
\and O.~Mansutti\orcid{0000-0001-5758-4658}\inst{\ref{aff19}}
\and O.~Marggraf\orcid{0000-0001-7242-3852}\inst{\ref{aff78}}
\and K.~Markovic\orcid{0000-0001-6764-073X}\inst{\ref{aff13}}
\and M.~Martinelli\orcid{0000-0002-6943-7732}\inst{\ref{aff41},\ref{aff79}}
\and N.~Martinet\orcid{0000-0003-2786-7790}\inst{\ref{aff80}}
\and F.~Marulli\orcid{0000-0002-8850-0303}\inst{\ref{aff81},\ref{aff17},\ref{aff23}}
\and R.~Massey\orcid{0000-0002-6085-3780}\inst{\ref{aff82}}
\and S.~Maurogordato\inst{\ref{aff83}}
\and S.~Mei\orcid{0000-0002-2849-559X}\inst{\ref{aff84}}
\and Y.~Mellier\inst{\ref{aff85},\ref{aff12}}
\and M.~Meneghetti\orcid{0000-0003-1225-7084}\inst{\ref{aff17},\ref{aff23}}
\and E.~Merlin\orcid{0000-0001-6870-8900}\inst{\ref{aff41}}
\and G.~Meylan\inst{\ref{aff48}}
\and M.~Moresco\orcid{0000-0002-7616-7136}\inst{\ref{aff81},\ref{aff17}}
\and L.~Moscardini\orcid{0000-0002-3473-6716}\inst{\ref{aff81},\ref{aff17},\ref{aff23}}
\and E.~Munari\orcid{0000-0002-1751-5946}\inst{\ref{aff19},\ref{aff18}}
\and C.~Neissner\orcid{0000-0001-8524-4968}\inst{\ref{aff86},\ref{aff37}}
\and S.-M.~Niemi\inst{\ref{aff73}}
\and J.~W.~Nightingale\orcid{0000-0002-8987-7401}\inst{\ref{aff87},\ref{aff82}}
\and C.~Padilla\orcid{0000-0001-7951-0166}\inst{\ref{aff86}}
\and S.~Paltani\orcid{0000-0002-8108-9179}\inst{\ref{aff52}}
\and F.~Pasian\orcid{0000-0002-4869-3227}\inst{\ref{aff19}}
\and K.~Pedersen\inst{\ref{aff88}}
\and W.~J.~Percival\orcid{0000-0002-0644-5727}\inst{\ref{aff89},\ref{aff90},\ref{aff91}}
\and V.~Pettorino\inst{\ref{aff73}}
\and S.~Pires\orcid{0000-0002-0249-2104}\inst{\ref{aff56}}
\and G.~Polenta\orcid{0000-0003-4067-9196}\inst{\ref{aff92}}
\and M.~Poncet\inst{\ref{aff93}}
\and L.~A.~Popa\inst{\ref{aff94}}
\and L.~Pozzetti\orcid{0000-0001-7085-0412}\inst{\ref{aff17}}
\and F.~Raison\orcid{0000-0002-7819-6918}\inst{\ref{aff24}}
\and R.~Rebolo\inst{\ref{aff43},\ref{aff95}}
\and A.~Renzi\orcid{0000-0001-9856-1970}\inst{\ref{aff96},\ref{aff55}}
\and J.~Rhodes\inst{\ref{aff13}}
\and G.~Riccio\inst{\ref{aff30}}
\and E.~Romelli\orcid{0000-0003-3069-9222}\inst{\ref{aff19}}
\and M.~Roncarelli\orcid{0000-0001-9587-7822}\inst{\ref{aff17}}
\and E.~Rossetti\orcid{0000-0003-0238-4047}\inst{\ref{aff22}}
\and R.~Saglia\orcid{0000-0003-0378-7032}\inst{\ref{aff25},\ref{aff24}}
\and Z.~Sakr\orcid{0000-0002-4823-3757}\inst{\ref{aff97},\ref{aff98},\ref{aff99}}
\and D.~Sapone\orcid{0000-0001-7089-4503}\inst{\ref{aff100}}
\and R.~Scaramella\orcid{0000-0003-2229-193X}\inst{\ref{aff41},\ref{aff79}}
\and M.~Schirmer\orcid{0000-0003-2568-9994}\inst{\ref{aff67}}
\and P.~Schneider\orcid{0000-0001-8561-2679}\inst{\ref{aff78}}
\and T.~Schrabback\orcid{0000-0002-6987-7834}\inst{\ref{aff101}}
\and A.~Secroun\orcid{0000-0003-0505-3710}\inst{\ref{aff58}}
\and E.~Sefusatti\orcid{0000-0003-0473-1567}\inst{\ref{aff19},\ref{aff18},\ref{aff20}}
\and G.~Seidel\orcid{0000-0003-2907-353X}\inst{\ref{aff67}}
\and S.~Serrano\orcid{0000-0002-0211-2861}\inst{\ref{aff40},\ref{aff39},\ref{aff102}}
\and C.~Sirignano\orcid{0000-0002-0995-7146}\inst{\ref{aff96},\ref{aff55}}
\and G.~Sirri\orcid{0000-0003-2626-2853}\inst{\ref{aff23}}
\and L.~Stanco\orcid{0000-0002-9706-5104}\inst{\ref{aff55}}
\and J.~Steinwagner\inst{\ref{aff24}}
\and P.~Tallada-Cresp\'{i}\orcid{0000-0002-1336-8328}\inst{\ref{aff36},\ref{aff37}}
\and H.~I.~Teplitz\orcid{0000-0002-7064-5424}\inst{\ref{aff103}}
\and I.~Tereno\inst{\ref{aff50},\ref{aff104}}
\and R.~Toledo-Moreo\orcid{0000-0002-2997-4859}\inst{\ref{aff105}}
\and F.~Torradeflot\orcid{0000-0003-1160-1517}\inst{\ref{aff37},\ref{aff36}}
\and I.~Tutusaus\orcid{0000-0002-3199-0399}\inst{\ref{aff98}}
\and E.~A.~Valentijn\inst{\ref{aff106}}
\and L.~Valenziano\orcid{0000-0002-1170-0104}\inst{\ref{aff17},\ref{aff107}}
\and T.~Vassallo\orcid{0000-0001-6512-6358}\inst{\ref{aff25},\ref{aff19}}
\and G.~Verdoes~Kleijn\orcid{0000-0001-5803-2580}\inst{\ref{aff106}}
\and A.~Veropalumbo\orcid{0000-0003-2387-1194}\inst{\ref{aff16},\ref{aff28}}
\and Y.~Wang\orcid{0000-0002-4749-2984}\inst{\ref{aff103}}
\and J.~Weller\orcid{0000-0002-8282-2010}\inst{\ref{aff25},\ref{aff24}}
\and G.~Zamorani\orcid{0000-0002-2318-301X}\inst{\ref{aff17}}
\and E.~Zucca\orcid{0000-0002-5845-8132}\inst{\ref{aff17}}
\and M.~Bolzonella\orcid{0000-0003-3278-4607}\inst{\ref{aff17}}
\and A.~Boucaud\orcid{0000-0001-7387-2633}\inst{\ref{aff84}}
\and E.~Bozzo\orcid{0000-0002-8201-1525}\inst{\ref{aff52}}
\and C.~Burigana\orcid{0000-0002-3005-5796}\inst{\ref{aff108},\ref{aff107}}
\and D.~Di~Ferdinando\inst{\ref{aff23}}
\and J.~A.~Escartin~Vigo\inst{\ref{aff24}}
\and R.~Farinelli\inst{\ref{aff17}}
\and J.~Gracia-Carpio\inst{\ref{aff24}}
\and N.~Mauri\orcid{0000-0001-8196-1548}\inst{\ref{aff42},\ref{aff23}}
\and A.~A.~Nucita\inst{\ref{aff109},\ref{aff110},\ref{aff111}}
\and V.~Scottez\inst{\ref{aff85},\ref{aff112}}
\and M.~Tenti\orcid{0000-0002-4254-5901}\inst{\ref{aff23}}
\and M.~Viel\orcid{0000-0002-2642-5707}\inst{\ref{aff18},\ref{aff19},\ref{aff21},\ref{aff20},\ref{aff113}}
\and M.~Wiesmann\orcid{0009-0000-8199-5860}\inst{\ref{aff62}}
\and Y.~Akrami\orcid{0000-0002-2407-7956}\inst{\ref{aff114},\ref{aff115}}
\and V.~Allevato\orcid{0000-0001-7232-5152}\inst{\ref{aff30}}
\and S.~Anselmi\orcid{0000-0002-3579-9583}\inst{\ref{aff55},\ref{aff96},\ref{aff116}}
\and M.~Ballardini\orcid{0000-0003-4481-3559}\inst{\ref{aff117},\ref{aff17},\ref{aff118}}
\and M.~Bethermin\orcid{0000-0002-3915-2015}\inst{\ref{aff119},\ref{aff80}}
\and A.~Blanchard\orcid{0000-0001-8555-9003}\inst{\ref{aff98}}
\and L.~Blot\orcid{0000-0002-9622-7167}\inst{\ref{aff120},\ref{aff116}}
\and S.~Borgani\orcid{0000-0001-6151-6439}\inst{\ref{aff121},\ref{aff18},\ref{aff19},\ref{aff20}}
\and S.~Bruton\orcid{0000-0002-6503-5218}\inst{\ref{aff122}}
\and R.~Cabanac\orcid{0000-0001-6679-2600}\inst{\ref{aff98}}
\and A.~Calabro\orcid{0000-0003-2536-1614}\inst{\ref{aff41}}
\and A.~Cappi\inst{\ref{aff17},\ref{aff83}}
\and C.~S.~Carvalho\inst{\ref{aff104}}
\and T.~Castro\orcid{0000-0002-6292-3228}\inst{\ref{aff19},\ref{aff20},\ref{aff18},\ref{aff113}}
\and K.~C.~Chambers\orcid{0000-0001-6965-7789}\inst{\ref{aff1}}
\and R.~Chary\orcid{0000-0001-7583-0621}\inst{\ref{aff103}}
\and S.~Contarini\orcid{0000-0002-9843-723X}\inst{\ref{aff24},\ref{aff81}}
\and T.~Contini\orcid{0000-0003-0275-938X}\inst{\ref{aff98}}
\and A.~R.~Cooray\orcid{0000-0002-3892-0190}\inst{\ref{aff123}}
\and B.~De~Caro\inst{\ref{aff55},\ref{aff96}}
\and G.~Desprez\inst{\ref{aff11}}
\and A.~D\'iaz-S\'anchez\orcid{0000-0003-0748-4768}\inst{\ref{aff124}}
\and S.~Di~Domizio\orcid{0000-0003-2863-5895}\inst{\ref{aff27},\ref{aff28}}
\and H.~Dole\orcid{0000-0002-9767-3839}\inst{\ref{aff125}}
\and S.~Escoffier\orcid{0000-0002-2847-7498}\inst{\ref{aff58}}
\and A.~G.~Ferrari\orcid{0009-0005-5266-4110}\inst{\ref{aff42},\ref{aff23}}
\and I.~Ferrero\orcid{0000-0002-1295-1132}\inst{\ref{aff62}}
\and F.~Finelli\orcid{0000-0002-6694-3269}\inst{\ref{aff17},\ref{aff107}}
\and F.~Fornari\orcid{0000-0003-2979-6738}\inst{\ref{aff107}}
\and L.~Gabarra\orcid{0000-0002-8486-8856}\inst{\ref{aff54}}
\and K.~Ganga\orcid{0000-0001-8159-8208}\inst{\ref{aff84}}
\and J.~Garc\'ia-Bellido\orcid{0000-0002-9370-8360}\inst{\ref{aff114}}
\and E.~Gaztanaga\orcid{0000-0001-9632-0815}\inst{\ref{aff39},\ref{aff40},\ref{aff126}}
\and F.~Giacomini\orcid{0000-0002-3129-2814}\inst{\ref{aff23}}
\and G.~Gozaliasl\orcid{0000-0002-0236-919X}\inst{\ref{aff127},\ref{aff71}}
\and A.~Hall\orcid{0000-0002-3139-8651}\inst{\ref{aff44}}
\and W.~G.~Hartley\inst{\ref{aff52}}
\and H.~Hildebrandt\orcid{0000-0002-9814-3338}\inst{\ref{aff128}}
\and J.~Hjorth\orcid{0000-0002-4571-2306}\inst{\ref{aff129}}
\and M.~Huertas-Company\orcid{0000-0002-1416-8483}\inst{\ref{aff43},\ref{aff14},\ref{aff130},\ref{aff131}}
\and O.~Ilbert\orcid{0000-0002-7303-4397}\inst{\ref{aff80}}
\and A.~Jimenez~Mu\~noz\orcid{0009-0004-5252-185X}\inst{\ref{aff132}}
\and J.~J.~E.~Kajava\orcid{0000-0002-3010-8333}\inst{\ref{aff133},\ref{aff134}}
\and V.~Kansal\orcid{0000-0002-4008-6078}\inst{\ref{aff135},\ref{aff136}}
\and D.~Karagiannis\orcid{0000-0002-4927-0816}\inst{\ref{aff137},\ref{aff138}}
\and C.~C.~Kirkpatrick\inst{\ref{aff69}}
\and L.~Legrand\orcid{0000-0003-0610-5252}\inst{\ref{aff139}}
\and G.~Libet\inst{\ref{aff93}}
\and A.~Loureiro\orcid{0000-0002-4371-0876}\inst{\ref{aff140},\ref{aff141}}
\and J.~Macias-Perez\orcid{0000-0002-5385-2763}\inst{\ref{aff132}}
\and G.~Maggio\orcid{0000-0003-4020-4836}\inst{\ref{aff19}}
\and M.~Magliocchetti\orcid{0000-0001-9158-4838}\inst{\ref{aff53}}
\and C.~Mancini\orcid{0000-0002-4297-0561}\inst{\ref{aff35}}
\and F.~Mannucci\orcid{0000-0002-4803-2381}\inst{\ref{aff142}}
\and R.~Maoli\orcid{0000-0002-6065-3025}\inst{\ref{aff143},\ref{aff41}}
\and C.~J.~A.~P.~Martins\orcid{0000-0002-4886-9261}\inst{\ref{aff144},\ref{aff32}}
\and S.~Matthew\orcid{0000-0001-8448-1697}\inst{\ref{aff44}}
\and L.~Maurin\orcid{0000-0002-8406-0857}\inst{\ref{aff125}}
\and R.~B.~Metcalf\orcid{0000-0003-3167-2574}\inst{\ref{aff81},\ref{aff17}}
\and P.~Monaco\orcid{0000-0003-2083-7564}\inst{\ref{aff121},\ref{aff19},\ref{aff20},\ref{aff18}}
\and C.~Moretti\orcid{0000-0003-3314-8936}\inst{\ref{aff21},\ref{aff113},\ref{aff19},\ref{aff18},\ref{aff20}}
\and G.~Morgante\inst{\ref{aff17}}
\and Nicholas~A.~Walton\orcid{0000-0003-3983-8778}\inst{\ref{aff145}}
\and J.~Odier\orcid{0000-0002-1650-2246}\inst{\ref{aff132}}
\and L.~Patrizii\inst{\ref{aff23}}
\and A.~Pezzotta\orcid{0000-0003-0726-2268}\inst{\ref{aff24}}
\and M.~P\"ontinen\orcid{0000-0001-5442-2530}\inst{\ref{aff71}}
\and V.~Popa\inst{\ref{aff94}}
\and C.~Porciani\orcid{0000-0002-7797-2508}\inst{\ref{aff78}}
\and D.~Potter\orcid{0000-0002-0757-5195}\inst{\ref{aff146}}
\and P.~Reimberg\orcid{0000-0003-3410-0280}\inst{\ref{aff85}}
\and I.~Risso\orcid{0000-0003-2525-7761}\inst{\ref{aff147}}
\and P.-F.~Rocci\inst{\ref{aff125}}
\and M.~Sahl\'en\orcid{0000-0003-0973-4804}\inst{\ref{aff148}}
\and C.~Scarlata\orcid{0000-0002-9136-8876}\inst{\ref{aff122}}
\and A.~Schneider\orcid{0000-0001-7055-8104}\inst{\ref{aff146}}
\and M.~Sereno\orcid{0000-0003-0302-0325}\inst{\ref{aff17},\ref{aff23}}
\and A.~Silvestri\orcid{0000-0001-6904-5061}\inst{\ref{aff149}}
\and P.~Simon\inst{\ref{aff78}}
\and A.~Spurio~Mancini\orcid{0000-0001-5698-0990}\inst{\ref{aff150},\ref{aff151}}
\and S.~A.~Stanford\orcid{0000-0003-0122-0841}\inst{\ref{aff152}}
\and C.~Tao\orcid{0000-0001-7961-8177}\inst{\ref{aff58}}
\and G.~Testera\inst{\ref{aff28}}
\and R.~Teyssier\orcid{0000-0001-7689-0933}\inst{\ref{aff153}}
\and S.~Tosi\orcid{0000-0002-7275-9193}\inst{\ref{aff27},\ref{aff28}}
\and A.~Troja\orcid{0000-0003-0239-4595}\inst{\ref{aff96},\ref{aff55}}
\and M.~Tucci\inst{\ref{aff52}}
\and C.~Valieri\inst{\ref{aff23}}
\and J.~Valiviita\orcid{0000-0001-6225-3693}\inst{\ref{aff71},\ref{aff72}}
\and D.~Vergani\orcid{0000-0003-0898-2216}\inst{\ref{aff17}}
\and G.~Verza\orcid{0000-0002-1886-8348}\inst{\ref{aff154},\ref{aff155}}
\and I.~A.~Zinchenko\inst{\ref{aff25}}}
										   
\institute{Institute for Astronomy, University of Hawaii, 2680 Woodlawn Drive, Honolulu, HI 96822, USA\label{aff1}
\and
Cosmic Dawn Center (DAWN)\label{aff2}
\and
Niels Bohr Institute, University of Copenhagen, Jagtvej 128, 2200 Copenhagen, Denmark\label{aff3}
\and
Physics and Astronomy Department, University of California, 900 University Ave., Riverside, CA 92521, USA\label{aff4}
\and
Cosmic Dawn Center (DAWN), Denmark\label{aff5}
\and
Department of Astronomy, University of Massachusetts, Amherst, MA 01003, USA\label{aff6}
\and
Lawrence Berkeley National Laboratory, One Cyclotron Road, Berkeley, CA 94720, USA\label{aff7}
\and
Carnegie Observatories, Pasadena, CA 91101, USA\label{aff8}
\and
The University of Texas at Austin, Austin, TX, 78712, USA\label{aff9}
\and
NRC Herzberg, 5071 West Saanich Rd, Victoria, BC V9E 2E7, Canada\label{aff10}
\and
Department of Astronomy \& Physics and Institute for Computational Astrophysics, Saint Mary's University, 923 Robie Street, Halifax, Nova Scotia, B3H 3C3, Canada\label{aff11}
\and
Institut d'Astrophysique de Paris, UMR 7095, CNRS, and Sorbonne Universit\'e, 98 bis boulevard Arago, 75014 Paris, France\label{aff12}
\and
Jet Propulsion Laboratory, California Institute of Technology, 4800 Oak Grove Drive, Pasadena, CA, 91109, USA\label{aff13}
\and
Instituto de Astrof\'isica de Canarias (IAC); Departamento de Astrof\'isica, Universidad de La Laguna (ULL), 38200, La Laguna, Tenerife, Spain\label{aff14}
\and
ESAC/ESA, Camino Bajo del Castillo, s/n., Urb. Villafranca del Castillo, 28692 Villanueva de la Ca\~nada, Madrid, Spain\label{aff15}
\and
INAF-Osservatorio Astronomico di Brera, Via Brera 28, 20122 Milano, Italy\label{aff16}
\and
INAF-Osservatorio di Astrofisica e Scienza dello Spazio di Bologna, Via Piero Gobetti 93/3, 40129 Bologna, Italy\label{aff17}
\and
IFPU, Institute for Fundamental Physics of the Universe, via Beirut 2, 34151 Trieste, Italy\label{aff18}
\and
INAF-Osservatorio Astronomico di Trieste, Via G. B. Tiepolo 11, 34143 Trieste, Italy\label{aff19}
\and
INFN, Sezione di Trieste, Via Valerio 2, 34127 Trieste TS, Italy\label{aff20}
\and
SISSA, International School for Advanced Studies, Via Bonomea 265, 34136 Trieste TS, Italy\label{aff21}
\and
Dipartimento di Fisica e Astronomia, Universit\`a di Bologna, Via Gobetti 93/2, 40129 Bologna, Italy\label{aff22}
\and
INFN-Sezione di Bologna, Viale Berti Pichat 6/2, 40127 Bologna, Italy\label{aff23}
\and
Max Planck Institute for Extraterrestrial Physics, Giessenbachstr. 1, 85748 Garching, Germany\label{aff24}
\and
Universit\"ats-Sternwarte M\"unchen, Fakult\"at f\"ur Physik, Ludwig-Maximilians-Universit\"at M\"unchen, Scheinerstrasse 1, 81679 M\"unchen, Germany\label{aff25}
\and
INAF-Osservatorio Astrofisico di Torino, Via Osservatorio 20, 10025 Pino Torinese (TO), Italy\label{aff26}
\and
Dipartimento di Fisica, Universit\`a di Genova, Via Dodecaneso 33, 16146, Genova, Italy\label{aff27}
\and
INFN-Sezione di Genova, Via Dodecaneso 33, 16146, Genova, Italy\label{aff28}
\and
Department of Physics "E. Pancini", University Federico II, Via Cinthia 6, 80126, Napoli, Italy\label{aff29}
\and
INAF-Osservatorio Astronomico di Capodimonte, Via Moiariello 16, 80131 Napoli, Italy\label{aff30}
\and
INFN section of Naples, Via Cinthia 6, 80126, Napoli, Italy\label{aff31}
\and
Instituto de Astrof\'isica e Ci\^encias do Espa\c{c}o, Universidade do Porto, CAUP, Rua das Estrelas, PT4150-762 Porto, Portugal\label{aff32}
\and
Dipartimento di Fisica, Universit\`a degli Studi di Torino, Via P. Giuria 1, 10125 Torino, Italy\label{aff33}
\and
INFN-Sezione di Torino, Via P. Giuria 1, 10125 Torino, Italy\label{aff34}
\and
INAF-IASF Milano, Via Alfonso Corti 12, 20133 Milano, Italy\label{aff35}
\and
Centro de Investigaciones Energ\'eticas, Medioambientales y Tecnol\'ogicas (CIEMAT), Avenida Complutense 40, 28040 Madrid, Spain\label{aff36}
\and
Port d'Informaci\'{o} Cient\'{i}fica, Campus UAB, C. Albareda s/n, 08193 Bellaterra (Barcelona), Spain\label{aff37}
\and
Institute for Theoretical Particle Physics and Cosmology (TTK), RWTH Aachen University, 52056 Aachen, Germany\label{aff38}
\and
Institute of Space Sciences (ICE, CSIC), Campus UAB, Carrer de Can Magrans, s/n, 08193 Barcelona, Spain\label{aff39}
\and
Institut d'Estudis Espacials de Catalunya (IEEC),  Edifici RDIT, Campus UPC, 08860 Castelldefels, Barcelona, Spain\label{aff40}
\and
INAF-Osservatorio Astronomico di Roma, Via Frascati 33, 00078 Monteporzio Catone, Italy\label{aff41}
\and
Dipartimento di Fisica e Astronomia "Augusto Righi" - Alma Mater Studiorum Universit\`a di Bologna, Viale Berti Pichat 6/2, 40127 Bologna, Italy\label{aff42}
\and
Instituto de Astrof\'isica de Canarias, Calle V\'ia L\'actea s/n, 38204, San Crist\'obal de La Laguna, Tenerife, Spain\label{aff43}
\and
Institute for Astronomy, University of Edinburgh, Royal Observatory, Blackford Hill, Edinburgh EH9 3HJ, UK\label{aff44}
\and
Jodrell Bank Centre for Astrophysics, Department of Physics and Astronomy, University of Manchester, Oxford Road, Manchester M13 9PL, UK\label{aff45}
\and
European Space Agency/ESRIN, Largo Galileo Galilei 1, 00044 Frascati, Roma, Italy\label{aff46}
\and
Universit\'e Claude Bernard Lyon 1, CNRS/IN2P3, IP2I Lyon, UMR 5822, Villeurbanne, F-69100, France\label{aff47}
\and
Institute of Physics, Laboratory of Astrophysics, Ecole Polytechnique F\'ed\'erale de Lausanne (EPFL), Observatoire de Sauverny, 1290 Versoix, Switzerland\label{aff48}
\and
UCB Lyon 1, CNRS/IN2P3, IUF, IP2I Lyon, 4 rue Enrico Fermi, 69622 Villeurbanne, France\label{aff49}
\and
Departamento de F\'isica, Faculdade de Ci\^encias, Universidade de Lisboa, Edif\'icio C8, Campo Grande, PT1749-016 Lisboa, Portugal\label{aff50}
\and
Instituto de Astrof\'isica e Ci\^encias do Espa\c{c}o, Faculdade de Ci\^encias, Universidade de Lisboa, Campo Grande, 1749-016 Lisboa, Portugal\label{aff51}
\and
Department of Astronomy, University of Geneva, ch. d'Ecogia 16, 1290 Versoix, Switzerland\label{aff52}
\and
INAF-Istituto di Astrofisica e Planetologia Spaziali, via del Fosso del Cavaliere, 100, 00100 Roma, Italy\label{aff53}
\and
Department of Physics, Oxford University, Keble Road, Oxford OX1 3RH, UK\label{aff54}
\and
INFN-Padova, Via Marzolo 8, 35131 Padova, Italy\label{aff55}
\and
Universit\'e Paris-Saclay, Universit\'e Paris Cit\'e, CEA, CNRS, AIM, 91191, Gif-sur-Yvette, France\label{aff56}
\and
School of Physics, HH Wills Physics Laboratory, University of Bristol, Tyndall Avenue, Bristol, BS8 1TL, UK\label{aff57}
\and
Aix-Marseille Universit\'e, CNRS/IN2P3, CPPM, Marseille, France\label{aff58}
\and
Istituto Nazionale di Fisica Nucleare, Sezione di Bologna, Via Irnerio 46, 40126 Bologna, Italy\label{aff59}
\and
FRACTAL S.L.N.E., calle Tulip\'an 2, Portal 13 1A, 28231, Las Rozas de Madrid, Spain\label{aff60}
\and
INAF-Osservatorio Astronomico di Padova, Via dell'Osservatorio 5, 35122 Padova, Italy\label{aff61}
\and
Institute of Theoretical Astrophysics, University of Oslo, P.O. Box 1029 Blindern, 0315 Oslo, Norway\label{aff62}
\and
Leiden Observatory, Leiden University, Einsteinweg 55, 2333 CC Leiden, The Netherlands\label{aff63}
\and
Department of Physics, Lancaster University, Lancaster, LA1 4YB, UK\label{aff64}
\and
Felix Hormuth Engineering, Goethestr. 17, 69181 Leimen, Germany\label{aff65}
\and
Technical University of Denmark, Elektrovej 327, 2800 Kgs. Lyngby, Denmark\label{aff66}
\and
Max-Planck-Institut f\"ur Astronomie, K\"onigstuhl 17, 69117 Heidelberg, Germany\label{aff67}
\and
Department of Physics and Astronomy, University College London, Gower Street, London WC1E 6BT, UK\label{aff68}
\and
Department of Physics and Helsinki Institute of Physics, Gustaf H\"allstr\"omin katu 2, 00014 University of Helsinki, Finland\label{aff69}
\and
Universit\'e de Gen\`eve, D\'epartement de Physique Th\'eorique and Centre for Astroparticle Physics, 24 quai Ernest-Ansermet, CH-1211 Gen\`eve 4, Switzerland\label{aff70}
\and
Department of Physics, P.O. Box 64, 00014 University of Helsinki, Finland\label{aff71}
\and
Helsinki Institute of Physics, Gustaf H{\"a}llstr{\"o}min katu 2, University of Helsinki, Helsinki, Finland\label{aff72}
\and
European Space Agency/ESTEC, Keplerlaan 1, 2201 AZ Noordwijk, The Netherlands\label{aff73}
\and
NOVA optical infrared instrumentation group at ASTRON, Oude Hoogeveensedijk 4, 7991PD, Dwingeloo, The Netherlands\label{aff74}
\and
Centre de Calcul de l'IN2P3/CNRS, 21 avenue Pierre de Coubertin 69627 Villeurbanne Cedex, France\label{aff75}
\and
Dipartimento di Fisica "Aldo Pontremoli", Universit\`a degli Studi di Milano, Via Celoria 16, 20133 Milano, Italy\label{aff76}
\and
INFN-Sezione di Milano, Via Celoria 16, 20133 Milano, Italy\label{aff77}
\and
Universit\"at Bonn, Argelander-Institut f\"ur Astronomie, Auf dem H\"ugel 71, 53121 Bonn, Germany\label{aff78}
\and
INFN-Sezione di Roma, Piazzale Aldo Moro, 2 - c/o Dipartimento di Fisica, Edificio G. Marconi, 00185 Roma, Italy\label{aff79}
\and
Aix-Marseille Universit\'e, CNRS, CNES, LAM, Marseille, France\label{aff80}
\and
Dipartimento di Fisica e Astronomia "Augusto Righi" - Alma Mater Studiorum Universit\`a di Bologna, via Piero Gobetti 93/2, 40129 Bologna, Italy\label{aff81}
\and
Department of Physics, Institute for Computational Cosmology, Durham University, South Road, DH1 3LE, UK\label{aff82}
\and
Universit\'e C\^{o}te d'Azur, Observatoire de la C\^{o}te d'Azur, CNRS, Laboratoire Lagrange, Bd de l'Observatoire, CS 34229, 06304 Nice cedex 4, France\label{aff83}
\and
Universit\'e Paris Cit\'e, CNRS, Astroparticule et Cosmologie, 75013 Paris, France\label{aff84}
\and
Institut d'Astrophysique de Paris, 98bis Boulevard Arago, 75014, Paris, France\label{aff85}
\and
Institut de F\'{i}sica d'Altes Energies (IFAE), The Barcelona Institute of Science and Technology, Campus UAB, 08193 Bellaterra (Barcelona), Spain\label{aff86}
\and
School of Mathematics, Statistics and Physics, Newcastle University, Herschel Building, Newcastle-upon-Tyne, NE1 7RU, UK\label{aff87}
\and
Department of Physics and Astronomy, University of Aarhus, Ny Munkegade 120, DK-8000 Aarhus C, Denmark\label{aff88}
\and
Waterloo Centre for Astrophysics, University of Waterloo, Waterloo, Ontario N2L 3G1, Canada\label{aff89}
\and
Department of Physics and Astronomy, University of Waterloo, Waterloo, Ontario N2L 3G1, Canada\label{aff90}
\and
Perimeter Institute for Theoretical Physics, Waterloo, Ontario N2L 2Y5, Canada\label{aff91}
\and
Space Science Data Center, Italian Space Agency, via del Politecnico snc, 00133 Roma, Italy\label{aff92}
\and
Centre National d'Etudes Spatiales -- Centre spatial de Toulouse, 18 avenue Edouard Belin, 31401 Toulouse Cedex 9, France\label{aff93}
\and
Institute of Space Science, Str. Atomistilor, nr. 409 M\u{a}gurele, Ilfov, 077125, Romania\label{aff94}
\and
Departamento de Astrof\'isica, Universidad de La Laguna, 38206, La Laguna, Tenerife, Spain\label{aff95}
\and
Dipartimento di Fisica e Astronomia "G. Galilei", Universit\`a di Padova, Via Marzolo 8, 35131 Padova, Italy\label{aff96}
\and
Institut f\"ur Theoretische Physik, University of Heidelberg, Philosophenweg 16, 69120 Heidelberg, Germany\label{aff97}
\and
Institut de Recherche en Astrophysique et Plan\'etologie (IRAP), Universit\'e de Toulouse, CNRS, UPS, CNES, 14 Av. Edouard Belin, 31400 Toulouse, France\label{aff98}
\and
Universit\'e St Joseph; Faculty of Sciences, Beirut, Lebanon\label{aff99}
\and
Departamento de F\'isica, FCFM, Universidad de Chile, Blanco Encalada 2008, Santiago, Chile\label{aff100}
\and
Universit\"at Innsbruck, Institut f\"ur Astro- und Teilchenphysik, Technikerstr. 25/8, 6020 Innsbruck, Austria\label{aff101}
\and
Satlantis, University Science Park, Sede Bld 48940, Leioa-Bilbao, Spain\label{aff102}
\and
Infrared Processing and Analysis Center, California Institute of Technology, Pasadena, CA 91125, USA\label{aff103}
\and
Instituto de Astrof\'isica e Ci\^encias do Espa\c{c}o, Faculdade de Ci\^encias, Universidade de Lisboa, Tapada da Ajuda, 1349-018 Lisboa, Portugal\label{aff104}
\and
Universidad Polit\'ecnica de Cartagena, Departamento de Electr\'onica y Tecnolog\'ia de Computadoras,  Plaza del Hospital 1, 30202 Cartagena, Spain\label{aff105}
\and
Kapteyn Astronomical Institute, University of Groningen, PO Box 800, 9700 AV Groningen, The Netherlands\label{aff106}
\and
INFN-Bologna, Via Irnerio 46, 40126 Bologna, Italy\label{aff107}
\and
INAF, Istituto di Radioastronomia, Via Piero Gobetti 101, 40129 Bologna, Italy\label{aff108}
\and
Department of Mathematics and Physics E. De Giorgi, University of Salento, Via per Arnesano, CP-I93, 73100, Lecce, Italy\label{aff109}
\and
INAF-Sezione di Lecce, c/o Dipartimento Matematica e Fisica, Via per Arnesano, 73100, Lecce, Italy\label{aff110}
\and
INFN, Sezione di Lecce, Via per Arnesano, CP-193, 73100, Lecce, Italy\label{aff111}
\and
Junia, EPA department, 41 Bd Vauban, 59800 Lille, France\label{aff112}
\and
ICSC - Centro Nazionale di Ricerca in High Performance Computing, Big Data e Quantum Computing, Via Magnanelli 2, Bologna, Italy\label{aff113}
\and
Instituto de F\'isica Te\'orica UAM-CSIC, Campus de Cantoblanco, 28049 Madrid, Spain\label{aff114}
\and
CERCA/ISO, Department of Physics, Case Western Reserve University, 10900 Euclid Avenue, Cleveland, OH 44106, USA\label{aff115}
\and
Laboratoire Univers et Th\'eorie, Observatoire de Paris, Universit\'e PSL, Universit\'e Paris Cit\'e, CNRS, 92190 Meudon, France\label{aff116}
\and
Dipartimento di Fisica e Scienze della Terra, Universit\`a degli Studi di Ferrara, Via Giuseppe Saragat 1, 44122 Ferrara, Italy\label{aff117}
\and
Istituto Nazionale di Fisica Nucleare, Sezione di Ferrara, Via Giuseppe Saragat 1, 44122 Ferrara, Italy\label{aff118}
\and
Universit\'e de Strasbourg, CNRS, Observatoire astronomique de Strasbourg, UMR 7550, 67000 Strasbourg, France\label{aff119}
\and
Kavli Institute for the Physics and Mathematics of the Universe (WPI), University of Tokyo, Kashiwa, Chiba 277-8583, Japan\label{aff120}
\and
Dipartimento di Fisica - Sezione di Astronomia, Universit\`a di Trieste, Via Tiepolo 11, 34131 Trieste, Italy\label{aff121}
\and
Minnesota Institute for Astrophysics, University of Minnesota, 116 Church St SE, Minneapolis, MN 55455, USA\label{aff122}
\and
Department of Physics \& Astronomy, University of California Irvine, Irvine CA 92697, USA\label{aff123}
\and
Departamento F\'isica Aplicada, Universidad Polit\'ecnica de Cartagena, Campus Muralla del Mar, 30202 Cartagena, Murcia, Spain\label{aff124}
\and
Universit\'e Paris-Saclay, CNRS, Institut d'astrophysique spatiale, 91405, Orsay, France\label{aff125}
\and
Institute of Cosmology and Gravitation, University of Portsmouth, Portsmouth PO1 3FX, UK\label{aff126}
\and
Department of Computer Science, Aalto University, PO Box 15400, Espoo, FI-00 076, Finland\label{aff127}
\and
Ruhr University Bochum, Faculty of Physics and Astronomy, Astronomical Institute (AIRUB), German Centre for Cosmological Lensing (GCCL), 44780 Bochum, Germany\label{aff128}
\and
DARK, Niels Bohr Institute, University of Copenhagen, Jagtvej 155, 2200 Copenhagen, Denmark\label{aff129}
\and
Universit\'e PSL, Observatoire de Paris, Sorbonne Universit\'e, CNRS, LERMA, 75014, Paris, France\label{aff130}
\and
Universit\'e Paris-Cit\'e, 5 Rue Thomas Mann, 75013, Paris, France\label{aff131}
\and
Univ. Grenoble Alpes, CNRS, Grenoble INP, LPSC-IN2P3, 53, Avenue des Martyrs, 38000, Grenoble, France\label{aff132}
\and
Department of Physics and Astronomy, Vesilinnantie 5, 20014 University of Turku, Finland\label{aff133}
\and
Serco for European Space Agency (ESA), Camino bajo del Castillo, s/n, Urbanizacion Villafranca del Castillo, Villanueva de la Ca\~nada, 28692 Madrid, Spain\label{aff134}
\and
ARC Centre of Excellence for Dark Matter Particle Physics, Melbourne, Australia\label{aff135}
\and
Centre for Astrophysics \& Supercomputing, Swinburne University of Technology,  Hawthorn, Victoria 3122, Australia\label{aff136}
\and
School of Physics and Astronomy, Queen Mary University of London, Mile End Road, London E1 4NS, UK\label{aff137}
\and
Department of Physics and Astronomy, University of the Western Cape, Bellville, Cape Town, 7535, South Africa\label{aff138}
\and
ICTP South American Institute for Fundamental Research, Instituto de F\'{\i}sica Te\'orica, Universidade Estadual Paulista, S\~ao Paulo, Brazil\label{aff139}
\and
Oskar Klein Centre for Cosmoparticle Physics, Department of Physics, Stockholm University, Stockholm, SE-106 91, Sweden\label{aff140}
\and
Astrophysics Group, Blackett Laboratory, Imperial College London, London SW7 2AZ, UK\label{aff141}
\and
INAF-Osservatorio Astrofisico di Arcetri, Largo E. Fermi 5, 50125, Firenze, Italy\label{aff142}
\and
Dipartimento di Fisica, Sapienza Universit\`a di Roma, Piazzale Aldo Moro 2, 00185 Roma, Italy\label{aff143}
\and
Centro de Astrof\'{\i}sica da Universidade do Porto, Rua das Estrelas, 4150-762 Porto, Portugal\label{aff144}
\and
Institute of Astronomy, University of Cambridge, Madingley Road, Cambridge CB3 0HA, UK\label{aff145}
\and
Department of Astrophysics, University of Zurich, Winterthurerstrasse 190, 8057 Zurich, Switzerland\label{aff146}
\and
Dipartimento di Fisica, Universit\`a degli studi di Genova, and INFN-Sezione di Genova, via Dodecaneso 33, 16146, Genova, Italy\label{aff147}
\and
Theoretical astrophysics, Department of Physics and Astronomy, Uppsala University, Box 515, 751 20 Uppsala, Sweden\label{aff148}
\and
Institute Lorentz, Leiden University, Niels Bohrweg 2, 2333 CA Leiden, The Netherlands\label{aff149}
\and
Department of Physics, Royal Holloway, University of London, TW20 0EX, UK\label{aff150}
\and
Mullard Space Science Laboratory, University College London, Holmbury St Mary, Dorking, Surrey RH5 6NT, UK\label{aff151}
\and
Department of Physics and Astronomy, University of California, Davis, CA 95616, USA\label{aff152}
\and
Department of Astrophysical Sciences, Peyton Hall, Princeton University, Princeton, NJ 08544, USA\label{aff153}
\and
Center for Cosmology and Particle Physics, Department of Physics, New York University, New York, NY 10003, USA\label{aff154}
\and
Center for Computational Astrophysics, Flatiron Institute, 162 5th Avenue, 10010, New York, NY, USA\label{aff155}}    

%
%
   \abstract{
   The Cosmic Dawn Survey (DAWN survey) provides multiwavelength (UV/optical to mid-IR) data across the combined 59 deg$^{2}$ of the Euclid Deep and Auxiliary fields (EDFs and EAFs). Here, the first public data release (DR1) from the DAWN survey is presented. DR1 catalogues are made available for a subset of the full DAWN survey that consists of two Euclid Deep fields: Euclid Deep Field North (EDF-N) and Euclid Deep Field Fornax (EDF-F). The DAWN survey DR1 catalogues do not include \Euclid data as they are not yet public for these fields. Nonetheless, each field has been covered by the ongoing Hawaii Twenty Square Degree Survey (H20), which includes imaging from CFHT MegaCam in the new $u$ filter and from Subaru Hyper Suprime-Cam (HSC) in the $griz$ filters. Each field is further covered by \textit{Spitzer}/IRAC 3.6-4.5$\mu$m imaging spanning 10 deg$^{2}$ and reaching $\sim$25 mag AB (5$\sigma$). All present H20 imaging and all publicly available imaging from the aforementioned facilities are combined with the deep \textit{Spitzer}/IRAC data to create source catalogues spanning a total area of 16.87 deg$^{2}$ in EDF-N and 2.85 deg$^{2}$ in EDF-F for this first release. Photometry is measured from these multiwavelength data using \farmer{}, a novel and well-validated model-based photometry code. Photometric redshifts and stellar masses are computed using two independent codes for modeling spectral energy distributions: \eazy{} and \lephare{}. Photometric redshifts show good agreement with spectroscopic redshifts ($\sigma_{\rm NMAD} \sim  0.5, \eta < 8\%$ at $i < 25$). Number counts, photometric redshifts, and stellar masses are further validated in comparison to the COSMOS2020 catalogue. The DAWN survey DR1 catalogues are designed to be of immediate use in these two EDFs and will be continuously updated and made available as both new ground-based data and spaced-based data from \Euclid are acquired and made public. Future data releases will provide catalogues of all EDFs and EAFs and include \Euclid data.
   }
%
%
%
%
   \titlerunning{\Euclid Preparation: Cosmic Dawn Survey DR1}
   \authorrunning{Euclid Collaboration: L. Zalesky et al.}
   
   \maketitle
%
%
%
%
   
\section{\label{sec:Intro}Introduction}

The \Euclid mission \citep{laureijs_euclid_2011, Mellier2024} has the potential to revolutionize cosmology through its survey of \num{14000}\,deg$^2$ of the extragalactic sky. Imaging in optical and near-infrared wavelengths will be obtained by \Euclid for billions of galaxies in addition to spectroscopy for roughly 50 million \citep[the Euclid Wide Survey: EWS;][]{Scaramella22}. The primary science objectives of \Euclid are to constrain the properties of dark matter and dark energy through weak lensing and galaxy-clustering measurements. The Euclid Wide Survey will reach an expected (5$\sigma$) limiting magnitude of 24.3 mag (AB) for point sources in the near-infrared imaging. At these depths, \Euclid will primarily probe the low-redshift ($z < 2$) Universe. 

Over the expected six years of the \Euclid mission, roughly 20\% of \Euclid observation time will be devoted to targeting six Euclid Auxilliary fields (EAFs) and three Euclid Deep Fields (EDFs). The EAFs and EDFs serve the mission in different ways. The EAFs comprise six extensively studied fields of scale 0.5--2  deg$^{2}$, including CDFS, COSMOS-Wide, SXDS, VVDS, AEGIS, and GOODS-N (see \citealt{Scaramella22} and \citealt{McPartland2024} for details). The auxiliary fields support photometric redshift and colour-gradient calibration and host extensive spectroscopic samples of galaxies. The EDFs comprise three fields easily accessible year round, given \Euclid's orbit, and include Euclid Deep Field North (EDF-N; 20 deg$^{2}$), Euclid Deep Field Fornax (EDF-F; 10 deg$^{2}$), and Euclid Deep Field South (EDF-S; 23 deg$^{2}$). The deep fields assist in characterizing the galaxy population of the wide survey, calibrating the noise-bias for weak lensing analyses, and quantifying completeness and purity (in EDF-N and EDF-S, specifically) for the EWS spectroscopic observations. The resulting \Euclid data in the EAFs and EDFs will be between 4--8 times deeper than the EWS data. Accordingly, the deep \Euclid data in the EAFs and EDFs enable tremendous legacy science at high redshift while simultaneously supporting the wide survey. 

The primary \Euclid science objectives, including weak lensing and galaxy clustering analyses, as well as legacy science endeavors, require supplemental ground-based data to establish quality photometric redshifts and calibrate colour gradients affecting chromatic (i.e., wavelength-dependent) point-spread functions \citep{laureijs_euclid_2011,Scaramella22}. Furthermore, the intrinsic properties of galaxies, such as stellar mass and star-formation rate, cannot be fully studied with the \Euclid data alone. Emission of star-forming galaxies at low redshift are dominant in wavelengths shorter than those covered by \Euclid. In addition, with increasing distance, significant spectral features, especially in rest-frame optical light, are shifted towards wavelengths longer than those covered by \Euclid. Ultimately, complementary depth-matched imaging and self-consistent photometry in the UV/optical and mid-infrared are important additions to the \Euclid data in order to constrain the full detailed shapes of galaxy spectral energy distributions. In the Euclid Deep and Auxilliary Fields, these data are provided by the Cosmic Dawn Survey (DAWN survey; \citealt{McPartland2024}). The DAWN survey is a 59 deg$^{2}$ multiwavelength survey of the EAFs and EDFs.   
The DAWN survey catalogues are complementary to the official \Euclid survey catalogues and are primarily distinguished from the official \Euclid survey catalogues by wavelength coverage. The DAWN survey catalogues include deep \textit{Spitzer}/IRAC imaging and photometry measured self-consistently via a model-based method described in greater detail below. Accordingly, the DAWN catalogues are optimized for galaxy evolution science beyond redshifts $z > 4$, where the \textit{Spitzer}/IRAC photometry probes rest-frame optical emission. In this paper we present the first public release (DR1) of catalogues from the DAWN survey, consisting entirely of pre-launch data. Future DAWN data releases (including EDF-S and the EAFs) will follow each of the \Euclid data releases. See \cite{McPartland2024} for a description the fields, observations, and science goals of the DAWN survey.

The DR1 catalogues from the DAWN survey provide multiwavelength photometry and galaxy properties across two EDFs, EDF-N and EDF-F. The DAWN survey DR1 catalogues do not yet include \Euclid data for these fields, as they are still being acquired. However, future data releases will provide catalogues including \Euclid photometry for all Euclid Deep Fields, including Euclid Deep Field South, and the EAFs. As described by \cite{McPartland2024}, the DAWN survey incorporates UV/optical imaging from multiple ground-based surveys and mid-infrared imaging from \Spitzer to complement and support the EDFs and EAFs. Across EDF-N and EDF-F, ultraviolet (UV) and optical coverage is provided by the Hawaii Twenty Square Degree Survey (H20). H20 utilizes the MegaCam instrument on the Canada-France-Hawaii Telescope (CFHT) and the Subaru telescope's Hyper Suprime-Cam (HSC). Mid-infrared coverage over EDF-N and EDF-F is provided by the DAWN survey \textit{Spitzer}/IRAC data \citep{Moneti2022}, where the primary contribution is from the \textit{Spitzer} Legacy Survey \cite[SLS;][]{Capak2016}. Both the H20 and the SLS surveys were designed to obtain imaging of comparable depth to the near-infrared observations that will be conducted by \Euclid in the EDFs. Notably, the SLS data represent the single largest allocation of \textit{Spitzer} time ever awarded. While the H20 survey targets the full twenty square degrees of EDF-N with UV/optical coverage, the \textit{Spitzer} mid-infrared imaging only covers the central ten square degrees. In EDF-F, both the H20 and SLS surveys target the full ten square degrees of the field. The combination of wavelength coverage, spanning the UV through mid-infrared, area, targeting more than twenty square degrees, and depth, reaching 5$\sigma$ depths of $\sim$27 AB mag in optical bands and 25 AB mag at 3.6--4.5$\mu$m, is unique across extragalactic fields. 

Since 2019, the H20 survey has been obtaining Subaru HSC imaging in the $griz$ filters and CFHT MegaCam $u$-band imaging across EDF-N and EDF-F. In order to produce the most complete co-added images to be paired with the deep \textit{Spitzer}/IRAC data, all archival imaging in EDF-N and EDF-F from the same listed facilities are included and processed alongside the data taken by H20. The DAWN survey DR1 catalogue of EDF-N spans a total of 16.87 deg$^{2}$, with 9.37 deg$^{2}$ reaching final survey depths (see Sect.~\ref{subsec:area_coverage}). The DAWN survey DR1 catalogue of EDF-F contains 2.85 deg$^{2}$ of the deepest presently available data, with 1.77 deg$^{2}$ reaching final survey depths in all but one band. Additional imaging is currently being acquired to expand EDF-N to final survey depths across 20 deg$^{2}$ and to complete EDF-F to its final survey depths across 10 deg$^{2}$. Although the ground-based data acquisition is ongoing, the DAWN survey DR1  catalogues are presented now in order to support both pre-launch and early science objectives in \Euclid Deep Fields. 

With only limited near-infrared imaging from \Euclid currently acquired over EDF-N and EDF-F, the present DAWN survey catalogues are selected using optical imaging, though future catalogues will be selected from the near-infrared \Euclid data. The creation of the DAWN survey catalogues benefits from the experience and insight garnered via the recent reprocessing and photometric extraction of all publicly available data in the COSMOS field \citep{Scoville2007}, which culminated in the release of the \enquote{COSMOS2020} catalogue \citep{Weaver2021}. Already the COSMOS2020 catalogue has proved a valuable resource for extragalactic science \citep{Ito_2022, Shuntov_2022, Davidzon_2022, Kauffmann_2022, Gould2023,taamoli2024}. Accordingly, many of the choices made in building the DAWN survey catalogues are motivated by the strategies developed during the construction of the COSMOS2020 catalogue. The similarity in depth, utilized facilities, and wavelength coverage mark COSMOS2020 as a forerunner to H20 and the DAWN survey, although the latter span more than an order of magnitude in larger volume. The total volume of the DAWN survey out to $z \sim$ 7 will be $\sim$3 Gpc$^3$, and roughly one half of this volume is contained by EDF-N and EDF-F alone. Thus, the unique data in EDF-N and EDF-F, and the DAWN survey generally, enable high-redshift studies where Poisson uncertainties and cosmic variance are not the dominant sources of error. By contrast, even in the 2 deg$^{2}$ COSMOS field, cosmic variance and Poisson uncertainties dominate the error budget for the abundance of massive ($M_{\star} > 10^{10.5}$M$_{\odot}$) galaxies \citep{Weaver2023SMF}. In addition, the unique data of the H20 and the DAWN survey enable exploration of diverse environments and significant cosmic volumes at high-redshift ($z > 3$). These volumes contain several tens to hundreds of massive dark matter haloes ($M_\textrm{halo} > 10^{12}$\,M$_{\odot}$) as well as voids, such that the variation of galaxy properties in cosmically distinct environments can be directly measured. In comparison, fewer than ten such massive halos are expected in a survey like COSMOS \citep{Despali2016} at these redshifts. 

\begin{figure*}
    \centering
    \includegraphics[width=\textwidth]{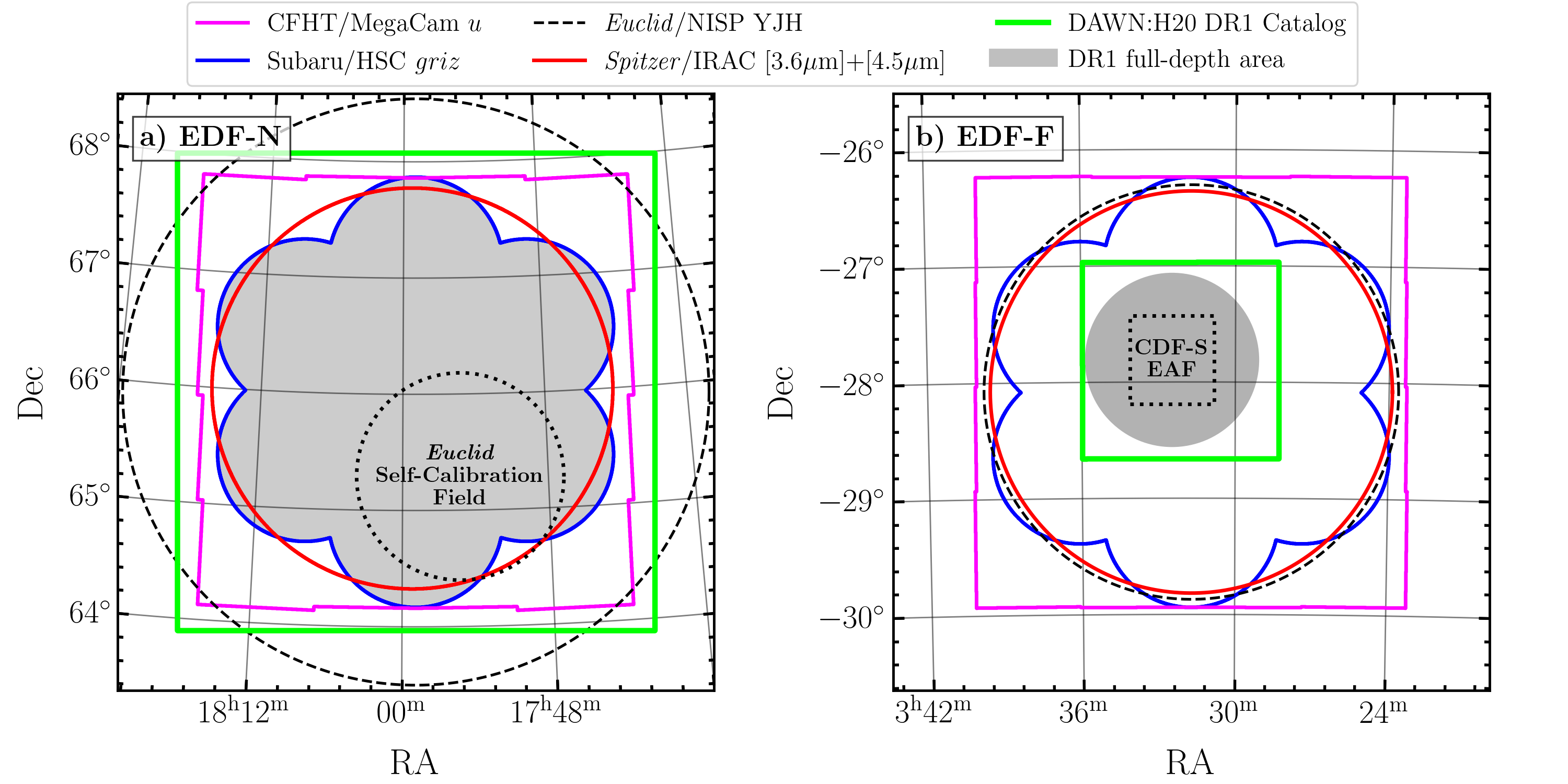}
    \caption{Layouts and facility coverage for the two fields included in the DAWN survey DR1 catalogues, EDF-N (left) and EDF-F (right). Also highlighted are areas of particular importance to \Euclid, namely, the \Euclid self-calibration field in EDF-N, and the Chandra Deep Field South (CDF-S) Euclid Auxiliary Field (EAF) in EDF-F. The regions spanned by the DR1 catalogues are illustrated by the green squares. The areas of the DR1 catalogues reaching approximately final integration times by all facilities (or covered to full depth; see Sect.~\ref{subsec:area_coverage} and Table~\ref{tab:exptimes}) are shown in grey. A future release from the DAWN survey will include catalogues with coverage from all facilities at full-depth spanning the entirety of EDF-N and EDF-F areas are targeted by \Euclid.}
    \label{fig:field_layouts}
\end{figure*}

\farmer{} \citep{WeaverFarmer} is used to measure multiwavelength photometry from the DAWN survey images. \farmer{} is an open-source package built around \tractor{} \citep{Lang2016} that derives photometry by fitting galaxy surface brightness profiles. While \tractor{} provides a library of models and optimization routines, \farmer{} handles organizational tasks including appropriate model selection for source parameterization, highly parallelized multiprocessing, and catalogue creation. Together, they yield self-consistent total flux and flux uncertainties across wide ranges of wavelength and spatial resolution. In total, \num{5286829} objects are detected over the 16.87 deg$^{2}$ area of the DR1 EDF-N catalogue, where \num{3513211} of the detected objects are in the 9.37 deg$^{2}$ full-depth region. In EDF-F, \num{1062645} objects are detected over the DR1 2.85 deg$^{2}$ area, where \num{727678} are 
 detected over the 1.77 deg$^{2}$ full-depth region. In the presentation of the COSMOS2020 catalogue, \cite{Weaver2021}  demonstrated the utility of obtaining measurements of \photoz{}, stellar mass, and star-formation rate from multiple independent codes. The same approach is adopted in this work, and \photoz{}s and physical properties of galaxies are measured with both \eazy{}~\citep{Brammer2008} and \lephare{}~\citep{Arnouts2002,Ilbert2006}.

This paper is structured as follows. In Sect.~\ref{sec:Observations} the imaging data and their reduction are presented. An overview is provided of the methods for source detection and photometry in Sect.~\ref{sec:SD_Photometry}. Section~\ref{sec:photoz} describes the \photoz{} measurements using the measured photometry, while Sect.~\ref{sec:properties} presents the physical properties of the galaxies. The paper and first data release are summarized in Sect.~\ref{sec:summary}.

This work adopts a standard $\Lambda$CDM cosmology with $H_0=70$\,km\,s$^{-1}$\,Mpc$^{-1}$, $\Omega_{\rm m}=0.3$ and $\Omega_{\Lambda}=0.7$. All magnitudes are expressed in the AB system \citep{Oke1974}, for which a flux $f_\nu$ in $\mu$Jy
($10^{-29}$\,erg\,cm$^{-2}$s$^{-1}$Hz$^{-1}$) corresponds to AB$_\nu=23.9-2.5\,\log_{10}(f_\nu/\mu{\rm Jy})$. SED fitting codes assume a Chabrier \citep{Chabrier2003} initial mass function. 

\section{\label{sec:Observations}Observations and data reduction}

The creation of the Cosmic Dawn Survey DR1 catalogue begins with the collection of multiwavelength data spanning the UV/optical to mid-infrared obtained across EDF-N and EDF-F. UV/optical imaging is provided by the H20 survey, specifically acquired from CFHT MegaCam in the new $u$ filter, and Subaru HSC $griz$. These data are paired with the deep \textit{Spitzer}/IRAC covering EDF-N and EDF-F from the DAWN survey \citep{Moneti2022}. Data from all facilities are sampled from their native pixel scales to the pixel scale of HSC (\ang{;;0.168}/pixel). The coverage according to each facility, along with bounding regions indicating the area spanned by each catalogue, is presented in maps of the two fields in Fig.~\ref{fig:field_layouts}. Below, the acquisition of data from the various observatories and their reduction is described. 

\subsection{\label{subsec:uband} Ultraviolet data}

The H20 survey has carried out an extensive campaign to obtain deep ultraviolet (UV) imaging in the $u$ band using CFHT and the MegaCam instrument \citep{Boulade2003} across EDF-N and EDF-F. MegaCam has a square field of view with an area of 1 deg$^{2}$. In both EDF-N and EDF-F, only imaging obtained with the instrument's new $u$ filter is considered, which replaced the old ($u^{*}$) filter in 2015 and has a more uniform transmission. Each field was observed in a square grid of $4\times4$ pointings (16 total), where each pointing overlaps by \ang{;;180} with its neighbors. A five point \enquote{large dithering pattern}, as defined by CFHT, is used for the majority of our exposures. The large dithering pattern covers an ellipse with a major axis of \ang{;;180} and minor axis of \ang{;;30}. Further, exposure times of 324\,s are primarily used for individual frames. In some cases, the dither pattern, number of dithers, and integration times were adjusted slightly in order to fully make use of the queue time awarded each semester.

To create the $u$-band mosaics, all available data in the new Megacam $u$ filter were gathered across EDF-N and EDF-F. 
Within the EDF-N field, approximately equal contribution is made by both the H20 survey and the Deep \Euclid U-band Survey \citep[DEUS; designed after the success of][]{Sawicki_2019}, while other archival imaging makes a smaller contribution. In EDF-F, only H20 imaging is utilized, as H20 provides the only data in the new $u$ filter within the field. Extreme outlier images with bad seeing, tracking, or transparency are initially removed. Detrending begins with the raw data from CFHT. For every observing run, new flat fields are built, where \textit{Gaia} DR3 \citep{Gaia2016,Gaia2022} was used as an astrometric reference. Proper motions of stars are accounted for at the epoch each image was taken before calibration, and each input image is calibrated separately. This calibration is accurate to approximately $\sim$20 mas.

\begin{figure*}
    \centering
    \includegraphics[width=\columnwidth]{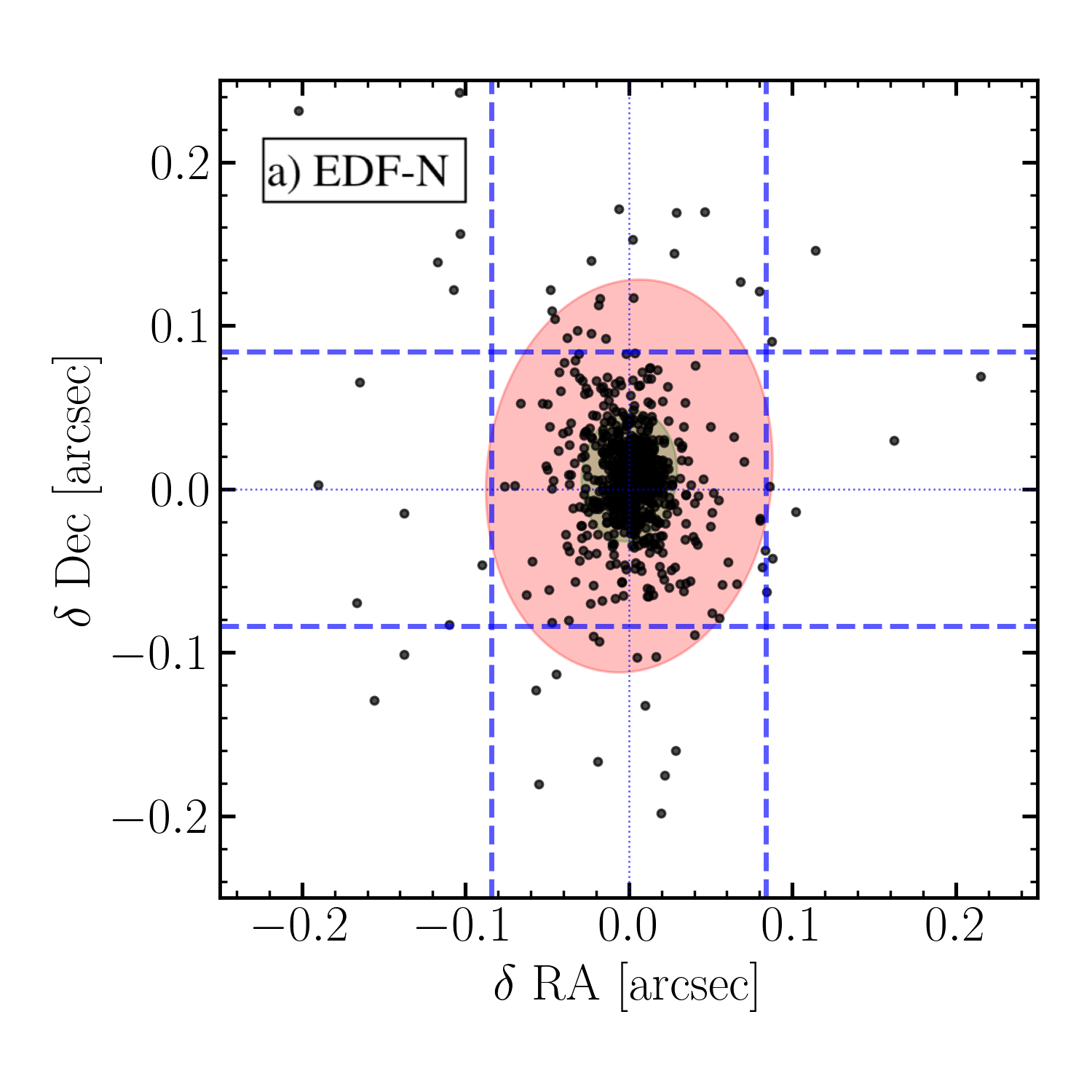}
    \includegraphics[width=\columnwidth]{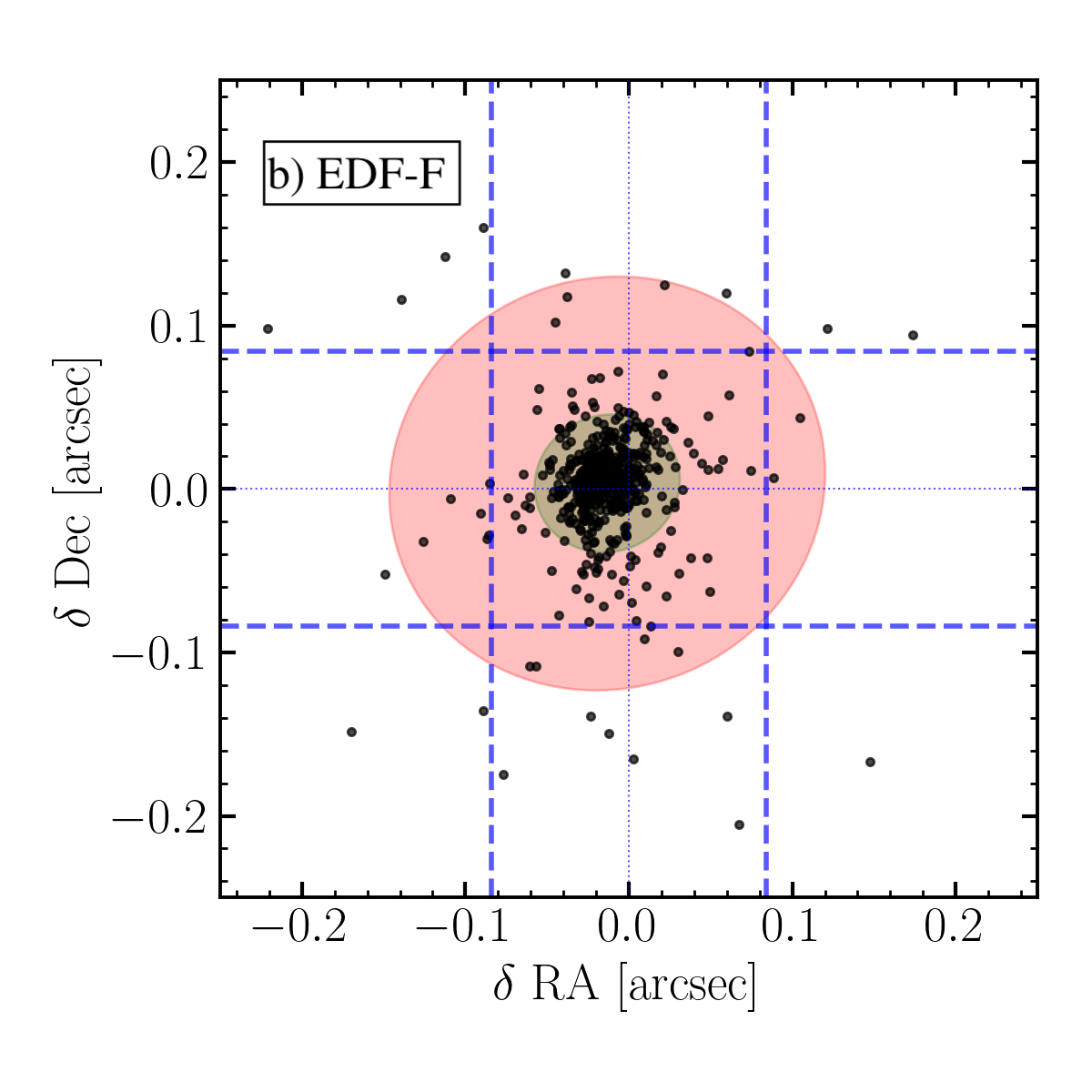}
    \caption{Astrometric comparison between \textit{Gaia} DR3 matched to the DAWN survey DR1 catalogues for objects with HSC $g$ magnitudes between 16 and 21 (\textit{left}: EDF-N, \textit{right}: EDF-F). The spacing between the dashed lines represent the dimensions of HSC pixels, while the green and red shaded regions represent the 1 and $3\sigma$ covariance ellipses, respectively. For clarity, only 2.5\% and 10\% of the matched sources are shown for EDF-N and EDF-F, respectively, where fewer sources are shown for EDF-N due to the higher stellar density (lower Galactic latitude) and greater catalogue extent.}
    \label{fig:astrometry}
\end{figure*}

Regarding photometric calibration, first a photometric \enquote{superflat} is applied to each input image to correct for the illumination of the focal plane. The superflat is built for each MegaCam run, using all the images that overlap with the Sloan Digital Sky Survey \citep[SDSS;][]{Kollmeier2017}. This process typically achieves a photometric flatness on the order of 0.005 mag. For the absolute calibration of image zero-points, \textit{Gaia} DR3 is used. The \textit{Gaia} spectra are multiplied by the appropriate filter passbands to create synthetic photometry, which is used to calibrate each image. A significant challenge is that the \textit{Gaia} spectra are only available for relatively bright stars, some of which are saturated. To mitigate the random noise and increase the sample of usable stars, all the catalogues from each image are merged to produce a deeper secondary photometric catalogue for calibration. This process makes individual image photometric calibration accurate to 0.01--0.02 mag internally.

Pixel masks for the $u$-band images are created with \texttt{WeightWatcher} \citep{Marmo2010}. This code identifies bad columns, bad pixels, and cosmic rays; good pixels are set to 1 while bad pixels are set to 0. Final image stacking is performed with \texttt{SWarp} \citep{Bertin2002,Bertin2010} using the clipped mean \enquote{combine type},  which provides a balance of outlier rejection when combined with the masking from the previous step,  and only minimally reduces SNR. During this step, the $u$-band images are also resampled to the scale and tangent point of the HSC data (Sect.~\ref{subsec:optical}). Due to the contribution from several observing programs with different observation patterns, the resulting $u$-band data in EDF-N is roughly 0.3 mag deeper than EDF-F and shows greater spatial consistency. Both fields are among the deepest $u$-band data available over such large areas.

\subsection{\label{subsec:optical}Optical data}

The central component of the H20 survey is deep optical imaging covering EDF-N and EDF-F. This is supplied by HSC \citep{Miyazaki2018} on the Subaru telescope. Subaru HSC has a circular field of view with an area of 1.8 deg$^{2}$. EDF-N and EDF-F are circular fields spanning 20 and 10 deg$^{2}$, respectively. To cover the central 10 deg$^{2}$ of each field, a flower petal observation pattern was designed with a single central pointing surrounded by a circle of six pointings with radius of \ang{1.1}. For the outer 10 deg$^{2}$ annulus of EDF-N, additional pointings are planned. Imaging with HSC was acquired in the $griz$ broad bands with exposure times for individual frames of 200\,s, 210\,s, 260\,s, and 300\,s, respectively. Throughout our observations, a standard five-point dither pattern with a throw of \ang{;;120} was employed. 

Just as is done for the CFHT $u$ band, data reduction begins by first gathering all existing public HSC imaging data over EDF-N and EDF-F from the Subaru archive \citep[SMOKA;][]{SMOKA}.\footnote{\url{https://smoka.nao.ac.jp/}} Programs with public data in EDF-N include HEROES \citep{Taylor2023} and \textit{AKARI} \citep{Oi2021}. In addition to the $griz$ bands, archival narrow-band imaging in the NB0816 and NB0921 filters is gathered in both fields, and in EDF-N archival HSC $y$ imaging is also gathered. All HSC data are reduced using the public data reduction pipeline \texttt{hscPipe} version 8.4 \citep{Bosch2018}. The default reduction routines of \texttt{hscPipe} are applied with the following modifications:
\begin{itemize}
    \item The older \texttt{jointcal} algorithm is used for astrometric calibration instead of the new FGCM algorithm \citep{SSP-DR3}, as the latter is more memory intensive and becomes too time consuming for deep data with many individual frames.
    \item Sigma clipping is applied for coadd images, which significantly reduces scattered light, satellite trails, and cosmic rays, among other spurious objects in the images.
    \item The internal parameters of \texttt{hscPipe} area is adjusted to enable extraction of PSF models much larger than the default size, as the default models were too small for the model-based photometry (see Sect. ~\ref{sec:SD_Photometry}).
\end{itemize}

Another significant feature of the HSC reduction is the photometric and astrometric calibration. These calibrations are applied by matching objects to the Pan-STARRS1 3$\pi$ survey \citep{Chambers2016} and deriving the appropriate colour and absolute photometric brightness corrections. For its astrometric calibration, Pan-STARRS uses \textit{Gaia} DR1 \citep{GaiaDR1,GaiaDR1_astrometry}, so the HSC imaging inherits this reference system for its astrometry. The quality of this calibration is validated by re-matching our detected objects to \textit{Gaia} DR3 \citep{Gaia2022} as demonstrated in Fig.~\ref{fig:astrometry}. In EDF-N, a standard deviation of $<3$ mas between our final measured coordinates and the \textit{Gaia} DR3 coordinates is observed, with slightly greater variation in Dec. For EDF-F, a standard deviation of $\sim$4\,mas with approximately equal variation in both RA and Dec is observed, and an additional offset $< 2$\,mas in RA. The smaller area considered for the present EDF-F data results in smaller sample size and thus a larger measured statistical variation in astrometry, in comparison to EDF-N. The tangent point and pixel-scale of the final stacked HSC images form the reference world-coordinate system (WCS) against which all imaging from other facilities are sampled to match.

\begin{table} \centering
\footnotesize
\caption{H20 target integration times for each instrument and band combination. Target integration times are quantified per pointing (i.e., not summed across the field) and listed in units of hours. Imaging in Subaru HSC $y$ and the narrow bands were not targeted under the Cosmic Dawn or H20 surveys, but are included as archival data and thus have no \enquote{target} integration time. For \textit{Spitzer} integration times, the reader is referred to \cite{Moneti2022}.}
\begin{tabular}{lr}
\hline\hline
Instrument / Band & Target integration  \\
 & time [hours] \\
\hline
CFHT MegaCam / $u$  & 2.5  \\
Subaru HSC / $g$    & 1.1  \\
Subaru HSC / $r$    & 2.5  \\
Subaru HSC / $i$    & 4.1  \\
Subaru HSC / $z$    & 4.8  \\

\hline
\end{tabular}
\label{tab:exptimes}
\end{table}

\begin{table} \centering
\footnotesize
\caption{Areas defining the extents of the DAWN survey DR1 catalogue regions. \Euclid footprint: the total area targeted by \Euclid; DAWN survey DR1: total extent of the DR1 catalogue areas; Masked by stars: regions excluded by bright star masks (see Sect.~\ref{subsec:masking}); Failed models: area spanned by objects for which photometry could not be measured (see Sect.~\ref{subsec:models}); Full-depth: catalogue area reaching final integration times (see Sect.~\ref{subsec:image_depths}); Full-depth masked by stars: regions excluded by stars in full-depth area; Full-depth failed models: area spanned by objeects for which photometry could not be measured in full-depth area; Effective full-depth area: total area of full-depth region minus the area masked by stars and failed models. }
\begin{tabular}{lrr}
\hline\hline
Region & EDF-N area &  EDF-F area \\
 & [deg$^{2}$] & [deg$^{2}$]   \\
\hline
\Euclid footprint  & 20 & 10  \\
DAWN Survey DR1    & 16.865 &  2.854 \\
Masked by stars    & 1.687 & 0.088 \\
Failed models    & 0.074 & 0.014 \\
Full-depth    & 9.373 & 1.767 \\
Full-depth masked by stars & 0.898 & 0.047 \\
Full-depth failed models & 0.055 & 0.010 \\
\hline
Effective full-depth & 8.420 & 1.710 \\
\hline
\end{tabular}
\label{tab:cat_areas}
\end{table}

\subsection{\label{subsec:infrared}Mid-infrared data}
The mid-infrared data in the DAWN survey DR1 catalogues come from the Spitzer Legacy Survey \citep{Capak2016,Moneti2022}. These data distinguish the EDF-N and EDF-F fields from other deep and wide extragalactic survey fields by providing the deepest \textit{Spitzer} imaging available over such large areas. The acquisition and reduction of these data are fully described in \cite{Moneti2022}. The images produced by that effort are sampled to a scale of \ang{;;0.6}/pixel and co-added using linear interpolation of the individual frames. As for the CFHT $u$ band, the \textit{Spitzer}/IRAC data are resampled in this work to the scale and tangent point of the HSC data using \texttt{SWARP}.

\subsection{\label{subsec:area_coverage}Area coverage}

The DAWN survey DR1 catalogues are provided to be of immediate use to science in EDF-N and EDF-F, though some areas of each field are not yet covered to their final target exposure times in every instrument and bandpass combination at the time of writing. The H20 survey with CFHT MegaCam and Subaru HSC is ongoing. Here, the status of data acquisition, at the time of writing, is defined. Recall that the total area of the \Euclid EDF-N is 20 deg$^{2}$, while the total area of EDF-F is 10 deg$^{2}$, as defined by \cite{Scaramella22}. Completed coverage, or \enquote{full-depth}, is defined as having acquired a total integration time equal to the target integration time and is considered per-pointing. A table of the target integration time of each facility is provided in Table~\ref{tab:exptimes}, and a table summarizing the regions of the DAWN survey DR1 catalogues is provided in Table~\ref{tab:cat_areas}.

The target integration time with CFHT MegaCam in the $u$ filter is 2.5 hours and is calculated to achieve a 5$\sigma$ point-source limiting magnitude of at least 26.4 mag assuming 1 arcsecond seeing. In practice, the integration time needed to reach the target value differs between EDF-N and EDF-F. The former hosts extensive archival imaging (predominantly provided by DEUS; see Sect.~\ref{subsec:uband}) whereas there is no previous CFHT MegaCam imaging in the new $u$ filter over EDF-F prior to H20. The CFHT MegaCam $u$-band imaging is complete over ten square degrees in both EDF-N and EDF-F, reaching approximately 14 deg$^{2}$ in both fields according to the tiling strategy described in Sect. \ref{subsec:uband}. The outer ten square degree annulus of EDF-N is expected to be completed with CFHT MegaCam $u$ in 2024, while no further CFHT MegaCam data are required in EDF-F.

The target integration times for the Subaru HSC $griz$ bands across EDF-N and EDF-F are 1.1, 2.5, 4.1, and 4.8 hours, respectively. These were calculated to achieve 5$\sigma$ limiting point-source magnitudes of 27.5, 27.5, 27, and 26.5, respectively, assuming 0.7 arcsecond seeing. EDF-N hosts significantly more complete coverage with Subaru HSC compared to EDF-F, given EDF-N is observable year-round from Hawaii whereas EDF-F is only observable in the second half of the calendar year. At the time of writing, EDF-N is completed across 9.37 deg$^{2}$ to target integration times in all $griz$ filters. Notably, the full-depth region of EDF-N essentially spans the entirety of the area covered by \textit{Spitzer}/IRAC. For EDF-F, an area of 1.77 deg$^{2}$, centered on Chandra Deep Field South (CDFS), is completed to the target integration time in all filters except HSC $z$ (lacking $\sim$20\% of the required time). Both fields have shallower coverage in all filters across the areas outside the respective full-depth regions. 

As noted in Sect.~\ref{subsec:optical}, we reduce all publicly available Subaru HSC imaging over EDF-N and EDF-F, including archival HSC $y$ imaging in EDF-N and HSC NB0816 and NB0921 imaging over both EDF-N and EDF-F. As these data were not targeted as part of the H20 Survey, they have no target integration time. The SLS has long been completed, and since then \Spitzer has been decommissioned. The reader is referred to the detailed description of the \textit{Spitzer}/IRAC integration times over EDF-N and EDF-F provided by \citealt{Moneti2022}.

The DAWN survey DR1 catalogues presented here have been created using all imaging processed as of January 2024, while additional data are currently being acquired and processed. The DAWN survey DR1 EDF-N catalogue spans 16.87 deg$^{2}$ total, extending beyond the area covered by \textit{Spitzer} and slightly beyond the area currently covered by CFHT MegaCam, while the DAWN survey DR1 EDF-F catalogue spans 2.85 deg$^{2}$ total. A future data release will include complete Subaru HSC and CFHT MegaCam imaging over the entire 20 deg$^{2}$ area of EDF-N and 10 deg$^{2}$ area of EDF-F with complete uniform coverage. The 9.37 deg$^{2}$ region in EDF-N and the 1.77 deg$^{2}$ region in EDF-F reaching full-depth in each $ugriz$ filter are indicated in Fig.~\ref{fig:field_layouts}. The respective areas are summarized in Table~\ref{tab:cat_areas}.  

\begin{table} \centering
\footnotesize
\caption{Point source depths and depths measured with 2\arcsec{} diameter apertures (both 5 $\sigma$). The measurement of the point source depth is described in Sect.~\ref{subsec:fphot}, while the measurement of the 2\arcsec{} aperture depth achieved is described in Sect.~\ref{subsec:image_depths}. As described in Sect.~\ref{subsec:fphot}, photometric uncertainies are underestimated for \textit{Spitzer}/IRAC and so a point source depth is not provided herein. Where two values are given, the first value applies to EDF-N and the second value applies to EDF-F. Future releases from the DAWN survey will include deeper HSC $z$ data in EDF-F. $^{\dagger}$Depth achieved is measured only across the \enquote{full depth} region of each field, 9.37 deg$^{2}$ for EDF-N and 2.85  deg$^{2}$ for EDF-F (see Sect.~\ref{subsec:area_coverage} for details).}
\begin{tabular}{lcc}
\hline\hline
Instrument / Band & Point source & 2\arcsec{} aperture \\
    & depth$^{\dagger}$ & depth$^{\dagger}$ \\
\hline
CFHT MegaCam / $u$                     & 26.7, 26.4 & 26.5, 26.4 \\
Subaru HSC / $g$                       & 27.2, 27.2 & 26.9, 27.2 \\
Subaru HSC / $r$                       & 27.4, 27.4 & 26.8, 26.9 \\
Subaru HSC / $i$                       & 26.8, 27.0 & 26.4, 26.6 \\
Subaru HSC / $z$                       & 26.2, 25.1 & 25.7, 25.1 \\
Subaru HSC / $y$                       & 24.5, --   & 24.2, --   \\
Subaru HSC / NB0816                    & 23.2, 24.6 & 23.1, 24.5 \\
Subaru HSC / NB0921                    & 24.7, 25.3 & 24.4, 25.1\\
\textit{Spitzer} IRAC / [3.6\,$\mu$m]  & -- & 24.9, 25.1 \\
\textit{Spitzer} IRAC / [4.5\,$\mu$m]  & -- & 24.8, 24.9\\
\hline
\end{tabular}
\label{tab:depths}
\end{table}

\subsection{\label{subsec:image_depths}Image depths}

\begin{figure*}
    \centering
    \includegraphics[width=\textwidth]{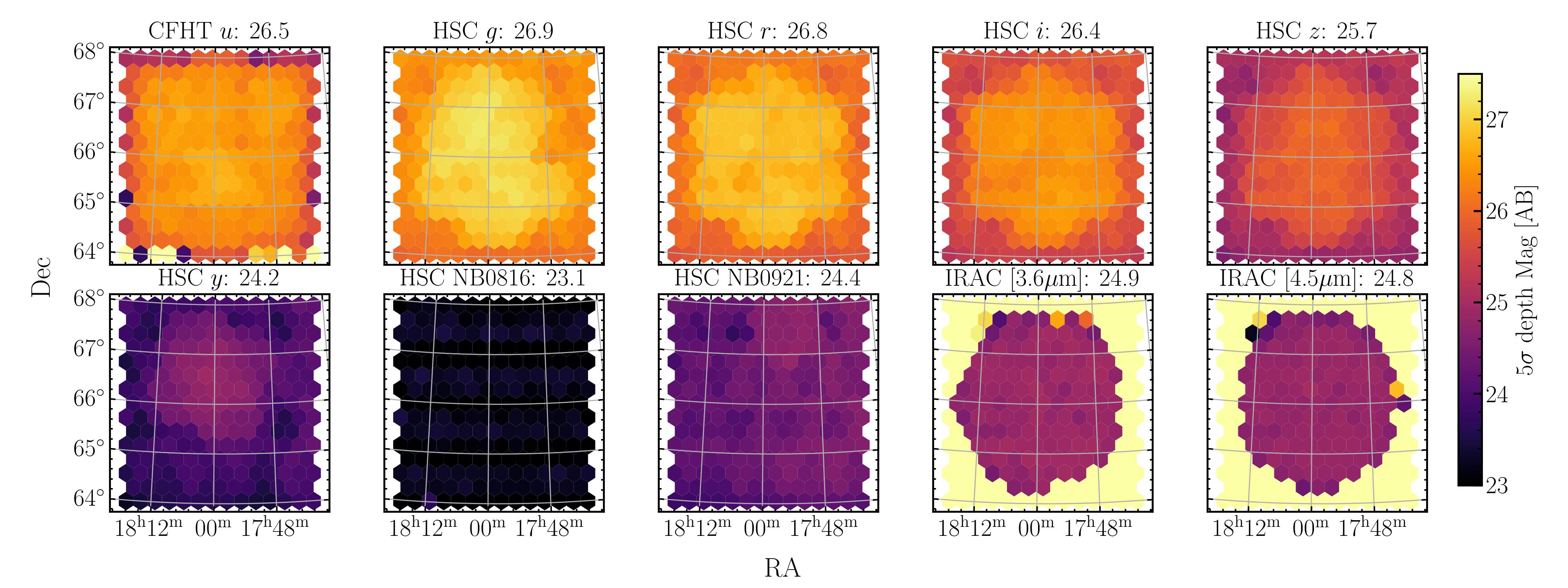}
    \includegraphics[width=\textwidth]{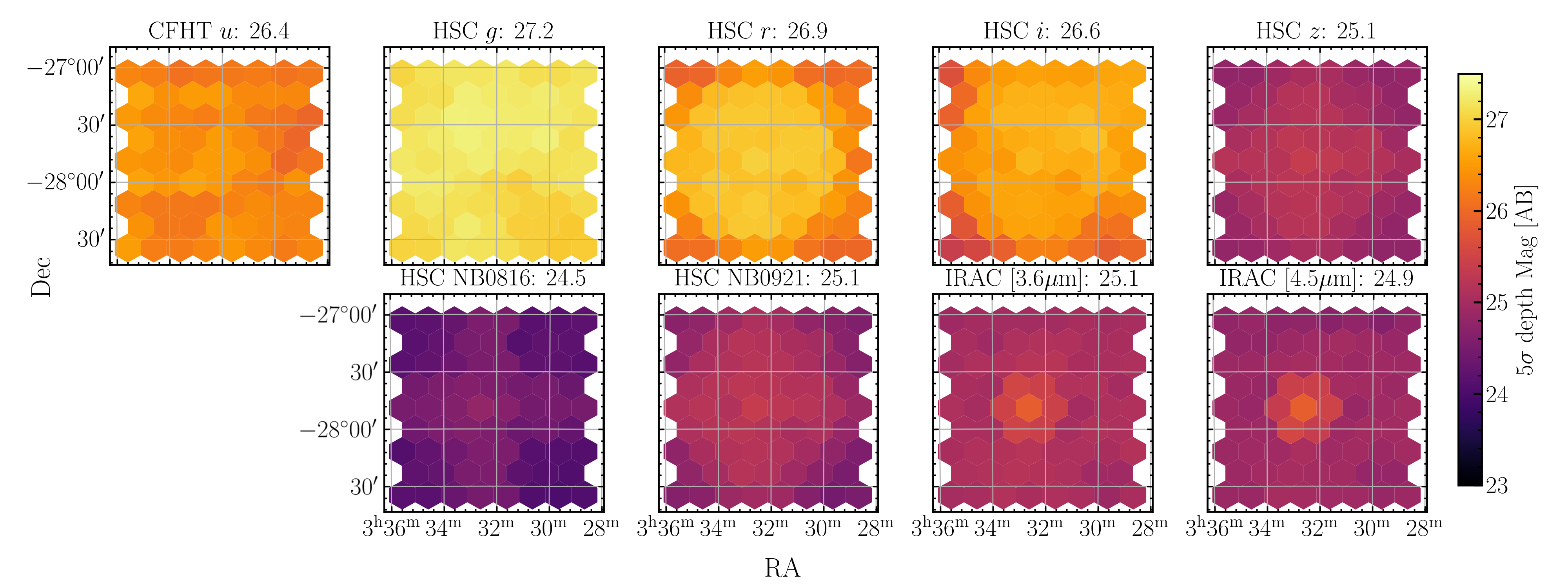}
    \caption{Depths and limiting magnitudes measured by the dispersion of empty aperture fluxes for each bandpass (see Sect.~\ref{subsec:image_depths} and Appendix~\ref{app:depths} for details). The top two rows correspond to EDF-N, and the bottom two rows correspond to EDF-F. The \enquote{full-depth} regions of EDF-N and EDF-F illustrated in Fig.~\ref{fig:field_layouts} is apparent by the areas of greater depth in the HSC $griz$ (see Sect.~\ref{subsec:area_coverage}, Table~\ref{tab:exptimes}, and Table~\ref{tab:cat_areas}). The quoted value along the top of each panel is the median 5$\sigma$ depth the full-depth region of each  field. Only the areas covered by the DAWN survey DR1 catalogues is shown, corresponding to the green rectangular regions in Fig.~\ref{fig:field_layouts}. Future data acquired from CFHT MegaCam and Subaru HSC will expand the areas of both fields and lead to more uniform coverage. The depths are summarized in Table~\ref{tab:depths}.}
    \label{fig:depths}
\end{figure*}

The limiting magnitude(s) of a survey is an essential characteristic for understanding the properties of galaxies detected therein and for comparing one survey to another. The amount of variation of noise in the sky background dictates the limiting magnitude by effectively establishing a minimum object flux that can be reliably measured. Many surveys (e.g., \citealt{Laigle2016,Weaver2021}) use the dispersion in flux measurements computed from many independent fixed-size apertures, placed away from astronomical sources, to describe the level of variation in the sky-background and thus the limiting magnitude. The dispersion is related to the \enquote{depth} of an image, where depth in this context relates to the limiting magnitude at detection or photometry. However, it is necessary to carefully consider the impact of undetected sources and the spatial sampling rate. Failing to address these challenges can bias the estimated limiting magnitudes up to $\sim$0.3 mag (see Appendix~\ref{app:depths}).

The depth and limiting magnitudes of the data used herein are measured from the dispersion of empty aperture fluxes according to the following method. First, apertures with 2\arcsec{} diameters are randomly placed in regions away from detected objects using the segmentation maps output by \SEP{} (see Sect.~\ref{subsec:source_detection}). Each image is sampled at a rate of one aperture per five square-arcseconds. Then, the flux is measured in each aperture and sigma-clipping is performed on the distribution of measured fluxes at the five-sigma level to limit the impact from undetected astronomical sources. To further mitigate the contribution of undetected astronomical objects, the next step is to model the distribution using a Gaussian function. A Gaussian function is iteratively fit to the data to extract the true profile of the empty aperture flux dispersion distribution. From the best-fit model, the standard deviation of the distribution may then be measured. The final quoted depths are given by the standard deviation of the final Gaussian fit, multiplied by five (i.e., 5$\sigma$ limiting magnitudes). Figure~\ref{fig:depths} depicts the variation in the 2\arcsec{} limiting magnitudes measured across the field. The limiting magnitudes are summarized in Table~\ref{tab:depths}. Further consideration regarding limiting magnitudes and the method described above are provided in Appendix~\ref{app:depths}. 

\subsection{\label{subsec:masking}Masking}

Bright foreground stars negatively affect photometry by obscuring galaxies directly and indirectly through internal reflection and scattered light within the telescope, saturation, and \enquote{ghosts.} Furthermore, reduction pipelines often struggle to accurately model the sky background in their vicinity, leading to significant fluctuation in the quality of background subtraction. Therefore, it is typically preferred to mask large regions surrounding bright stars in the images entirely. Bright star masks are created using the \textit{Gaia} DR3 catalogue \citep{Gaia2022}, masking all identified stars brighter than 17 mag in the \textit{Gaia} $G$ band, where the size of the masked region is proportional to the star's brightness. The masks are applied at all wavelengths. At present, these are the only masks used to reduce the impact of spurious objects, though future releases may include additional masks for other known sources of artifacts. The total areas affected by bright stars and excluded in the DAWN survey DR1 catalogues is given in Table~\ref{tab:cat_areas}.

\subsection{\label{subsec:spectroscopy}Spectroscopic data}

A number of programs with different instruments have targeted galaxies in EDF-N and EDF-F for spectroscopy. In EDF-N, the \textit{AKARI} team primarily observed infrared-selected galaxies and AGN \citep{Goto2017}, while in EDF-F, thousands of galaxies in the GOODS-S region have been targeted \citep{Garilli2021,Kodra2023}. The Texas Euclid Survey for Lyman-Alpha (TESLA) is conducting spectroscopic analysis of the EDF-N NEP field using the Visible Integral-field Replicable Unit Spectrograph (VIRUS) on the Hobby-Eberly Telescope \citep{ChavezOrtiz2023}. VIRUS is designed to be sensitive to Ly$\alpha$ emission from galaxies at 1.9 $< z <$ 3.5 above a flux limit of 5$\times$10$^{-17}$ erg s$^{-1}$ cm$^{-2}$ (5$\sigma$). In addition, H20 has been carrying out spectroscopic follow-up of objects selected from the DR1 catalogues using the Deep Extragalactic Imaging Multi-Object Spectrograph \citep[DEIMOS;][]{Faber2003} on the 10\,m Keck II telescope. The H20 efforts have been primarily to confirm galaxy protoclusters at $z > 4$ by targeting over-dense regions associated with Lyman-break galaxies. A paper describing the target selection for H20 spectroscopy is forthcoming (Chartab et al. in prep.). The use of spectroscopy in this work is limited to the validation of \photoz{}s (Sect.~\ref{subsec:photoz_valid}). As the H20 spectroscopic data are still being gathered and processed, only the external spectroscopic datasets are employed in this work. 

For the purposes of validating \photoz{}s, we use a \specz{} sample from the \textit{AKARI} team \citep{Goto2017} which includes a total of 1987 sources in EDF-N. In EDF-F we use the GOODS-S CANDELS \specz{} sample \citep{Kodra2023}, including 2697 objects as well as the public VANDELS \specz{} sources \citep{Garilli2021}, including 2085 sources. For the EDF-N sample of spectroscopic sources, we only use those labeled as galaxies within the \textit{AKARI} catalogue, removing 527 sources labeled as AGN or having X-ray activity. 


\section{\label{sec:SD_Photometry}Source detection and photometry}

Flux measurements and uncertainties in the DAWN survey DR1 catalogues are measured from the H20 and \textit{Spitzer} multiwavelength imaging using \farmer{}. In brief, \farmer{} is a pythonic wrapper, driver, and user-interface for the model-optimization code \tractor{} \citep{Lang2016}. \tractor{} provides a library of models to describe astrophysical light profiles and methods for fitting these models but requires customized code to employ them in any efficient way. \farmer{} was first introduced in \cite{Weaver2021}, where the model-based photometry method was used to create one of the two publicly available COSMOS2020 catalogues (the other being made with \enquote{classic} aperture photometry). Therein, the authors demonstrated the reliability of \farmer{} in producing accurate photometry through a detailed comparison with well-understood aperture photometry of the \enquote{classic} catalogue. \farmer{} flux measurements for COSMOS2020 were further validated through SED modeling and yielded excellent \photoz{}s, especially for faint sources. Lastly, the capabilities of \farmer{} were investigated and benchmarks were quantified in \cite{WeaverFarmer} using simulations of deep multiwavelength imaging. The authors validated various outputs of \farmer{}, including photometry, resulting number counts, galaxy shapes, and statistical metrics related to goodness-of-fit. The reader is referred to these works for a detailed explanation of the inner workings of \farmer{}. The remainder of this section includes a summary of the relevant steps to measuring photometry using \farmer{} and a discussion of features unique to H20. 

Similar to the \enquote{patches} used by \texttt{hsc\textunderscore pipe} \citep{Bosch2018}, \farmer{} breaks apart large survey mosaics into smaller images referred to as \enquote{bricks}. Bricks are used to be more easily handled in computer memory than full mosaics and can be processed in parallel. In general, it is advantageous to define the dimensions of bricks such that the ratios of the mosaic axes lengths to the brick axes lengths are integers, enabling straightforward comparisons and treatment across bricks. For this reason, the bricks in the EDF-N field are 4000 pixels on each side (11.2 arcmin), representing a 22$\times$22 grid of the EDF-N mosaic. Likewise, EDF-F bricks are 3620 pixels on a side (10.1 arcmin), representing a 10$\times$10 grid of the EDF-F mosaic, which, as previously stated, only includes the deepest region as of this publication. Slight differences in brick size do not impact any significant features of the photometry and are only used in accordance with the mosaic size (that is, to achieve an integer multiple of bricks).

\subsection{\label{subsec:source_detection}Source detection}

Source detection begins with designing the image from which sources are to be detected. A multiwavelength composite image is built as the detection image, where each pixel value corresponds to a probability of belonging to sky-noise, following the now widespread approach first introduced by \cite{Szalay1999}. In short, the pixel values of the multiwavelength composite image approximately follow a modified $\chi^{2}$ distribution with degrees of freedom equal to the number of input images. The probablity of belonging to sky noise may then be directly inferred from the pixel value. Being primarily interested in the high-redshift universe, images from the deepest and reddest bandpasses in DAWN survey DR1 catalogues are combined. These include the HSC $r+i+z$ bandpasses. Note that the assignment of a particular wavelength range spanned by the detection image directly influences the selection of galaxies (see Sect.~\ref{subsec:completeness}). Future catalogues of the DAWN survey will select galaxies from similarly deep near-infrared imaging from \Euclid. 

The images are combined using \texttt{SWARP} with the \texttt{CHI-MEAN} co-addition setting. This setting creates a multiwavelength composite where the pixel values follow a $\chi$-distribution. This distribution is re-centered on the mean value depending on the number of inputs (see appendix B in \citealt{Drlica-Wagner2018} for a comparison of the different combination settings in \texttt{SWARP}, including a version of the original method used by \citealt{Szalay1999}). This technique has been used by the CFHT Legacy Survey \citep{Cuillandre2012}, the COSMOS survey \citep{Ilbert2013,Laigle2016,Weaver2021}, the Dark Energy Survey \citep{Drlica-Wagner2018}, DECam images in the SHELA survey \citep{Stevans2018}, and recent work combining datasets from different HST campaigns \citep{Bouwens2021}. 

To carry out object detection and segmentation, \farmer{} utilizes the python library of Source Extraction and Photometry (\SEP{}; \citealt{Barbary2016}), a python interface wrapping many of the core functionalities of the widely used Source Extractor \citep{Bertin1996}. Thee source detection parameter settings used here are identical to those used in the COSMOS2020 catalogue \citep{Weaver2021}. 
Sources located within the bright star masks (Sect.~\ref{subsec:masking}) are removed after detection. All other detected sources are catalogued, and their properties measured by \SEP{} (e.g., position, shape) are stored for the modeling stage as initial conditions (Sect.~\ref{subsec:models}).

In total, \num{5286829} objects are detected over the 16.87 deg$^{2}$ area of the DR1 EDF-N catalogue, where \num{3513211} of the detected objects are in the 9.37 deg$^{2}$ full-depth region. In EDF-F, \num{1062645} objects are detected over the DR1 2.85 deg$^{2}$ area, where \num{727678} are in the 1.77 deg$^{2}$ full-depth region.

\subsection{\label{subsec:psfs}PSF handling}
\begin{figure*}
     \centering
     \includegraphics[width=\textwidth]{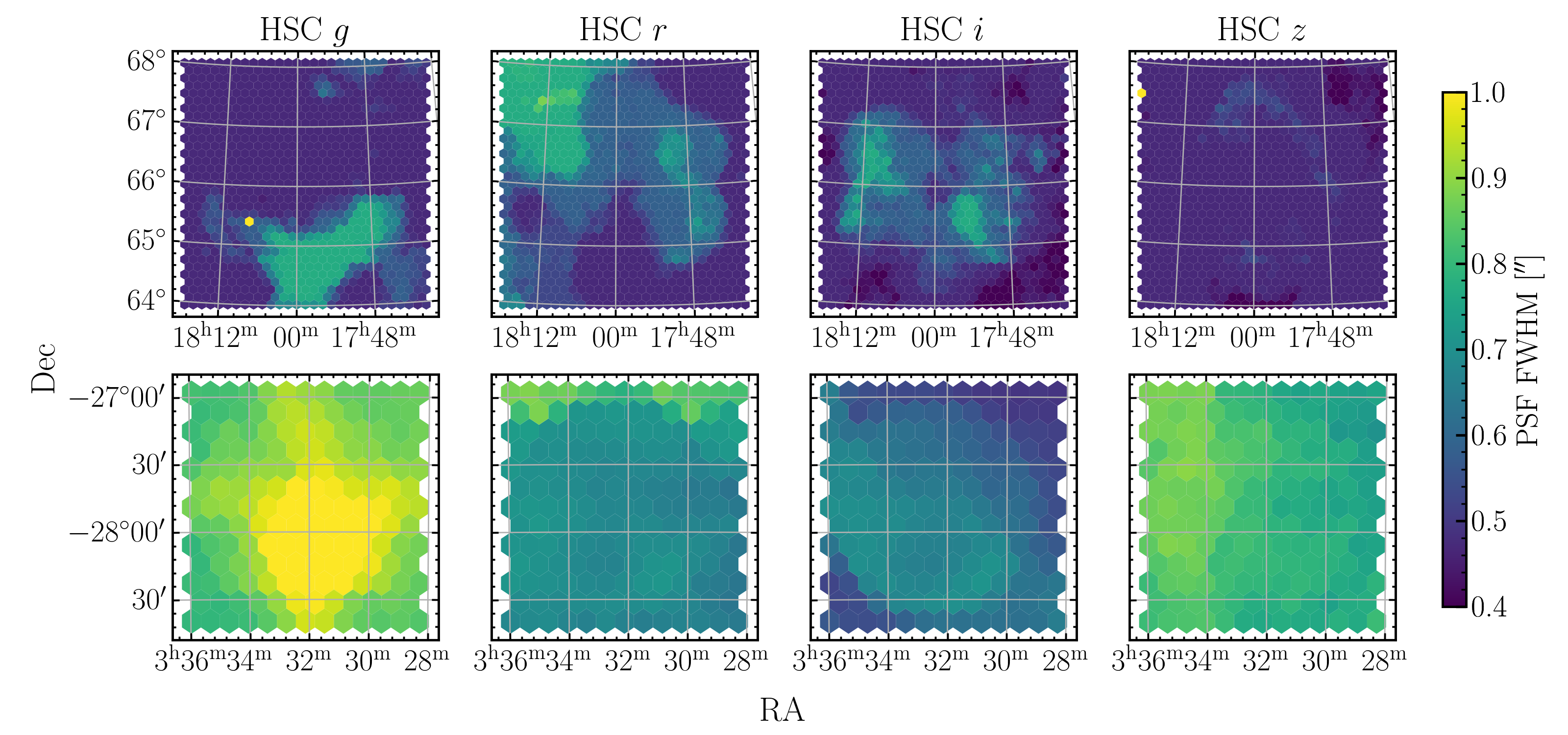}
     \includegraphics[width=\textwidth]{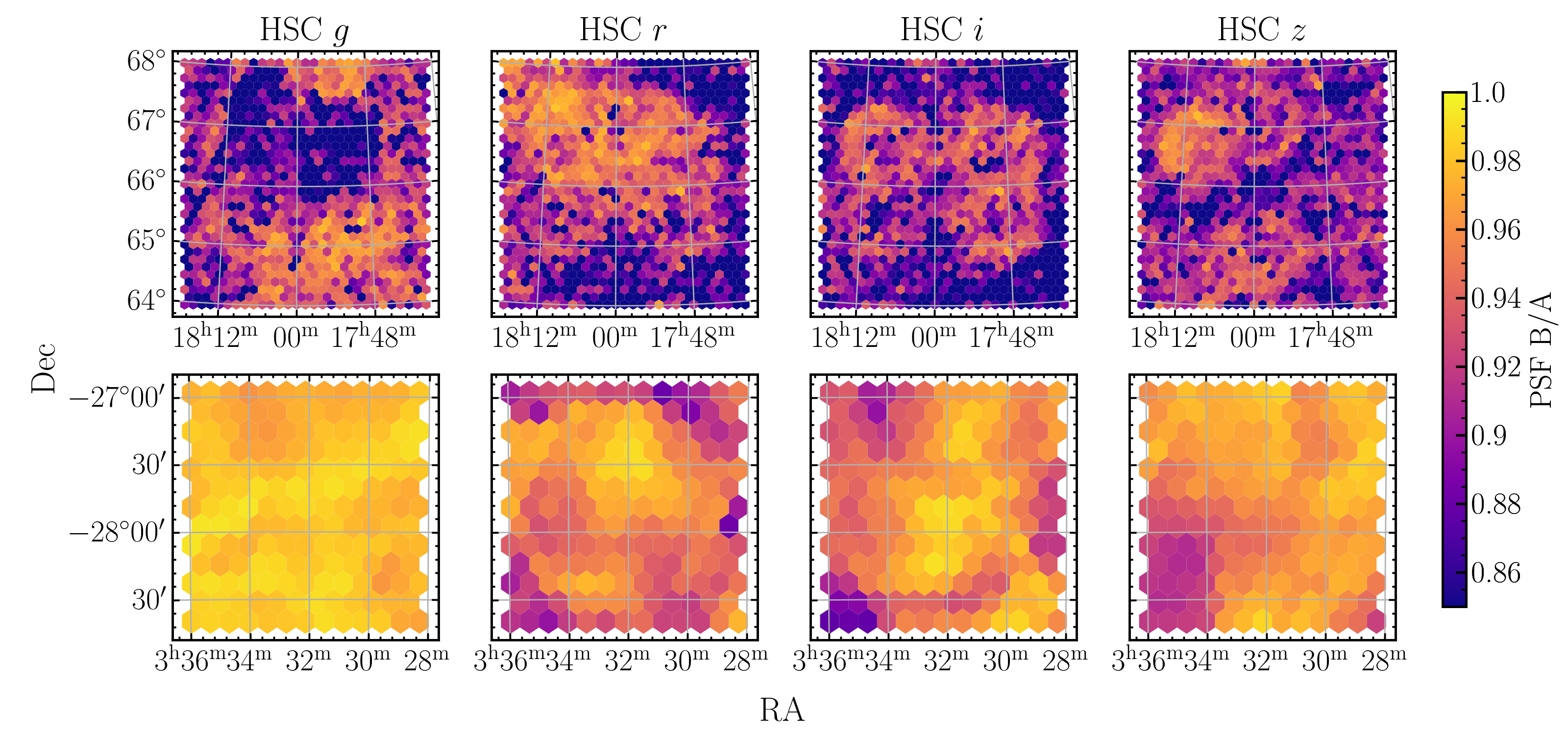}
     \caption{Variation of the HSC $griz$ PSF FWHM (top) and ellipticity ($B/A$, bottom). EDF-N is shown in the top row of each pair, while EDF-F is shown in the bottom. Only the areas covered by the DAWN survey DR1 catalogues are shown, corresponding to the green square regions in Fig.~\ref{fig:field_layouts} and consistent with Fig.~\ref{fig:depths}.}
     \label{fig:psf-variation}
 \end{figure*}

\begin{figure*}
    \centering
    \includegraphics[width=\textwidth]{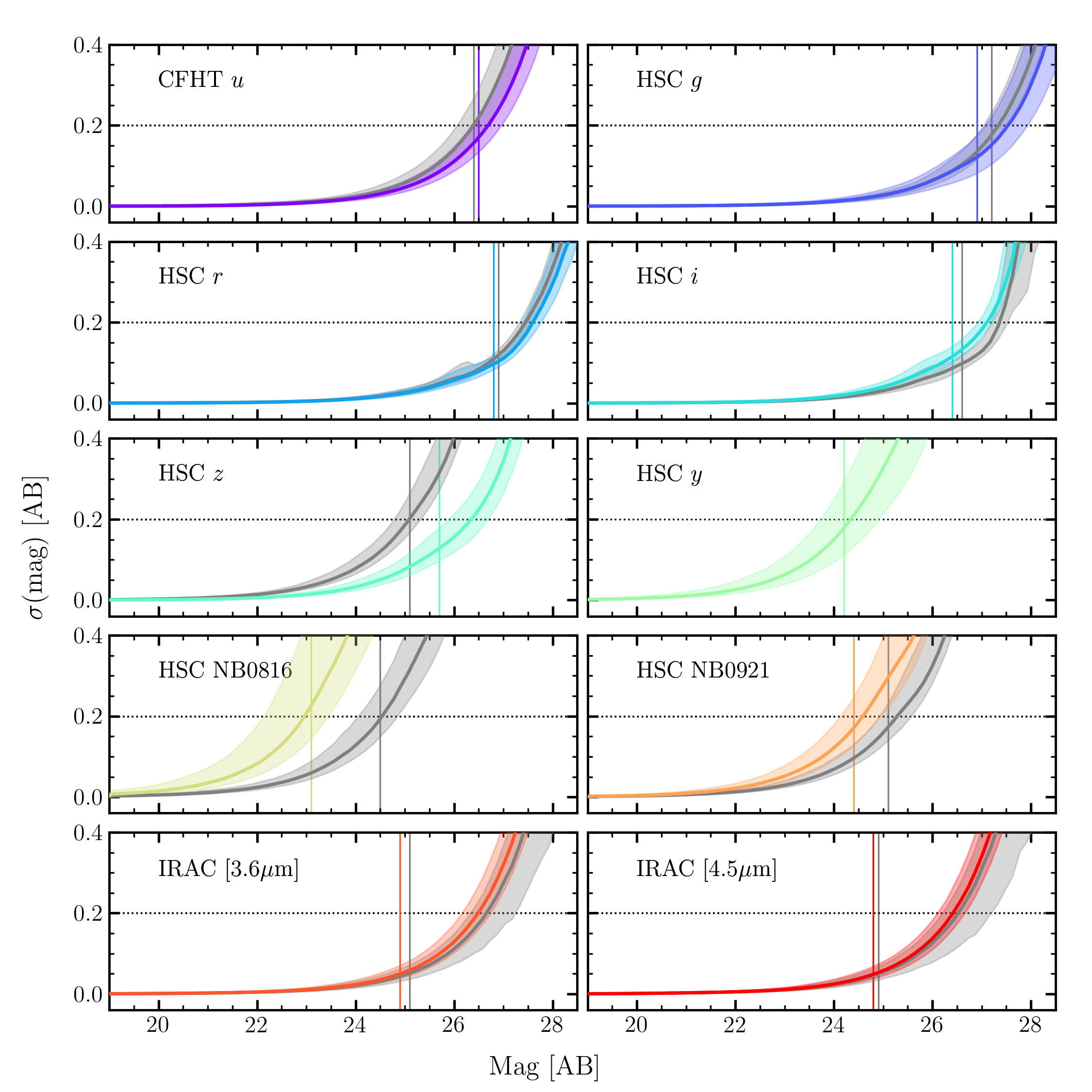}
    \caption{Magnitude vs. magnitude error for each unique facility/filter combination, with EDF-N shown in colour and EDF-F shown in grey. Each solid line represents the median magnitude uncertainty as a function of magnitude for a given band. The shaded regions are bounded by the 84th and 16th percentiles of the magnitude error distributions, enclosing 68\% of the objects. The 5$\sigma$ limiting magnitudes measured with 2\arcsec{} diameter apertures reported in Table~\ref{tab:depths} are shown by vertical lines. A horizontal dotted line indicates a 5$\sigma$ model-based photometric uncertainty. }
    \label{fig:mag_magerr}
\end{figure*}

Most methods of photometry, including both aperture photometry and model-based photometry, require accurate characterization of the point spread functions (PSFs) for the image. Aperture-based methods require PSF homogenization -- an intentional degradation of high-resolution information -- to obtain consistent measurements of total fluxes and colours across images of varying resolution. One of the benefits of some model-based photometry methods, including those used by \tractor{}, is that PSF homogenization is not necessary. \tractor{} uses parametric representations of sources which are independent of the image PSF. However, before constructing models of the detected sources with \farmer{}, representations of the PSF in each of the imaging sets must be obtained. Then, when a model is fit to a source observed in a given bandpass, the PSF corresponding to that image is simply convolved with the model, preserving the full information of each image. Thus, instead of homogenizing the PSF of all of the multiwavelength imaging to a common reference, each band is treated  independently according to its PSF model. 

Beginning with our bluest band, PSFex \citep{Bertin2013} is used to create models of the CFHT MegaCam $u$-band imaging. Bright, but not saturated, point-like objects are identified via their position in the magnitude-effective radius diagram. One spatially constant PSF model is created per mosaic brick in each field, providing a sampling of approximately 30 PSF models per deg$^{2}$ in EDF-N and 35 PSF models per deg$^{2}$ in EDF-F. As noted in \cite{Weaver2021}, \farmer{} works best when supplied with large PSF renderings, which can account for the light-profile of objects that may include significant flux in the wings of some sources. Therefore, PSF models with 201 pixels in diameter (\ang{;;33.77}) are created.

For the Subaru HSC bands, a grid of PSF models is constructed to describe and account for the variation of the PSF across the survey area. This is required because the sigma-clipping step in the image stacking (Sect.~\ref{subsec:optical}) deforms the uniformity of the PSF across each field. 
Furthermore, creating large images (several degrees on a side) with the same tangent point may also cause a non-negligible impact on the PSF. The initial grid spacing is \ang{;;29}, which matches the sampling scale for Spitzer IRAC (see below). The PSF models are built using routines within \texttt{hscpipe}. PSF models with radii of 103 pixels are extracted, manually overriding the default settings of \texttt{hscpipe}, which otherwise produces PSF models with radii of 43 pixels. PSF models with axis ratios less than 0.9 and those with first or second moments that could not be accurately measured are flagged.

A grid of PSF models across the survey area for \textit{Spitzer}/IRAC images is built in a similar manner as for HSC. For this operation, the software \texttt{PRFmap} (A. Faisst, private communication) is used. Across each \textit{Spitzer}/IRAC mosaic (in this case, \chOne{} and \chTwo{}), the code considers each of the individual frames that went into creating the final co-added image and builds a specific Point Response Function (PRF) model for each frame. Each PRF model is unique because the response function is not rotationally symmetric. Finally, the individual PRF models are stacked at each grid point. These PRF models are constructed on the same pixel scale as the native IRAC images, \ang{;;0.6}, before being resampled to the HSC pixel scale, \ang{;;0.168}.

\subsection{\label{subsec:models}Model determination}

Once the PSFs models are constructed for each set of imaging, parametric models may be determined for the detected sources' light profiles. The default configuration of \farmer{} is used, which includes consideration of five different parametric models to describe a given sources. The parametric models include options for both point-like and extended objects and are fully described in \cite{Weaver2021} and \cite{Weaver2023}. The Subaru HSC $r$, $i$, and $z$ images are used individually as the joint constraints for models, which are the same bands used to create the composite detection image. This ensures that the PSF can be properly handled in each image, that information utilized at the detection stage is carried through the photometry stage, and that all detected sources will have model constraints coming from at least one band.  Using the combined detection image is not advised, as the properties of the PSF for the detection image are not easily determined. Future releases from the DAWN survey will include both source detection and model-based photometry using \Euclid near-infrared imaging. 

Model parameters, such as position and flux, are initialized with values determined from the detection stage. Sources with approximately overlapping light profiles (described as \enquote{modeling groups} in \citealt{WeaverFarmer}) are fit simultaneously with one another to account for their overlapping light profiles. The model that best describes each source's light profile is selected according to a decision tree which proceeds from simple to more complex models in an approach described and validated in \cite{Weaver2021} and \citet{WeaverFarmer}. The final model is optimized according to the constraints imposed by the Subaru HSC $r$, $i$, and $z$ images, which include flux, position, and shape. However, the model includes only one value for flux and accordingly must be re-fitted to each individual band in the forced photometry stage (Sect.~\ref{subsec:fphot}).

A small subset of detected objects ($< 0.5\%$) in each field are not able to be fit by a model, likely due to contamination from a bright neighboring source. The positions of the objects are recorded in the catalogues and their cumulative area for each field is reported in Table~\ref{tab:cat_areas}.

\begin{figure}
    \centering
    \includegraphics[width=\columnwidth]{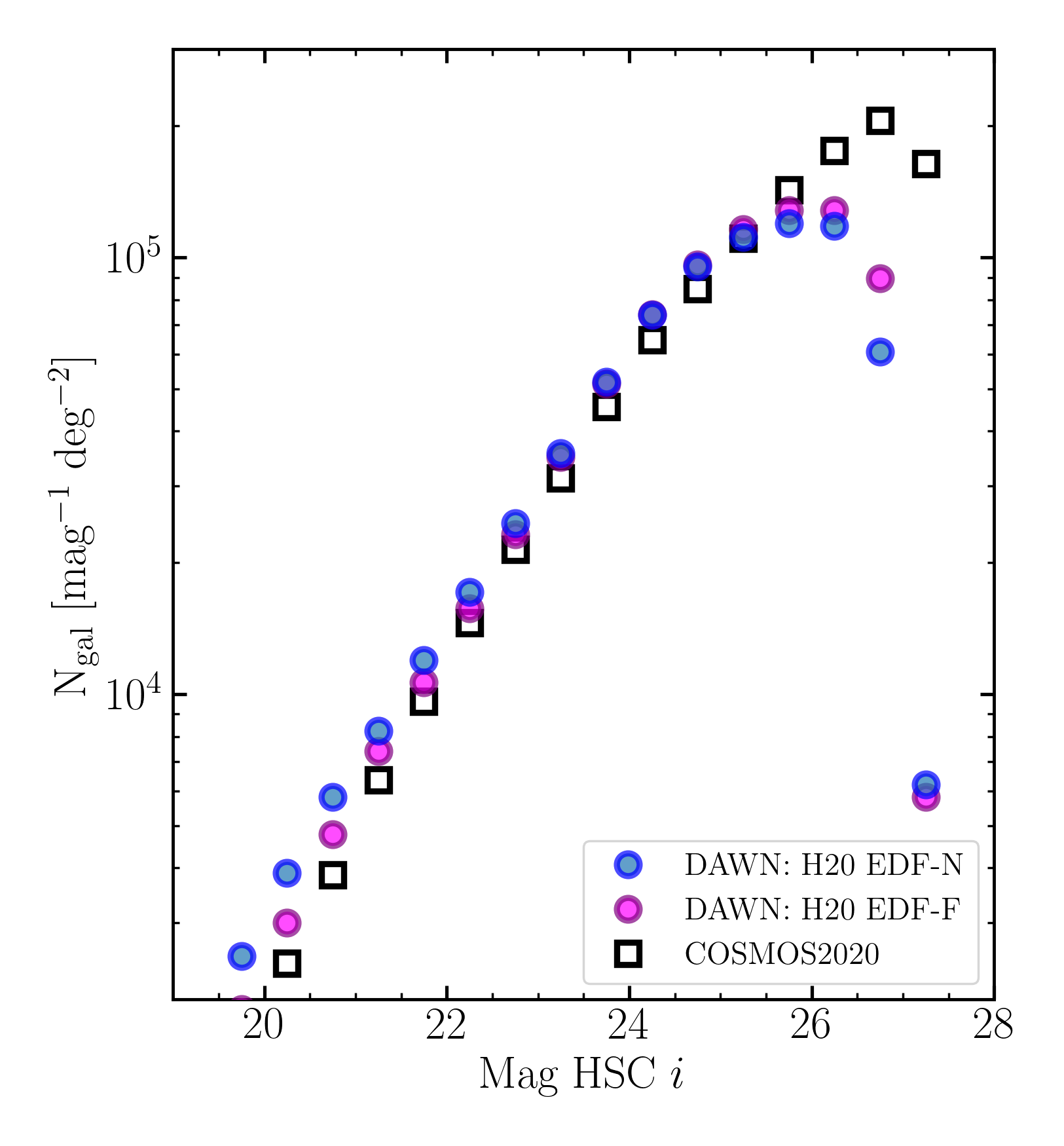}
    \caption{HSC $i$ number counts of galaxies detected in the DAWN survey DR1 HSC $r+i+z$ detection image. As a comparison, also shown are the HSC $i$ number counts from the COSMOS2020 \farmer{} catalogue (\citealt{Weaver2021}), which was created using many of the same methods used in the present work. Bin spaces are 0.5 mag in each case. A small offset in the measured number counts before the turnover is caused by differences in the methods for separating stars from galaxies.}
    \label{fig:number_counts}
\end{figure}

\subsection{\label{subsec:fphot}Forced photometry}

Total model fluxes are measured by re-fitting the final optimized models obtained during the model determination stage (Sect.~\ref{subsec:models}) at the locations of each detected source in each of the H20 and \textit{Spitzer}/IRAC images. This operation is commonly referred to as \enquote{forced} photometry. Here, morphological parameters of the models are held fixed while the flux is re-optimized in each band. Positions are also anchored to the HSC $r+i+z$ model values, but are left to vary within a strict Gaussian prior with a standard deviation of 0.3 pixels (note that as all images have been resampled to the same pixel scale, this corresponds to a constant angular scale across all images). This flexibility can overcome slight offsets in astrometry and prevent erroneous positions for faint objects. 

Total object fluxes are measured in the CFHT MegaCam $u$ band, the Subaru HSC $griz$ bands, and in the Spitzer IRAC \chOne{} and \chTwo{} bands. Where available, we also measure photometry from archival Subaru HSC $y$ (restricted to EDF-N), NB0816, and NB0921 imaging. These flux measurements, in addition to flux uncertainties, are recorded in the catalogue. To reiterate the description provided in \cite{Weaver2021}, \tractor{} computes flux uncertainties by summing the weight map pixels in quadrature, where each pixel is further weighted by the unit-normalized model profile (for point-sources is simply the PSF). This method prioritizes the per-pixel uncertainty directly under the peak of the model profile and gives less weight to the per-pixel uncertainty near the edges of the model. 

Band-specific relationships of the measured flux and flux uncertainty are presented in Fig.~\ref{fig:mag_magerr} after converting the flux and flux error measurements to magnitude and magnitude error, respectively. The curves representing this relationship for each facility and bandpass follow the expected distributions.  That is, they are smoothly and monotonically increasing for fainter objects and measured flux uncertainties representing a 5$\sigma$ measurement are near to the values measured via the dispersion of empty aperture fluxes in Sect. \ref{subsec:image_depths}. The exception to this is \textit{Spitzer}/IRAC, where the uncertainties measured by \farmer{} appear to be underestimated. A similar feature was noticed in \cite{Weaver2021}, wherein the authors attributed this underestimation to the difficulty of accounting for the contribution of pixel co-variance towards total photometric uncertainty, even for model-based methods like \tractor{}. As the \textit{Spitzer}/IRAC images have been significantly oversampled from their native pixel scale, from \mbox{\ang{;;0.6}/pixel} to \mbox{\ang{;;0.168}/pixel}, the amount of covariance in the resampled image plan is expected to be significant. 

When using \farmer{}, faint objects are predominantly modeled as point sources \citep{WeaverFarmer}. For the CFHT and HSC filters, the curves depicted in Fig.~\ref{fig:mag_magerr} may be used to infer the limiting magnitudes for point source photometry of the images, given the model of the PSF. The point source depth at 5$\sigma$ corresponds to the intersection of a given curve and the 5$\sigma$ uncertainty (dotted line). These values are presented in Table~\ref{tab:depths}. The limiting magnitude of point sources is fainter than for an aperture of fixed size when the FWHM of the PSF is more narrow than the aperture. An image with a PSF of similar scale to the fixed aperture should have a similar point source depth compared to the aperture depth. Accordingly, instrument and filter combinations with broad PSFs in Fig.~\ref{fig:mag_magerr} have similar point source depths compared to their corresponding aperture depths. Again, the exception is \textit{Spitzer}/IRAC, where the image properties preclude a proper comparison. 

\farmer{} provides  further information, in addition to fluxes and uncertainties, related to the model-fitting. \cite{WeaverFarmer} provides a detailed explanation of the different possible outputs from \farmer{}, but in short, the code also provides $\chi^{2}$ goodness of fit metrics as well as three metrics measured from the moments of the residuals weighted by the per-pixel variance, including the median, standard deviation, and D’Agostino’s $K^{2}$ test. 

\begin{table}
\caption{Tabulated logarithmic galaxy number counts of the DAWN catalogues as depicted by Fig.~\ref{fig:number_counts}. These counts, representing an HSC $r+i+z$ selection, are listed in units of mag$^{-1}$\,deg$^{-2}$ with bin spacing of 0.5\,mag.}
\centering\footnotesize
\begin{tabular*}{0.6\columnwidth}{lcc}
\hline\hline
Mag & EDF-N & EDF-F \\
\hline
19.25 & 3.22 & 3.07 \\
19.75 & 3.40 & 3.28 \\
20.25 & 3.59 & 3.48 \\
20.75 & 3.76 & 3.68 \\
21.25 & 3.92 & 3.87 \\
21.75 & 4.08 & 4.03 \\
22.25 & 4.23 & 4.20 \\
22.75 & 4.39 & 4.37 \\
23.25 & 4.55 & 4.54 \\
23.75 & 4.71 & 4.71 \\
24.25 & 4.87 & 4.87 \\
24.75 & 4.98 & 4.98 \\
25.25 & 5.05 & 5.07 \\
25.75 & 5.08 & 5.11 \\
26.25 & 5.07 & 5.11 \\
26.75 & 4.78 & 4.95 \\
27.25 & 3.79 & 3.76 \\
\hline
\end{tabular*}
\label{tab:gal_counts}
\end{table}

\subsection{\label{subsec:gcounts}Galaxy number counts}

The full-depth area of the EDF-N catalogue is 9.37 deg$^{2}$, and after accounting for masked regions (see Sect.~\ref{subsec:masking}) and failed models, the effective area is 8.42 deg$^{2}$. The full-depth area of the EDF-F catalogue is 1.77 deg$^{2}$, and after accounting for masked regions and failed models, the effective area is 1.71 deg$^{2}$. The number counts of each field are shown in Fig.~\ref{fig:number_counts}. The two fields show excellent consistency with EDF-F reaching slightly fainter sources due to greater HSC $r$ and  $i$ band depths (see Table~\ref{tab:depths} and Fig.~\ref{fig:depths}). Disagreement on the bright end may be explained by the significantly greater stellar density in EDF-N due to its low galactic latitude, perhaps indicating an incomplete removal of all stars. 

As an initial step towards validating the H20 photometry, the galaxy number counts in EDF-N and EDF-F are also compared with those of COSMOS2020 reported by \cite{Weaver2021}. Recall that COSMOS2020 shares many of the same methodologies employed here, most notably, the method of photometry, in addition to the wavelength coverage. For this presentation, the Subaru HSC $i$ band is  selected as it covers the central wavelengths of the detection image, although it is the bluest band included in the COSMOS2020 detection image. Stars have been identified and removed via SED-fitting, following the procedures described in Sects. \ref{subsec:lephare} and \ref{subsec:eazy}. Galaxies at magnitudes HSC $i < 25$ show good agreement with the well-established HSC $i$ counts of COSMOS2020. Slight variations within this magnitude limit are explained by differences in the methods used in each work to separate stars from galaxies. For example, COSMOS2020 uses morphology from the \textit{Hubble} Space Telescope in addition to SED-fitting to remove likely stars and also uses more bands in the SED-fitting. At magnitudes HSC $i > 25$, the disagreement is dominated by the difference in depths for the two surveys. The disagreement on the faint end is further exacerbated by the combination of near-infrared wavelengths in the COSMOS2020 detection image, which enables detection of optically faint galaxies. 


\section{\label{sec:photoz}Photometric redshifts}

Several works \citep{Weaver2021,Kodra2023,Pacifici2023} have demonstrated the utility of having multiple \photoz{} estimates from different codes for every source. Their approach is followed, and \photoz{}s are computed for the DAWN  survey DR1 catalogues using \lephare{}~\citep{Arnouts2002,Ilbert2006} and \eazy{}~\citep{Brammer2008}. HSC narrow bands are not included during SED fitting with either code because spurious photometric measurements in their limited wavelength ranges can drive systematic biases, for example, requiring an emission line at a given wavelength. 

\subsection{\label{subsec:lephare}\lephare{}}
\begin{figure}
    \centering
    \includegraphics[width=\columnwidth,trim={0, 0.2cm, 0, 0.2cm},clip]{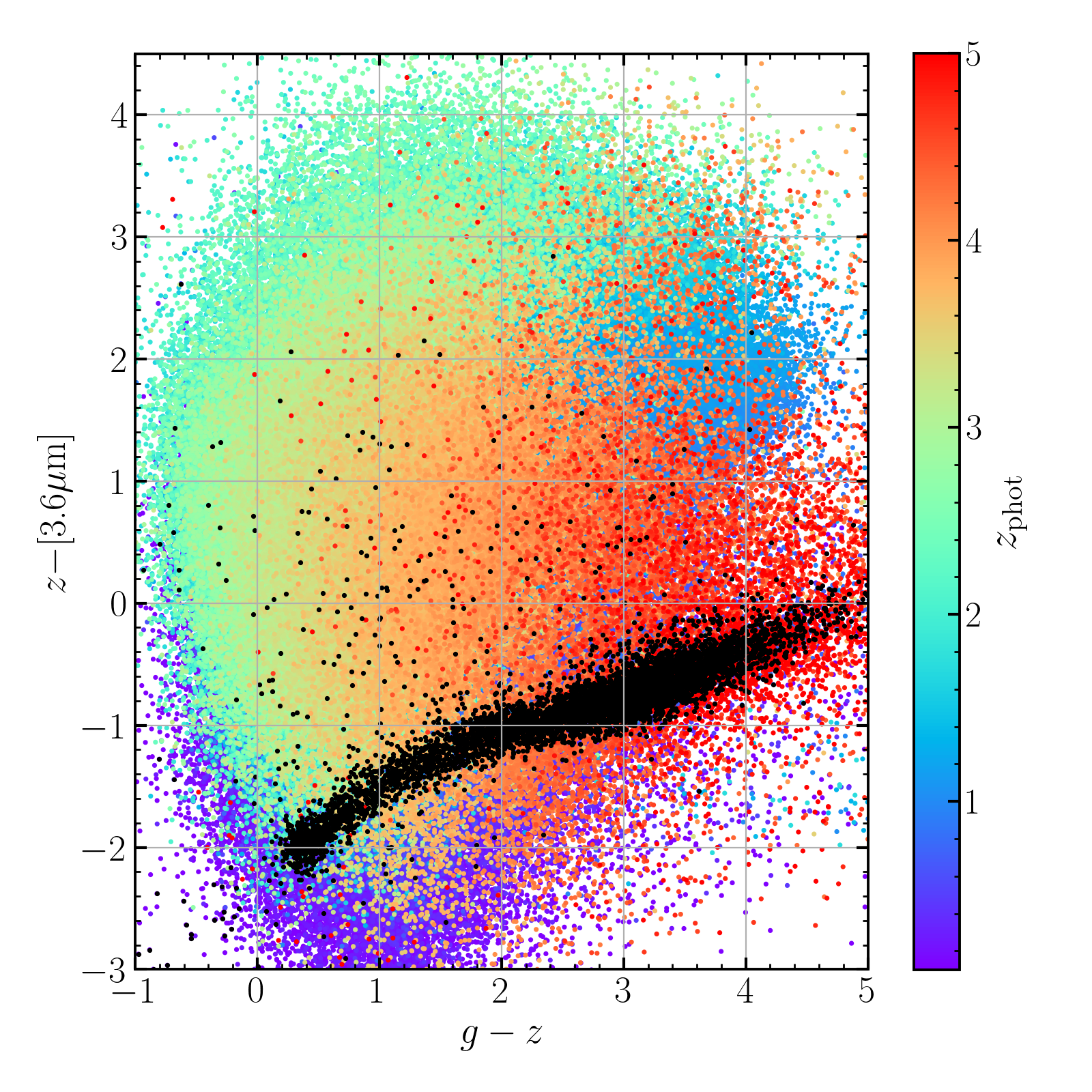}
    \caption{Colour-colour diagram for stars and galaxies identified via SED-fitting. Only sources with $\mathrm{S/N} > 3$ in the listed bands are included. Galaxies are coloured according to their \photoz{} measured by \lephare{}, and sources with $z_{\rm phot} > 5$ are all shown in red. Stars are shown as black points following a well-defined sequence in colour-colour space. A comparison validating the approximate locations of galaxies in colour-colour space as a function of redshift, as well as the size and extent of the stellar sequence, can be made with Fig. 12 of \cite{Weaver2021}.}
    \label{fig:star_galaxy}
\end{figure}

\begin{figure*}
    \centering
    \includegraphics[width=\textwidth]{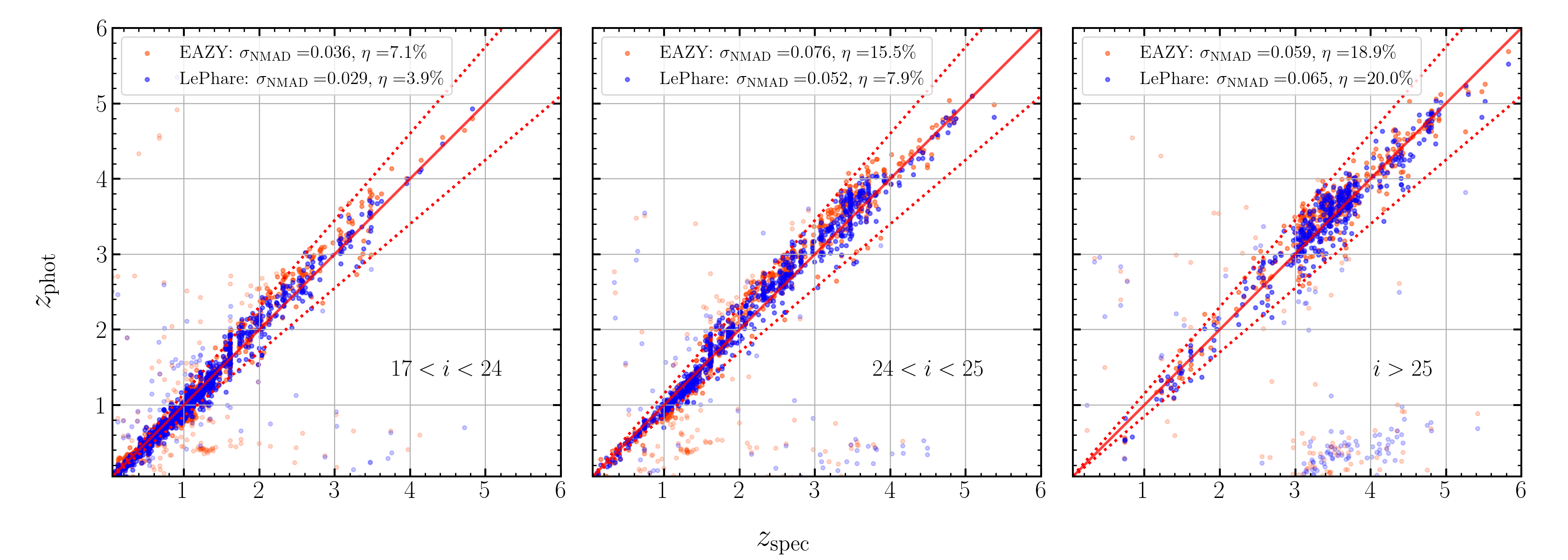}
    \includegraphics[width=\textwidth]{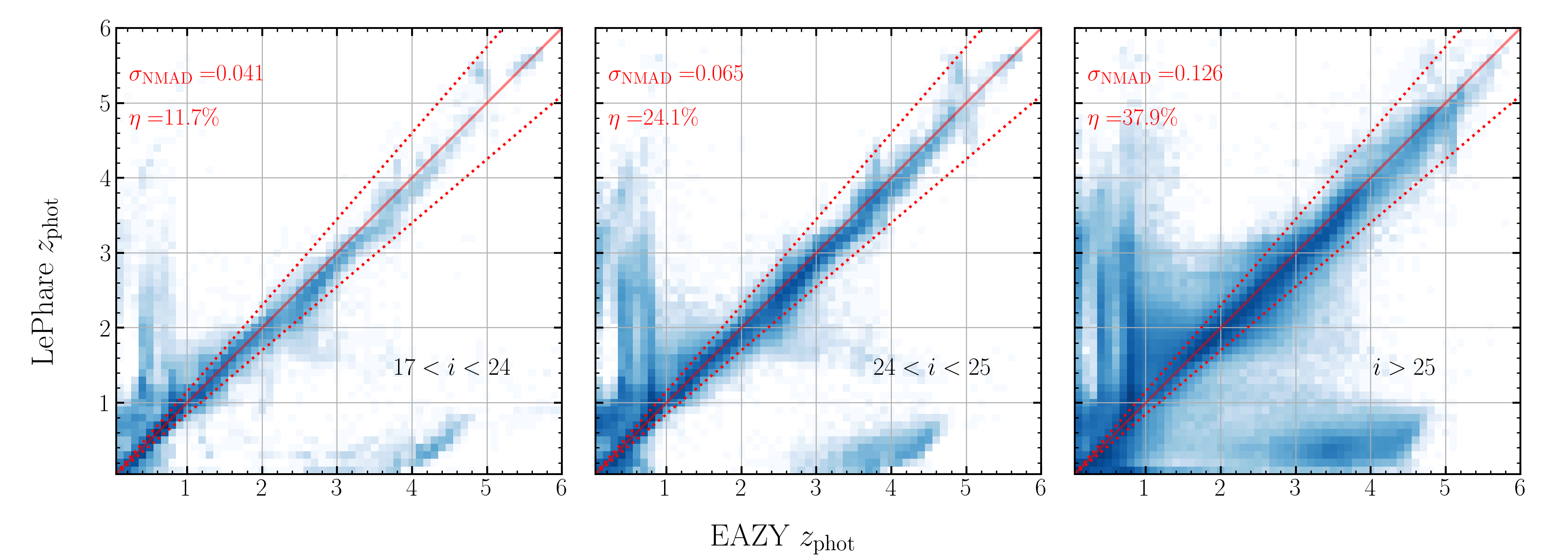}
    \caption{Validation of photometric redshifts (\photoz{}s) computed with two codes, \eazy{} and \lephare{}, in three bins of apparent HSC $i$ magnitude. The red solid line represents a one-to-one relation, and the dashed lines enclose the \photoz{} at $\pm 0.15(1+z_{\rm spec})$ representing galaxies typically considered non-outliers. The fraction of sources outside the dashed lines (denoted $\eta$) and the precision measured with the normalized absolute deviation (noted $\sigma_{\rm NMAD}$) are indicated in each panel.
    \textit{Top:} Comparison between \photoz{}s  and \specz{}s. Galaxies shown are those from the EDF-F spectroscopic sample from GOODS-S \citep{Garilli2021,Kodra2023}. The \photoz{} measurements of \eazy{}{} are displayed as orange circles while the \photoz{} measurements of \lephare{} are displayed as blue circles.
    \textit{Bottom:} Comparison between \photoz{}s as calculated with the two codes for the full sample of galaxies in EDF-F. The \eazy{} \photoz{}s are displayed along the $x$-axis while the \lephare{} \photoz{}s are displayed along the $y$-axis. Bin colouring increases on a logarithmic scale.}
    \label{fig:zphot_zspec}
\end{figure*}

\lephare{} is used to compute \photoz{}s closely following the procedure outlined in \cite{Ilbert2013}, \cite{Laigle2016}, and \cite{Weaver2021}. One objective of the procedure used in the aforementioned works was to create a SED-fitting configuration and SED template library that would be well-suited to describe a diverse range of galaxies across cosmic time. Having been well-validated in several works, their methods and template libraries are adopted here with little modification. The reader is referred to \cite{Weaver2021} for the most recent description of the \lephare{} configuration. A description of key differences with respect to our setup follows. 

\cite{Ilbert2006} introduced a method to use a subsample of galaxies with spectroscopic redshift measurements (\specz{}s) to improve \photoz{} measurements. To do so, offsets between the observed and predicted photometry from the template set are derived after fixing the redshift at the \specz{} value. This procedure is repeated over the template set and \specz{} sample until the offsets converge. This method is used here, employing the different spectroscopic samples for EDF-N and EDF-F described in Sect.~\ref{subsec:spectroscopy}.

Photometric uncertainties are modified prior to SED fitting in order to account for discrepancies between the theoretical templates and observed photometry, a step also taken by \cite{Ilbert2013}, \cite{Laigle2016}, and \cite{Weaver2021}. Offsets of 0.02 mag are added to the MegaCam and HSC broadband errors and 0.05 mag are added to the IRAC \chOne{} and \chTwo{} errors. All additions are done in quadrature. The range of redshifts explored is limited to $0 < z < 8$ with steps of 0.01, departing from the range of \cite{Weaver2021} wherein the authors allowed solutions out to $z = 10$. Given the set of detection bands considered in this work, the reddest being the HSC $z$ band, it is virtually guaranteed that galaxies beyond $z = 8$ will not be detected, and even galaxies beyond $z \sim 7$ should be extremely difficult to detect. Considerations regarding the set of galaxy templates, range of $E(B-V)$, dust attenuation curves, and treatment of emission lines are otherwise identical to those used by \cite{Weaver2021}. Both the \enquote{best fit} or maximum-likelihood redshift is recorded as well as the redshift corresponding to the median of the probability distribution function of redshift, $P(z)$, as measured by \lephare{}. The \photoz{} uncertainty 1$\sigma$ lower and upper bounds are given by the 16th and 84th percentiles of the $P(z)$, respectively. 

As in \cite{Weaver2021}, templates that describe active galactic nuclei (AGN) and stellar sources are considered in addition to the galaxy templates; the reader is referred to this work for a full description of various template sets employed. The goodness of fit of these alternative templates (and \photoz{} in the case of AGN) are recorded to aid in classifying each source as either star, galaxy, or AGN. As a demonstration and validation of their utility, stars are  separated from galaxies by simply requiring the reduced $\chi^{2}$ of the stellar template fit to be less than that of the best galaxy template fit. We further require the source to have $\mathrm{S/N} > 3$ in the IRAC \chOne{} band, as the infrared flux measurement is essential for accurately distinguishing stars from galaxies. The result of the star-galaxy separation is shown in Fig.~\ref{fig:star_galaxy}. Only stars with HSC $i$ magnitudes $< 21.5$ are labeled as such, following \cite{Weaver2021}. The majority of sources identified as stars fall on the expected sequence.

\subsection{\label{subsec:eazy}\eazy{}}

The most recent version of \eazy{} written in python (\texttt{eazy-py}; \citealt{Gould2023}) is used to measure \photoz{}s and physical parameters of galaxies. As with \lephare{}, SED-fitting with \eazy{} is carried out following the strategy laid out by \cite{Weaver2021} for the same motivations outlined above. \eazy{} and \lephare{} share many similarities in their approach to SED-fitting. However, the most significant difference between the two codes is that \lephare{} is typically used to fit a large library of many individual templates while \eazy{} is typically used to fit a small library of individual templates but allows for an unrestricted non-negative linear combination of templates to create a single model for each galaxy. This flexibility of \eazy{} is useful for efficiently describing a wide variety of galaxies, especially on the scale expected from a survey spanning tens of square degrees. However, the same flexibility is not guaranteed to be well-constrained in cases of limited wavelength coverage, which may lead to disagreements in the measurements of physical parameters when compared to \lephare{}. 

One departure from \cite{Weaver2021} is the specification of the \eazy{} template set. Recently, several sets of templates were added to the online repository\footnote{\url{https://www.github.com/gbrammer/eazy-py}} that allow some of the physical attributes of the templates to evolve with redshift. For example, some templates include redshift-dependent star-formation histories and require the maximum attenuation of the reddened templates to evolve with redshift as well. In some works (e.g., \citealt{Weaver2023}), these template sets have been shown to outperform previous template sets from \eazy{}. Specifically, the template set described by the file \texttt{\enquote{corr\textunderscore
sfhz\textunderscore13.param}} is used. 

Similar to \lephare{}, \eazy{} has methods for determining photometric offsets between observed and predicted photometry from the template set in specific bands. In contrast to \lephare{}, galaxies with spectroscopic redshifts are not needed. Instead, a user-defined fraction or subset of galaxies is selected from the catalogue, their \photoz{}s are computed, and the differences between the observed and predicted photometry from the best-fit templates are recorded. This photometric offset is then applied to the sample, and the procedure iterates five times, after which the change in the derived offset is $< 1\%$. There is no guarantee for the \photoz{} measurements to improve for spectroscopically confirmed galaxies according to this method, although often they do. The strength of this method is that a large and unbiased sample of galaxies can be used to correct for systematics in observed photometric bands or in specific wavelength regimes in the theoretical template set. By contrast, spectroscopic samples tend to be biased in one way or another, over-representing galaxies from which redshifts can easily be measured.

Unlike with \lephare{}, photometric uncertainties are not manually adjusted in specific bands when using \eazy{}. This is because \eazy{} uses a \enquote{template error function} \citep{Brammer2008} that serves a similar purpose. The template error function is designed to account for many of the causes of disagreement between the observed photometry and the theoretically predicted photometry from the template set. \eazy{} also provides two options for redshift priors, an observed $K$-band magnitude prior and an observed $r$-band magnitude prior. While the $r$-band is included in our wavelength coverage, high-redshift ($z > 3$) solutions are too strongly disfavored when it is in use. Therefore a magnitude-based redshift prior is not used.

To assist in star/galaxy separation, the built-in routines of \eazy{} are used to fit stellar templates provided with the code in the same manner described as in \cite{Weaver2021}. The catalogues include the goodness of fit and effective temperature for the best-fit stellar template for each source.

\subsection{\label{subsec:photoz_valid}Photometric redshift validation}
\begin{figure}
    \centering
    \includegraphics[width=\columnwidth]{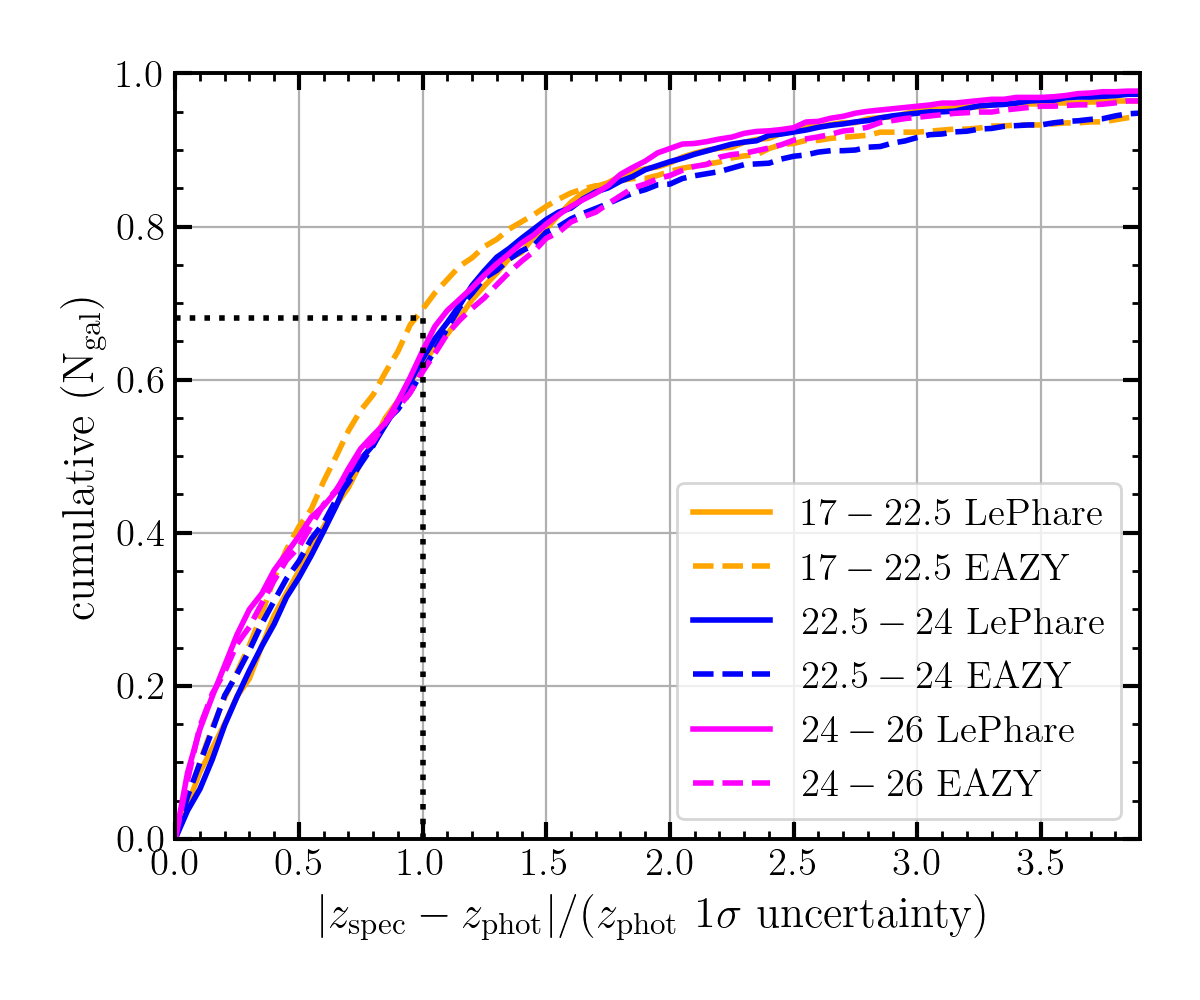}
    \caption{Cumulative fraction of galaxies between $|z_{\rm spec} - z_{\rm phot}|$ divided by the 1$\sigma$ uncertainty for the EDF-F spectroscopic sample. For well-calibrated \photoz{} uncertainties, the enclosed fraction ($y$-axis) should be equal to 0.68 when this ratio ($x$-axis) is equal to 1, highlighted by the dotted black line. The 1$\sigma$ uncertainty is defined as the maximum between $(z_{\rm phot}-z_{\rm phot}^{\rm min})$ and $(z_{\rm phot}^{\rm max}-z_{\rm phot})$. Solid lines correspond to the cumulative fraction of galaxies for \lephare{}{} while dashed lines correspond to \eazy{}{}. Samples are shown in three bins of apparent HSC $i$ magnitude.}
    \label{fig:zphot_cumulative}
\end{figure}

Perhaps the most common method for validating \photoz{}s is to directly compare the measurements from SED-fitting codes with reliable spectroscopic redshifts. The obvious advantage of this approach is that it involves a direct comparison with \enquote{ground truth} for spectroscopic sources. To this end, the spectroscopic samples described above in Sect. \ref{subsec:lephare} are employed for the respective fields. Galaxies are matched between the spectroscopic catalogues and our photometric catalogues using a matching radius of 0.5 arcseconds, yielding a total of 1460 spectroscopic matches in EDF-N and 3300 matches in EDF-F. To assess the quality of the \photoz{}s, summary statistics regularly used in the literature are calculated. The first quantifies the precision and is the normalized median absolute deviation (NMAD; \citealt{Hoaglin1983}), defined as
\begin{equation}
\sigma_{\mathrm{NMAD}}=1.48 \times \operatorname{median}\left(\left|\frac{\Delta z-\operatorname{median}(\Delta z)}{1+z_{\mathrm{spec}}}\right|\right),
\end{equation}
following \cite{Brammer2008}. The second statistic is a measure of purity, and it quantifies the rate of \enquote{catastrophic} outliers (given by $\eta$) as the fraction of galaxies that differ from their \specz{} by $|\Delta z| > 0.15(1+z_{\rm spec})$ \citep{Hildebrandt2012}. 

A comparison between \photoz{}s and \specz{}s is presented in Fig.~\ref{fig:zphot_zspec}, where galaxies have been separated into different intervals depending on their apparent magnitude in the HSC $i$ band. Here, the EDF-F photometry and matched-spectroscopic sample is highlighted, as it provides dense sampling across redshift and magnitude. Globally, we find excellent agreement between photometric and spectroscopic redshifts with both \eazy{} and \lephare{}, despite the lack of near-infrared photometry. For galaxies with HSC $i$ magnitudes between $17 < i < 24$, both codes achieve a strong precision of $\sigma_{\rm NMAD} \sim 0.035$ and an outlier fraction $\eta < 7\%$. As is to be expected, the \photoz{} performance generally declines for fainter objects, both in terms of $\sigma_{\rm NMAD}$ and the outlier fraction $\eta$. The performance of the two codes, based on comparison with spectroscopic galaxies, is broadly similar.

Another consideration for validating the output of the two \photoz{} codes is to compare their output with each other. This provides an opportunity to identify large-scale disagreements and biases between the two codes for the entire sample of galaxies. This comparison is shown in the bottom row of Fig.~\ref{fig:zphot_zspec}. Generally, there is good agreement between the two codes across redshift, despite the many differences in the respective templates considered. As to be expected, fainter galaxies disagree in their \photoz{} assignment more frequently. The majority of galaxies that are off-diagonal on either side of the 1:1 relation are due to disagreements in spectral \enquote{breaks} that cause strong colours and are typically indicative of a particular redshift. The two most prominent in the case of SED-fitting are the Lyman and Balmer breaks, and their confusion interchanges high- and low-redshift solutions. Further discussion of validating \photoz{} estimates is provided in Appendix ~\ref{app:c2020_more}.

A significant feature of \photoz{} measurement is the uncertainty associated with the measurement. Calculated correctly, the uncertainty is informative of the confidence of the \photoz{} measurement. One method for investigating the reliability of the \photoz{} uncertainties consists of measuring the cumulative fraction of galaxies with spectroscopic redshifts contained within the interval [$z_{\rm phot}^{\rm min}$, $z_{\rm phot}^{\rm max}$] as a function of the \photoz{} uncertainty. If the uncertainty is adequately measured, then the cumulative fraction of galaxies with spectroscopic redshifts enclosed within this interval should be approximately 0.68 when the \photoz{} uncertainty is $1\sigma$. If too few galaxies are found to be within this interval, then one possible explanation is that the \photoz{} errors are underestimated (and vice versa). One way to address this problem is to modify the flux uncertainties which propagate through to the \photoz{} uncertainty. 

The cumulative fraction of galaxies between $|z_{\rm spec} - z_{\rm phot}|$ divided by the 1$\sigma$ uncertainty is shown in Fig.~\ref{fig:zphot_cumulative}, again for the EDF-F spectroscopic sample. Here, the 1$\sigma$ uncertainty is defined as  the maximum between $(z_{\rm phot}-z_{\rm phot}^{\rm min})$ and $(z_{\rm phot}^{\rm max}-z_{\rm phot})$. Based on this exercise, both the the \eazy{} and \lephare{} \photoz{} uncertainties appear well calibrated. In \cite{Weaver2021}, only the brightest galaxies ($17 < i < 22.5$) in the COSMOS2020 catalogue created with aperture photometry (as opposed to with \farmer{}; see text for details) actually reach 0.68 when the value of the $x$-axis is 1, while all other samples enclose less. In this way, the observed cumulative distribution function of this work may reflect a reliable \photoz{} uncertainty, albeit greater when compared to the results of \cite{Weaver2021}. The exact shape and displacement of these curves relative to each other appears dependent on the spectroscopic sample, a feature also noted by \cite{Laigle2019}. Following \cite{Weaver2021}, we have applied a factor of $2\times$ scaling to the flux errors prior to SED fitting to improve the scaling of the \photoz{} uncertainties, a modification likewise necessary in both \cite{Laigle2016} and \cite{Weaver2021}. However, a factor of $2\times$ scaling is applied to the \textit{Spitzer}/IRAC flux errors, given their underestimation (see Fig.~\ref{fig:mag_magerr}).

\begin{figure}
    \centering
    \includegraphics[width=\columnwidth]{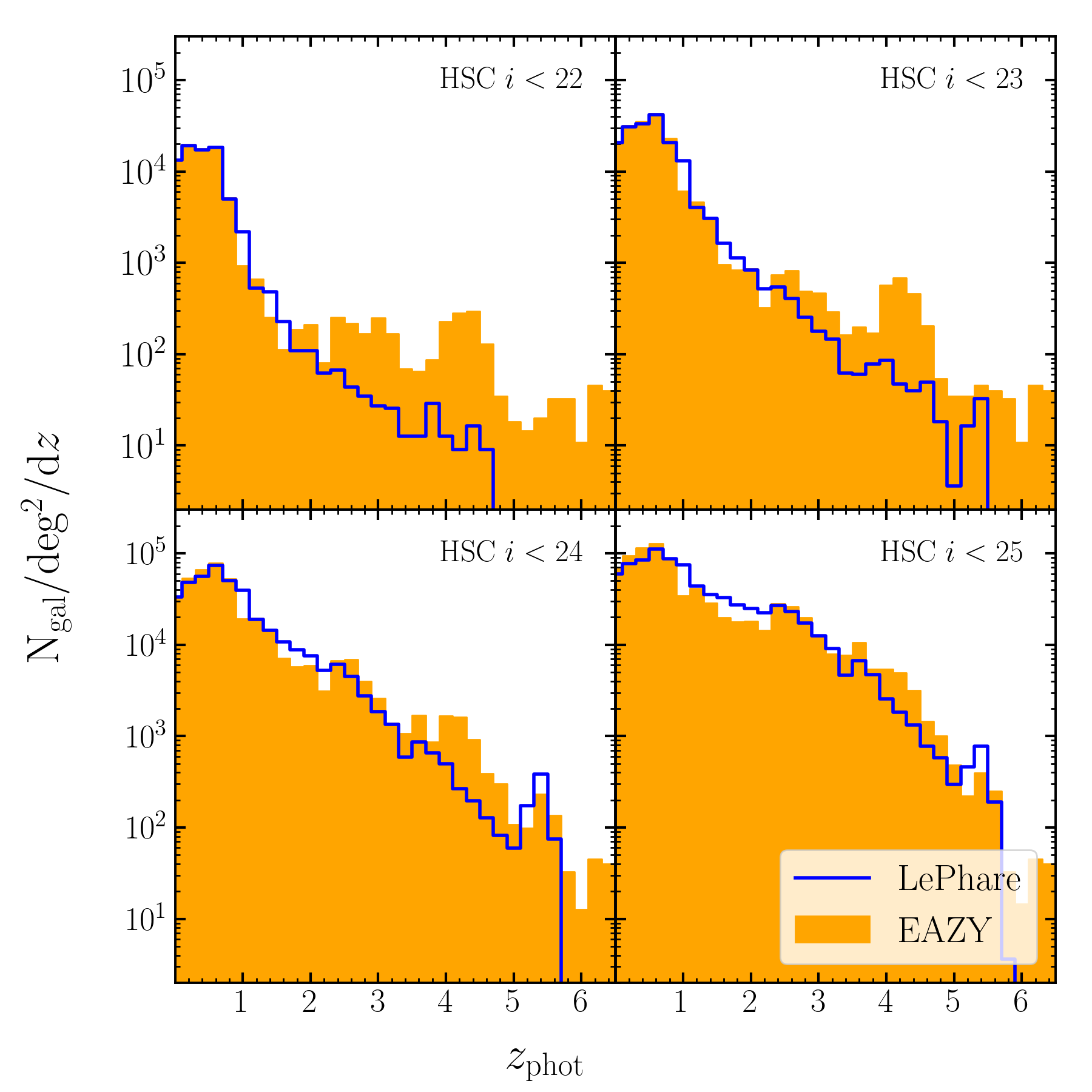}
    \caption{Redshift distribution of galaxies included in the EDF-F catalogue according to \eazy{} (orange filled histogram) and \lephare{} (blue transparent histogram). Each panel considers a different selection of galaxies depending on their HSC $i$ magnitude.}
    \label{fig:zphot_hist}
\end{figure}

The full \photoz{} distributions according to \eazy{} and \lephare{}, are shown as histograms in Fig.~\ref{fig:zphot_hist} in four ranges of HSC $i$ apparent magnitude. As expected, the distributions tend towards higher redshift with decreasing flux density. Greater \photoz{} bias may be present in the \eazy{} measurements, according to the noticeable structure at $z > 3$ in the brightest bins. A significant feature to note is the absence of galaxies above $z = 6$. This is consistent with the detection bands of this catalogue (HSC~$r+i+z$) and the implied selection function, as galaxies above $z = 6$ are mostly detected in (observed-frame) near-infrared wavelengths. The absence of galaxies at $z > 6$ in this catalogue is therefore a further affirmation of the \photoz{} methods utilized in this work. 

\section{\label{sec:properties}Physical properties of galaxies}

The SED fitting codes employed for photometric redshift estimation in Sect. \ref{sec:photoz} are also capable of providing estimates of the physical properties of galaxies. At present, the primary interest is towards constraining basic physical properties, including absolute magnitudes in particular broadband filters and galaxy stellar mass. Measurements of additional physical quantities from the DAWN survey data are deferred to future work. 

For \lephare{}, the procedure used here follows both \cite{Laigle2016} and \cite{Weaver2021}. The reader is referred to these works for a more detailed explanation for the estimation of physical parameters. In brief, a template library of \texttt{BC03} \citep{Bruzual2003} stellar population synthesis (SPS) models is generated and compared to the measured photometry after fixing the redshift to the derived \photoz{} for each source, in this case, the median \photoz{} of the $P(z)$ distribution. Unlike the template library used in the \photoz{} estimation, which includes empirical SEDs from which physical properties cannot be derived, the \texttt{BC03} SSP templates are fully synthetic and enable estimates of all the physical parameters that define the templates. The variable parameters among of the \texttt{BC03} templates include stellar mass, metallicity, age, two parameterizations of star-formation history (exponentially declining and delayed), two dust attenuation curves, and a range of $E(B-V)$ values. 

As for \eazy{}, physical parameters of galaxies are measured simultaneously alongside redshifts during the SED fitting since the templates used in the \photoz{} estimation (described above in Sect. \ref{subsec:eazy}) are themselves fully synthetic. \cite{Weaver2021} found \eazy{} suitable for measuring physical parameters in addition to redshifts. However, the lack of near-infrared imaging in the DAWN survey DR1 catalogues present a challenge in constraining the entire shape of galaxy SEDs when using non-linear combinations of basis templates. More specifically, the large gap in the wavelength coverage between HSC $z$ and IRAC \chOne{}, and the lack of constraints redder than IRAC \chTwo{}, enable unphysical solutions at times. Examples include solutions with too-strong Balmer breaks and/or unrealistic observed-frame mid-infrared colours. For this reason, a detailed analysis of the quality of physical parameters as measured by \eazy{} is deferred to future work is. However, the physical properties of galaxies as measured by \eazy{} may be made available upon request.  

The future combination of near-infrared data from \Euclid with the UV/optical and infrared data from the DAWN survey will yield significantly improved physical parameter measurements from both \eazy{} and \lephare{}. In the remainder of this work, only the physical properties of galaxies as provided by the \lephare{} measurements are considered. 

\begin{figure}
    \centering
    \includegraphics[width=\columnwidth]{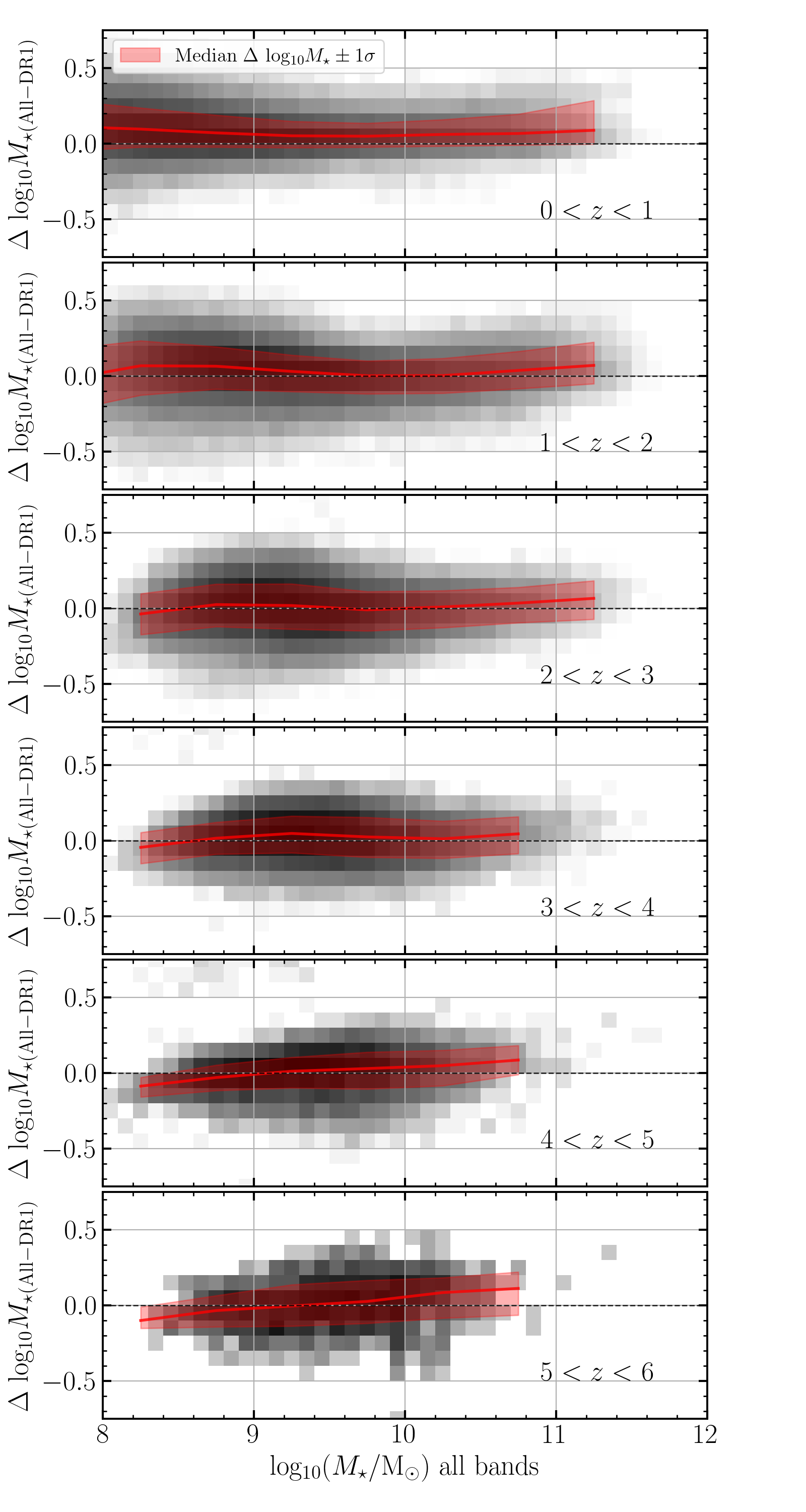}
    \caption{Validation of stellar masses as a function of redshift measured using photometry in the bandpasses included by the present DAWN survey DR1 catalogues. Each panel shows the difference in stellar mass measured using the COSMOS2020 catalogue \citep{Weaver2021} when using only the present DAWN bandpasses (\enquote{DR1}) and all forty (\enquote{All}) of the original bandpasses of COSMOS2020. The medians are indicated by the red lines, and the shaded envelopes enclose 68\% of the sources corresponding to $1\sigma$.}
    \label{fig:mass_comp}
\end{figure}

\begin{figure*}
    \centering
    \includegraphics[width=0.73\textwidth,trim={0.7cm, 0.7cm, 1.cm, 0.7cm},clip]{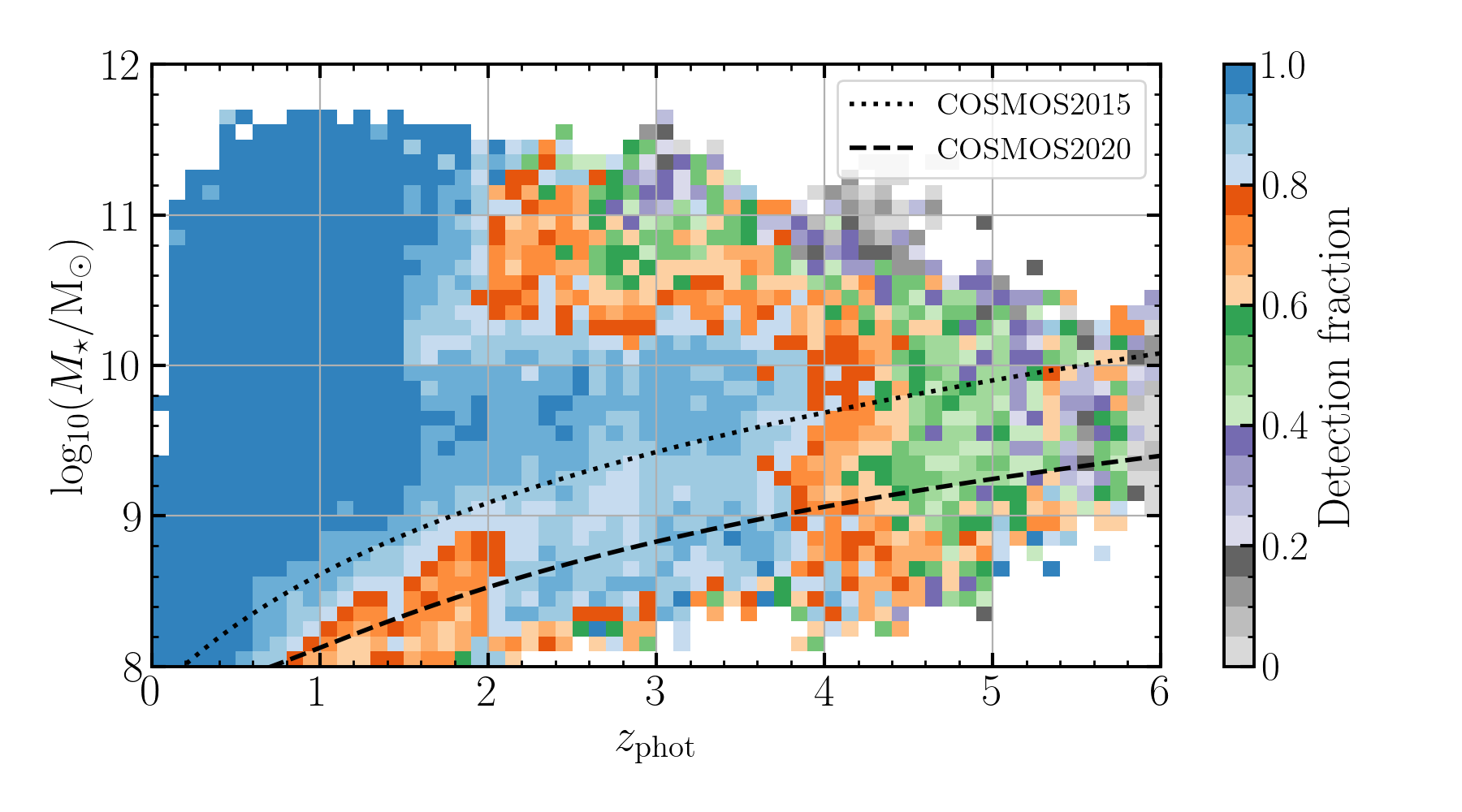}
    \caption{Empirical description of the DAWN survey DR1 selection function as viewed through the COSMOS2020 dataset. Each cell represents the fraction of galaxies detected from the original COSMOS2020 catalogue \citep{Weaver2021} according to the DAWN survey DR1 selection function. The dotted line represents the COSMOS2015 stellar mass completeness limit \citep{Laigle2016}, while the dashed line represents the COSMOS2020 stellar mass completeness limit \citep{Weaver2023SMF}. Future catalogues selected on near-infrared imaging provided by \Euclid will enhance the mass completeness of galaxies beyond $z\sim2$.}
    \label{fig:mass_det}
\end{figure*}

\subsection{\label{subsec:validation}Stellar mass reliability}
A large body of work has been devoted to validating galaxy stellar masses measured from broadband photometry (e.g., \citealt{Mobasher2015,Pacifici2023} and citations therein). The objective here is simply to demonstrate that the stellar masses presented in the DAWN survey DR1 catalogues, as measured with \lephare{}, are useful and reliable. However, unlike photometric redshifts, which may be compared to spectroscopic redshifts, there are no \enquote{ground truth} measurements of the stellar masses of observed distant galaxies. Empirically, one straightforward option is to compare a set of stellar mass measurements with another reference set of measurements that has been validated in its own right. There is no such reference set available in the EDF-N and EDF-F for extensive comparison. Instead, the following test has been devised using the COSMOS2020 catalogue. Having been measured with some forty photometric bands and carefully vetted, the COSMOS2020 stellar masses are considered reliable. In addition, the COSMOS2020 catalogue includes all the photometric bands also contained by the DAWN survey DR1 catalogues presented in this work. With this in mind, stellar masses are measured from the COSMOS2020 catalogue using only photometry in bands present in both catalogues. From there, a comparison of the resulting stellar masses may be made with the original stellar masses presented in \cite{Weaver2021}. 

COSMOS2020 does not only provide greater sampling in wavelength than the DAWN survey DR1 catalogues but is also slightly deeper in the overlapping bands. Therefore, the test described above is made more realistic by further inflating the original photometric uncertainties of COSMOS2020 to broadly match the relationship between magnitude and magnitude error (i.e., the relationships of Fig.~\ref{fig:mag_magerr}) of the DAWN survey DR1 catalogues. In practice, this consists of modeling the relationship between magnitude and magnitude error with an exponential function of the form $y = A{\rm e}^{bx}$, where $y$ is the magnitude error, $x$ is the magnitude, and $A$ and $b$ are free parameters. The model is fit using least-squares optimization for both the present catalogues and the COSMOS2020 catalogue, obtaining a  functional relationship for each dataset. The magnitude errors of the COSMOS2020 catalogue are then rescaled by the ratio of the two functions, effectively applying a magnitude-dependent scaling factor, to match the relationship between magnitude and magnitude error of the DAWN survey DR1 catalogues. The corresponding modification is finally propagated to the flux errors. 

Correctly measuring the stellar mass of a given galaxy depends on first correctly determining its redshift. To this end, \lephare{} is used to first fit for photometric redshifts and then for stellar masses using the modified COSMOS2020 dataset. We measure the stellar mass at the newly derived photometric redshift following the exact methods as described above (in Sect.~\ref{subsec:lephare}). We achieve good agreement between the newly estimated photometric redshifts and those presented in \cite{Weaver2021} and even with the COSMOS spectroscopic sample. At bright magnitudes (HSC $i < 25$), fewer than 6\% of objects strongly disagree in their redshift determination.  We address this comparison of \photoz{} more fully in Appendix~\ref{app:c2020_more}.

A comparison between stellar masses computed only with the bands overlapping between the present DAWN survey DR1 catalogues (\enquote{DR1}) and of COSMOS2020 (\enquote{All}) is presented in six redshift bins in Fig.~\ref{fig:mass_comp}. Here galaxies with $|\Delta z| < 0.15(1+z_{\rm phot, all})$ are selected (see Fig.~\ref{fig:COSMOSzphot_zspec} for a comparison of \photoz{}). This selection effectively removes disagreements in stellar mass that are driven by disagreements in the assumed redshift. An additional requirement is signal-to-noise of at least 3 in \textit{Spitzer}/IRAC \chTwo{} and either the HSC $z$ band or the HSC $i$ band. The agreement is strong across both redshift and mass: there appears to be a small, variable offset of $<$ 0.1 dex, and a spread that varies between 0.1--0.2 dex ($1\sigma$). The variable offset in stellar mass is consistent with the uncertainties of stellar masses measured from SED-fitting, which are typically of order 0.1--0.3 dex. The spread, on the other hand, is driven mostly by differences in photometric redshift. Selecting samples with with smaller differences in measured redshifts decreases the spread. Indeed, as discussed in Appendix~\ref{app:c2020_more}, disagreement in \photoz{} is virtually entirely responsible for disagreement in stellar mass estimates.

It is emphasized that the same template set was used in both the work of \cite{Weaver2021} and the present work and fit to the same photometry, although only a subset of the photometry (with increased flux uncertainties) is used herein. Accordingly, some amount of agreement is to be expected. However, the test presented here demonstrates that both \photoz{}s and stellar masses are very reliably constrained using the filter set of the DAWN survey DR1 catalogues.

\subsection{\label{subsec:completeness}Stellar-mass completeness}

A key characteristic of every galaxy survey is its selection function. The selection function directly relates to various completeness limits (e.g., flux, colour, stellar mass, and intersections of such qualities). For many science investigations, the stellar mass completeness limit is of primary interest. In an ideal case, the mapping between the selection function and the completeness limit is roughly linear. This is the basis for the empirical method of measuring stellar mass completeness limits presented by \cite{Pozzetti2010}. The method consists of converting the detection limit of a given survey to a stellar mass completeness limit by first inferring a mass-to-light ratio, applying a transformation to the measured stellar masses given the difference between their measured flux and the limiting flux, and using the rescaled stellar masses to describe the completeness limit. Many works \citep{Ilbert2013,Laigle2016,Weaver2021} have used this method to arrive at an analytical description of the stellar mass completeness limit. The crucial assumption of this method is that the selection function can be reduced to a detection limit and that the detection limit maps linearly to the stellar mass limit. 

The challenge of describing the stellar mass completeness limit of the present DAWN survey DR1 catalogues is that the above assumption does not hold for all galaxies. In general, galaxy stellar masses are most directly correlated with rest-frame optical emission. Therefore, selection functions defined by a domain of wavelength mostly trace stellar mass (directly) within redshift ranges where rest-frame optical emission is included in that domain, with the exception of galaxies with very young and intrinsically blue stellar populations. For the DAWN survey DR1 catalogues, the selection function is defined by the wavelength domain represented by the HSC $r+i+z$ filters. Accordingly, rest-frame optical emission falls out of this wavelength domain by $z \sim 1.5$. As demonstrated by Fig.~\ref{fig:zphot_hist}, included in the DAWN survey DR1 catalogues are many galaxies at $z > 1.5$, necessitating an alternative method to the one presented by \cite{Pozzetti2010}. However, as previously stated, future catalogues from the DAWN survey will include galaxies selected from the near-infrared imaging of \Euclid, which will overcome some of these limitations and significantly improve mass-completeness.

One alternative to the \cite{Pozzetti2010} method is to use a reference survey with well-understood characteristics that is deeper than the survey at hand, matching detections from the latter to the former, and quantifying the fraction of galaxies that are missed. In general, many works combine the \cite{Pozzetti2010} method with the one just described \citep{Davidzon2017,Weaver2021,Weaver2023SMF}. As previously stated in Sect.~\ref{subsec:validation}, there is no such reference survey overlapping with the EDF-N and EDF-F fields with extensive and well-vetted stellar mass measurements. The solution presented here is to perform a comparison test similar to the one used to validate the method for measuring stellar masses. In this case, the test begins with creating a detection image with the same properties of the detection image used for the  DAWN survey DR1 catalogues (Sect.~\ref{subsec:source_detection}), but using the COSMOS2020 images. By construction, the galaxies detected on the modified image will be defined by the selection function of the DAWN survey DR1 catalogues. As such, a comparison may be made between the newly detected galaxies and those originally included in \cite{Weaver2021} to obtain an empirical description of the selection function. 

\begin{figure*}
    \centering
    \includegraphics[width=\textwidth]{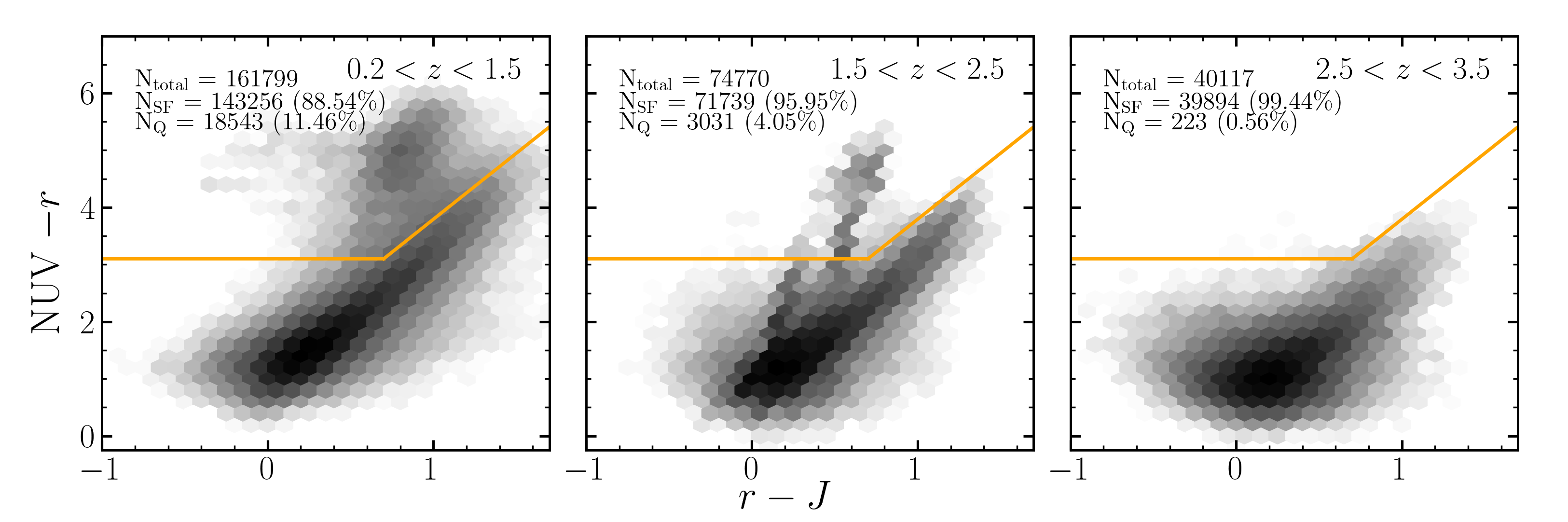}
    \caption{Rest-frame colour-colour diagram classifying galaxies as either star forming (SF) or quiescent (Q) for galaxies in the DAWN survey  DR1 EDF-F catalogue. The COSMOS2015 \citep{Laigle2016} stellar mass completeness has been assumed (see Fig.~\ref{fig:mass_det}). The bounding region discriminating the two populations is provided by \cite{Ilbert2013}. The fraction of quiescent galaxies decreases with increasing redshift due to both physical and observational effects driven by the present DAWN selection function. Future catalogues selected on near-infrared imaging provided by \Euclid will enhance both the detection and identification of quiescent galaxies out to $z\sim3$.}
    \label{fig:sf_qg}
\end{figure*}

To adequately represent the DAWN survey DR1 selection function, the modified COSMOS2020 detection image must share the characteristics of the DAWN survey DR1 detection image, including wavelength domain and sky noise. Achieving an equivalent wavelength domain solely requires limiting the included images to the HSC $r+i+z$ images. As the original COSMOS2020 images are deeper than the H20 images, each of these images must then be individually modified to share the same level of noise across the image plane to its corresponding H20 counterpart. This is achieved by measuring the per-pixel RMS variation in both the original COSMOS2020 images and the H20 images and adding random noise (drawn from a Gaussian distribution) to the former such that the median resulting RMS agrees with the H20 RMS. This operation is performed on each of the three filters. The modified COSMOS2020 images are finally combined and sources are detected following the procedure presented in Sect.~\ref{subsec:source_detection}.

Finally, the present DAWN survey DR1 catalogue selection function, as viewed through the COSMOS2020 dataset, is presented. First, a two-dimensional histogram describing the number of galaxies as a function of redshift and stellar mass is measured from the original COSMOS2020 catalogue. Then, a second two-dimensional histogram is measured according to the same properties, but limited to include only the galaxies that are detected using the modified COSMOS2020 detection image. The ratio of these two histograms describes the influence of the present selection function on the stellar mass completeness as a function of redshift. This result is depicted in Fig.~\ref{fig:mass_det}. For comparison, also included are the analytical stellar mass completeness curves for the COSMOS2015 catalogue \citep{Laigle2016} and the COSMOS2020 catalogue \citep{Weaver2021}. The fraction of galaxies detected is essentially 100\% within the COSMOS2015 stellar mass completeness limit out to $z = 1.5$. Beyond this redshift, the DAWN survey DR1 selection function does not include rest-frame optical emission, so the fraction of detected galaxies drops to between 80--90\% until $z \sim 4$. Many galaxies are detected at $z > 4$, but the fraction decreases with increasing redshift as galaxies continue to fall out of the detection bands. There are further two notable features that stand out in the redshift range $2 < z < 4$ corresponding to massive and low-mass galaxies. Regarding massive galaxies, there is a subset of galaxies within this redshift range with $\mathrm{log}_{10}(M_{\star}/\mathrm{M}_{\odot}) > 10.2$ that are not detected, comprising $\sim$25\% of all galaxies with those qualities. According to their rest-frame colours as presented in \cite{Weaver2021}, these galaxies are red, dusty, and only detectable with near-infrared coverage. The fraction of detected low-mass galaxies (below the COSMOS2015 completeness limit), on the other hand, increases in the range $2 < z < 4$. At these redshifts, our detection bands probe bluer wavelengths, becoming more sensitive to star formation. These low-mass galaxies are likely UV-dominated making them easily detected. 

In its present state, the DAWN survey DR1 catalogues are well-suited for characterizing galaxies as a function of stellar mass, at least at $z < 2$, though careful efforts to account for missing objects may be taken to extend analyses to $z \sim 4$. Future catalogues produced by the DAWN survey in the EDFs and EAFs will yield samples of galaxy populations exceeding the mass-completeness achieved by COMSOS2020 through detection on the even deeper \Euclid near-infrared imaging.

\subsection{\label{subsec:classification}Galaxy classification}

Galaxies are classified as either star-forming or quiescent using absolute magnitudes measured by \lephare{}. More specifically quiescent and star-forming galaxies are identified via their location in the rest-frame ${\rm NUV}-r$ vs. $r-J$ diagram following the approach described in \cite{Ilbert2013}. Here, quiescent galaxies are defined as those with $M_{\rm NUV} - M_{r} > 3(M_{r} - M_{J}) + 1$ and $M_{\rm NUV} - M_{r} > 3.1$. Absolute magnitudes are calculated according to the procedure given in the appendix of \cite{Ilbert2005}, where the absolute magnitude in a given filter $\lambda_{\rm abs}$ is constrained by the observed-frame filter nearest to  $\lambda_{\rm abs}(1+z)$. This minimizes the dependence of the $k$-correction on the assumed galaxy template.

It is important to note that without the near-infrared photometry provided by \Euclid, there is not any redshift with direct overlapping wavelength coverage between the current observed filters and all of the rest-frame diagnostic filters. Beyond redshift $z = 1.5$, the 4000\AA \,break is redshifted out of our detection filters, suggesting that the completeness of our quiescent galaxy sample will drop off (analogous to stellar mass completeness). At this point, only rest-frame NUV and rest-frame $J$ are directly constrained by the DAWN survey DR1 catalogues, until $z = 3.5$. Beyond $z =  3.5$ direct overlap with the rest-frame NUV decreases. Therefore, the best-fit template is relied on to infer absolute magnitude in at least one of the three diagnostic filters at all redshifts.

Figure~\ref{fig:sf_qg} shows the breakdown of quiescent and star-forming galaxies in the rest-frame NUV$rJ$ diagram separated in three redshift bins from $z = 0.2$ to $z = 3.5$ assuming the COSMOS2015 stellar mass completeness (see Fig.~\ref{fig:mass_det}). The fraction of quiescent galaxies increases dramatically from high to low-redshift for well-documented physical reasons \citep{Ilbert2010,2013ApJ...770...57B,Weaver2023SMF} in addition to the observational effects described above. Though the uncertainties in the rest-frame colours may be substantial, the large cosmic volume covered by the H20 dataset should provide legitimate quiescent galaxies, and future work will investigate their rate of false positive classification. Finally, the imminent addition of \Euclid near-infrared photometry and detection will dramatically enhance the identification of quiescent galaxies throughout the redshift ranges considered here.

\section{Summary}\label{sec:summary}

This paper describes the creation and contents of the first public multiwavelength catalogues from the Cosmic Dawn Survey (DAWN). The DAWN Survey provides depth-matched multiwavelength data to the Euclid Deep and Auxiliary Fields. In this first release (DR1), photometry is measured from the Hawaii Twenty Square Degree Survey (H20) data, which includes deep ultraviolet and optical ground-based imaging from CFHT MegaCam and Subaru HSC, respectively. These data are paired with deep infrared imaging across EDF-N and EDF-F provided by the DAWN survey \citep{Moneti2022} \textit{Spitzer}/IRAC data, where the primary contribution is from the SLS \citep{Capak2016}. In addition to photometry and photometric uncertainties, photometric redshifts and estimates of galaxy physical parameters are provided as measured from two SED fitting codes, \eazy{} and \lephare{}. These catalogues represent the deepest available UV/optical photometry covering EDF-N and EDF-F, as well as the deepest \textit{Spitzer} photometry over any such area of this extent. 

The EDF-N DAWN survey DR1 catalogue spans 16.87 deg$^{2}$ with 9.37 deg$^{2}$ completed to final survey depth in all bands. The EDF-F DAWN survey DR1 catalogue spans 2.85 deg$^{2}$ with 1.77 deg$^{2}$ completed to final survey depth. Sources are detected in the HSC $r+i+z$ bands. In total, \num{5286829} objects are detected over the 16.87 deg$^{2}$ area of the DR1 EDF-N catalogue, where \num{3513211} of the detected objects are in the 9.37 deg$^{2}$ full-depth region. In EDF-F, \num{1062645} objects are detected over the DR1 2.85 deg$^{2}$ DR1 area, where \num{727678} are in the 1.77 deg$^{2}$ full-depth region. Model-based photometry is measured using \farmer{} from all publicly available CFHT Megacam $u$ band, Subaru HSC, and \textit{Spitzer}/IRAC imaging overlapping EDF-N and EDF-F. The two catalogues are then used to measure photometric redshifts and galaxy properties with the SED fitting codes \eazy{} and \lephare{}. The two codes show general agreement for the majority of sources with high signal-to-noise, but disagree for fainter objects. Photometric redshifts achieve strong performance compared to our spectroscopic sample, with an outlier fraction of less than 10\% for galaxies brighter than HSC $i$ = 25, and a precision of $\sigma_{\rm NMAD} \sim 0.06(1+z)$ for the same sample. Stellar masses are found to be reliable and rest-frame colours are sufficiently constrained for identifying galaxies as star-forming and quiescent to $z \sim 3$. However, rest-frame optical emission falls out of the selection function of the present DAWN catalogues by $z \sim 1.5$, impacting both stellar mass completeness and the numbers of quiescent galaxies detected.

As further Subaru HSC imaging is acquired, the full-depth areas in both EDF-N and EDF-F will continue to grow to their respective final goals of 20 deg$^{2}$ for EDF-N and 10 deg$^{2}$ for EDF-F. Even more significantly, \Euclid will soon provide deep near-infrared imaging across these fields and the other EDFs and EAFs. Upon the acquisition of \Euclid data, the \Euclid near-infrared imaging will be used to create a new detection image resulting in a near-infrared selected catalogue. Together with \Euclid, the Cosmic Dawn Survey data will enable scientific investigations out to $z \sim 10$, firmly through the epoch of reionization, as described by \cite{McPartland2024}. The Cosmic Dawn Survey catalogues will be updated continually as new ground-based and space-based data is processed, and new releases will be made publicly available on our website, further described in Appendix~\ref{app:release}. 

%
%

\begin{acknowledgements}

The Cosmic Dawn Center (DAWN) is funded by the Danish National Research Foundation under grant DNRF140. The Hyper Suprime-Cam (HSC) collaboration includes the astronomical communities of Japan and Taiwan, and Princeton University. The HSC instrumentation and software were developed by the National Astronomical Observatory of Japan (NAOJ), the Kavli Institute for the Physics and Mathematics of the Universe (Kavli IPMU), the University of Tokyo, the High Energy Accelerator Research Organization (KEK), the Academia Sinica Institute for Astronomy and Astrophysics in Taiwan (ASIAA), and Princeton University. Funding was contributed by the FIRST program from Japanese Cabinet Office, the Ministry of Education, Culture, Sports, Science and Technology (MEXT), the Japan Society for the Promotion of Science (JSPS), Japan Science and Technology Agency (JST), the Toray Science Foundation, NAOJ, Kavli IPMU, KEK, ASIAA, and Princeton University. 

This work was made possible by utilizing the CANDIDE cluster at the Institut d’Astrophysique de Paris,which was funded through grants from the PNCG,CNES, DIM-ACAV, and the Cosmic Dawn Center and maintained by S. Rouberol.

This paper makes use of software developed for the Large Synoptic Survey Telescope. We thank the LSST Project for making their code available as free software at  http://dm.lsst.org

The Pan-STARRS1 Surveys (PS1) have been made possible through contributions of the Institute for Astronomy, the University of Hawaii, the Pan-STARRS Project Office, the Max-Planck Society and its participating institutes, the Max Planck Institute for Astronomy, Heidelberg and the Max Planck Institute for Extraterrestrial Physics, Garching, The Johns Hopkins University, Durham University, the University of Edinburgh, Queen’s University Belfast, the Harvard-Smithsonian Center for Astrophysics, the Las Cumbres Observatory Global Telescope Network Incorporated, the National Central University of Taiwan, the Space Telescope Science Institute, the National Aeronautics and Space Administration under Grant No. NNX08AR22G issued through the Planetary Science Division of the NASA Science Mission Directorate, the National Science Foundation under Grant No. AST-1238877, the University of Maryland, and Eotvos Lorand University (ELTE) and the Los Alamos National Laboratory.

Based in part on data collected at the Subaru Telescope and retrieved from the HSC data archive system, which is operated by Subaru Telescope and Astronomy Data Center at National Astronomical Observatory of Japan.

Some of The data presented herein were obtained at the W. M. Keck Observatory, which is operated as a scientific partnership among the California Institute of Technology, the University of California and the National Aeronautics and Space Administration. The Observatory was made possible by the generous financial support of the W. M. Keck Foundation.

Based on observations obtained with MegaPrime/MegaCam, a joint project of CFHT and CEA/DAPNIA, at the Canada-France-Hawaii Telescope (CFHT) which is operated by the National Research Council (NRC) of Canada, the Institut National des Science de l'Univers of the Centre National de la Recherche Scientifique (CNRS) of France, and the University of Hawaii. The observations at the Canada-France-Hawaii Telescope were performed with care and respect from the summit of Maunakea which is a significant cultural and historic site.

This work is based in part on observations made with the Spitzer Space Telescope, which was operated by the Jet Propulsion Laboratory, California Institute of Technology under a contract with NASA. Support for this work was provided by NASA through an award issued by JPL/Caltech.

The authors wish to recognize and acknowledge the very significant cultural role and reverence that the summit of Maunakea has always had within the indigenous Hawaiian community.  We are most fortunate to have the opportunity to conduct observations from this mountain.
\AckEC
\end{acknowledgements}

%
%

\bibliography{lukas_bib_h20}

%

\begin{appendix}
  \onecolumn 
\section{\label{app:release}Data release}
The DAWN survey DR1 catalogues are currently hosted in an online repository accessible witth a username and password\footnote{\url{https://exchg.calet.org/dawn_edfn_edff_dr1/}}. Please contact the authors of this work for access. As further ground-based data are obtained and reduced, updated catalogues will be produced and shared therein. Included alongside the catalogues are \texttt{README} files explaining the contents of each catalogue (e.g., column naming conventions). Table~\ref{tab:colnames} provides a description of the main columns used to produce the figures of this work. As described above (e.g., Fig.~\ref{fig:field_layouts}, Sect.~\ref{sec:Observations}, Sect.~\ref{subsec:area_coverage}) \textit{Spitzer}/IRAC imaging is not available across the entirety of the 20 deg$^{2}$ survey area of EDF-N but only the innermost 10 deg$^{2}$ area. Accordingly, there is significant variation in the quality of galaxy properties measured from SED-fitting for sources that lack \textit{Spitzer}/IRAC photometry. Although these sources may be appropriately studied via their CFHT and HSC photometry, caution is advised if considering SED-inferred propreties for sources lacking \textit{Spitzer}/IRAC coverage.

\begin{table} \centering
\footnotesize
\caption{DAWN survey DR1 catalogue columns used for the creation of figures in this work.}
\begin{tabular}{ll}
\hline\hline
Column name & Description  \\
\hline
ID  & Source identifier, unique for each field \\
ALPHA\_J2000    & Right ascension  \\
DELTA\_ J2000   & Declination  \\
\enquote{BAND\_NAME}\_FLUX(ERR)    & Flux and flux error for each band: [CFHT\_u, HSC\_g,r,i,z,y,NB0816,NB0921, IRAC\_CH1,2]  \\
\enquote{BAND\_NAME}\_MAG(ERR)    & Magnitude (AB) and magnitude error for each band: [CFHT\_u, HSC\_g,r,i,z,y,NB0816,NB0921, IRAC\_CH1,2]  \\
lp\_zPDF & \lephare{} \photoz{} measured using the galaxy templates. Median of the likelihood distribution. \\
lp\_zPDF\_l68 & \lephare{} \photoz{} lower limit, 68\% confidence level \\
lp\_zPDF\_u68 & \lephare{} \photoz{} upper limit, 68\% confidence level \\
lp\_chi2best & \lephare{} reduced $\chi^{2}$ (-99 if less than 3 filters) for best-fit galaxy template \\
lp\_chis & \lephare{} reduced $\chi^{2}$ (-99 if less than 3 filters) for best-fit stellar template \\
lp\_MNUV & \lephare{} NUV absolute magnitude \\
lp\_MR & \lephare{} R absolute magnitude \\
lp\_MJ & \lephare{} J absolute magnitude \\
lp\_mass\_med & \lephare{} log stellar mass from BC03 best-fit template, median of the PDF. \\
ez\_z500 & \eazy{} 50th percentile of PDF($z$) \\
ez\_z160 & \eazy{} 16th percentile of PDF($z$) \\
ez\_z840 & \eazy{} 84th percentile of PDF($z$) \\

\hline
\end{tabular}
\label{tab:colnames}
\end{table}

\section{\label{app:depths}Image depths continued}

The following further describes the method and caveats of measuring limiting magnitudes (Sect.~\ref{app:depths}). The dispersion of empty aperture fluxes is expected to follow an approximately Gaussian distribution, assuming proper image processing (i.e., flat fielding and background subtraction) and that the apertures do not capture object flux. One important consideration is adequate sampling of the image on scales relevant to the expected variation. Using a small number of apertures spread across an image can only capture variation on the largest scales. Further care with respect to the sample size of the empty aperture fluxes is especially needed if an operation such as a sigma-clip is to be performed; with a small sample, a sigma-clip will remove measurements that are not actually outside the true underlying population distribution (i.e., the sample standard deviation will not approximate the population standard deviation). For example, in the measurements presented here, using only 1 aperture per 100 square arcseconds results in a limiting magnitude $\sim$0.4 mag deeper than using 1 aperture per 5 square arcseconds.

Another challenge in measuring limiting magnitudes and image depths using dispersion of empty aperture fluxes is properly accounting for the contribution to the dispersion by undetected astronomical sources. As previously described, the method relies on placing apertures away from astronomical sources because apertures placed near sources will bias the measurement towards greater dispersion. However, given a flux threshold for source detection, it is impossible to detect every single object. Every survey fails to detect some number of sources in a trade-off between completeness, detecting more sources, and purity, not mistakenly detecting spurious objects. Accordingly, apertures placed away from detected sources inevitably fall upon some number of sources that are not detected in the first place. The usual approach to account for undetected sources is to simply sigma-clip the measured distribution of aperture fluxes. However, the undetected sources are not all contained within the tail of the distribution (i.e., outside some standard deviation), and a sigma-clip does not solve the problem of the bias. The apertures landing on or near undetected sources create a non-Gaussian distribution of measured fluxes and applying a sigma-clip may not lead to informative statistics. 

In general, faint undetected sources have the effect of shifting the entire distribution of empty aperture fluxes towards positive values (i.e., a translation), as the brightness of the missed objects approaches the sky background while exceeding intrinsic background variation. These sources are not problematic and are effectively accounted for in background subtraction during image processing. Brighter undetected objects, on the other hand, will broaden the distribution at values greater than the median (typically a value of 0) in addition to creating a tail towards even larger values and therefore a skewed distribution. Consequently, the values greater than the median (the positive component) are biased in a way that values less than the median (the negative component) are not. 

In light of the above, the true profile of the distribution of empty aperture fluxes may be extracted through careful consideration of the negative component of the distribution. This may be achieved by modeling the distribution with a Gaussian function in the the domain of fluxes spanned by [$\mu_{0} -3\sigma_{0}$,$\mu_{0}+\sigma_{0}$], where $\sigma_{0}$ is the standard deviation of the sigma-clipped fluxes and $\mu_{0}$ is the median. This domain effectively gives more weight to the negative component than to the positive component. The values of the fit are typically more robust than the point statistic measured on the full distribution. However, even this sigma-clipped distribution can be affected by undetected astronomical sources and accordingly broadened beyond the value corresponding to the variation of the sky background. With robust initial values provided by the first model fit, the impact of the contaminated fluxes is further limited by performing a second Gaussian fit, this time considering only data points in the domain [$-2\sigma_{\rm 1,fit}$,~$\mu_{\rm 1,fit} + {0.5\sigma_{\rm 1,fit}}$]. Here, $\sigma_{\rm 1,fit}$ is the standard deviation of the first Gaussian fit, and $\mu_{\rm 1,fit}$ is the mean of the first Gaussian fit. 

In general, measuring the standard deviation from the distribution of sigma-clipped fluxes, without modeling, results in a shallower limiting magnitude of order $\sim$0.2 mag. While it may not be strictly necessary to include more than one iteration of Gaussian model fitting, the first is useful for identifying a reliable domain for the model fit during the final measurement as well as initial values for the model. Further, the method appears to perform equally well for all bands from CFHT-$u$ through \textit{Spitzer}/IRAC \chTwo{}, as indicated by Fig.~\ref{fig:depths}.  The measured depth of the \textit{Spitzer}/IRAC  data matches the expecation of the SLS \citep{Capak2016}, and the significantly deeper region of \textit{Spitzer}/IRAC data in EDF-F pointed out by \cite{Moneti2022} is visible as well at the appropriate depth.  

\section{\label{app:c2020_more}Further validation with COSMOS2020}

\begin{figure*}
    \centering
    \includegraphics[width=\textwidth]{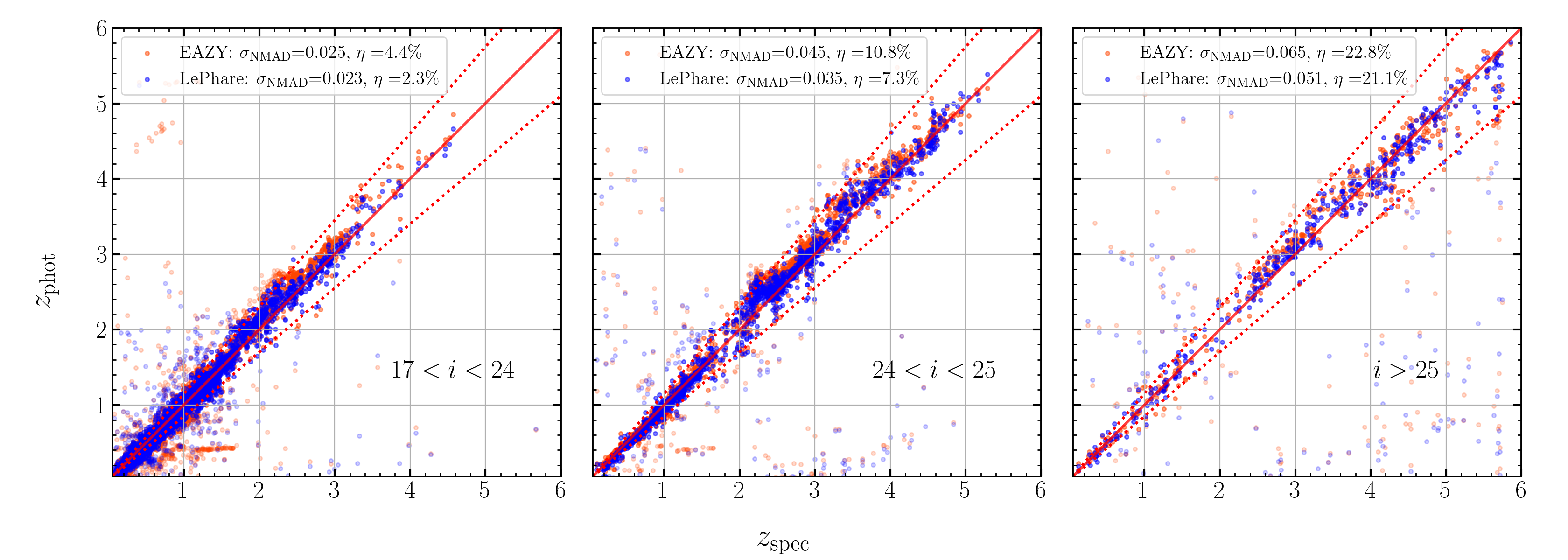}
    \includegraphics[width=\textwidth]{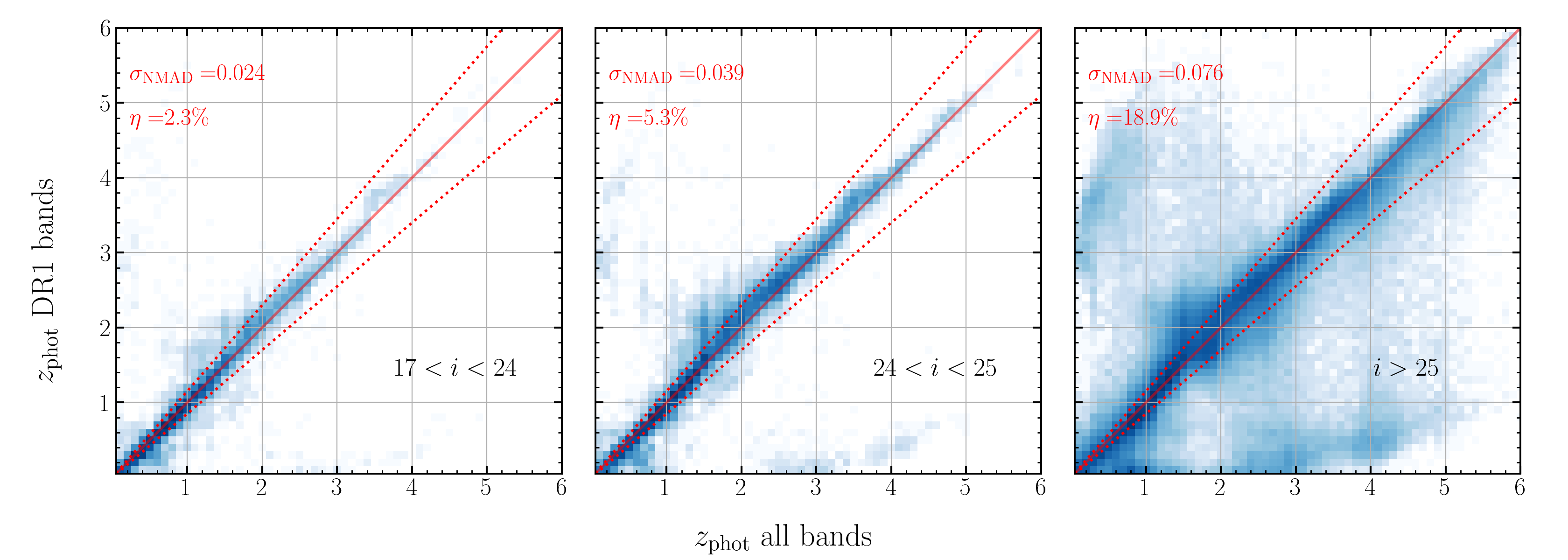}
    \caption{Photometric redshifts measured using the COSMOS2020 catalogue but limited to only the bands included in the present DAWN survey DR1 catalogues, namely, CFHT $u$, HSC $grizy$, and \textit{Spitzer}/IRAC \chOne{} and \chTwo{}. Further, the photometric uncertainties have been scaled to follow the relations for EDF-N depicted in Fig.~\ref{fig:mag_magerr}. This figure is analogous to Fig.~\ref{fig:zphot_zspec} except that the bottom panel compares only photometric redshifts measured with \lephare{}; the $x$-axis indicates the \photoz{}s measured with all forty bands of the COSMOS2020 catalogue, while the $y$-axis indicates the \photoz{}s measured with only the DAWN survey DR1 subset.
    }
    \label{fig:COSMOSzphot_zspec}
\end{figure*}

The validation of stellar masses presented in Sect.~\ref{subsec:validation} is supplemented here by a brief discussion of photometric redshifts (\photoz{}s). Correctly measuring the stellar mass of a galaxy requires first correctly measuring its redshift; changes in the redshift will be propagated to changes in the stellar mass. Thus, in order to compare stellar masses measured with photometry from the bandpasses included in the present DAWN survey DR1 catalogue, the \photoz{}s must first be measured. Measuring \photoz{}s for the modified COSMOS2020 catalogue follows the procedure for \lephare{} described in Sect.~\ref{subsec:lephare} and the procedure for \eazy{} described in Sect.~\ref{subsec:eazy}. Put differently, the methods for calculating \photoz{} are exactly the same as for the DAWN survey DR1 catalogue, but the input catalogue is changed to the modified COSMOS2020 catalogue containing only photometry measured in the DAWN survey DR1 bandpasses and with inflated flux errors corresponding to the measured relationship between magnitude and magnitude error in the  DAWN survey DR1 catalogues (i.e., Fig.~\ref{fig:mag_magerr}). The top panel of Fig.~\ref{fig:COSMOSzphot_zspec} presents a comparison between \photoz{}s measured with the two codes and \specz{}s. The bottom panel of Fig.~\ref{fig:COSMOSzphot_zspec} presents a comparison between \photoz{}s measured with \lephare{} using the modified COSMOS2020 catalogue and the original \photoz{} measurements presented in \cite{Weaver2021} using all forty bands. The agreement between \photoz{}s and \specz{}s is strong across both \eazy{} and \lephare{}, and the performance of both code agrees well with the comparison presented in Fig.~\ref{fig:zphot_zspec}. This provides a further confirmation that redshifts are generally well-constrained by the wavelength range spanned and sampled by the DAWN survey DR1 catalogues. In addition, considering the entire COSMOS2020 catalogue, the agreement between \photoz{}s measured with \lephare{} and only the DAWN survey DR1 catalogue filters compared to those of \cite{Weaver2021} is likewise strong, roughly mimicking the agreement between \photoz{}s and \specz{}s. At magnitudes HSC $i < 25$ the fraction of galaxies with correct redshifts is nearly 95\%. Even at HSC $i > 25$, more than 80\% of all galaxies agree in their \photoz{}s when they are measured with the subset of bands available the DAWN survey DR1 catalogues. 

Although, the quality of the \photoz{} estimates presented herein are generally strong, users of the DAWN survey DR1 catalogues may benefit from consideration of where the \photoz{} estimates are weaker. A nearly universal consequence of estimating \photoz{}s is the degradation of the performance for both the brightest and the faintest objects. The former, with HSC $i \sim $17--18 are predominantly at $z < 1$, and their brightness results in an exceedingly large signal-to-noise such that even small differences in the predicted photometry from the best-fit template result in large $\chi^{2}$ value. Galaxies belonging to the latter group, with HSC $i >> 25$, may be well-fit by many templates and hence their redshifts become more uncertain. Another nearly ubiquitous weakness of \photoz estimatation are the galaxies for which the Balmer and Lyman breaks are not easily distinguished from the photometry. Such a situation arises when there is not sufficient signal to reliably constrain the rest-frame wavelength of both the Balmer and Lyman breaks simultaneously and is consequently more common for fainter galaxies. However, as demonstrated by \cite{Weaver2021} (e.g., Sect. 5.3 and Fig. 14), deep photometry measured in some forty bands from FUV to mid-infrared wavelengths is not enough to circumvent this problem. 

One area of weakness unique to the current DAWN survey catalogs is the redshift range of approximately $1.5 < z < 2.6$. At these redshifts, the available filters do not sufficiently constrain either the Balmer break or the Lyman break. As illustrated by the lower row of Fig.~\ref{fig:COSMOSzphot_zspec}, an increased scatter is observed therein, primarily towards higher redshifts. The galaxies that appear to scatter from their true redshifts (or at least the redshifts reported by \cite{Weaver2021}) have weak UV/optical colours (e.g., $u-g$, $g-r$, $r-i$, $i-z$) but are red when comparing HSC $z$ to \textit{Spitzer}/IRAC \chOne{}. Accordingly, without a constraint on either of the two breaks, higher-redshift solutions are generally allowed because many solutions exist that are consistent with such colours. On the other hand, fewer lower-redshift solutions are allowed, as otherwise the Balmer break would have been observed thus producing a stronger UV/optical colour. It is not until $z \sim 2.6$ that the observed-frame wavelength of the Lyman break exceeds 10\% transmission in the CFHT $u$ band and the scatter decreases. Users of the DAWN survey DR1 catalogues should be aware of the uncertain redshifts for such galaxies. However, the imminent inclusion of the \Euclid NIR photometry will significantly improve \photoz{} estimation in this regime by constraining the Balmer break after it drops out of the HSC $z$ band, at least until $z \sim 4$.

\end{appendix}

\end{document}